\documentclass[a4paper]{ar-1col}
\usepackage{amsmath, amsfonts, amssymb}
\usepackage[comma]{natbib}
\usepackage{url}
\usepackage{graphicx}
\usepackage{xcolor}
\usepackage{txfonts}

\DeclareRobustCommand{\ion}[2]{%
\relax\ifmmode
\ifx\testbx\f@series
{\mathbf{#1\,\sc{#2}}}\else
{\mathrm{#1\,\sc{#2}}}\fi
\else\textup{#1\,{\mdseries\textsc{#2}}}%
\fi}

\DeclareRobustCommand{\ION}[2]{%
\relax\ifmmode
\ifx\testbx\f@series
{\mathbf{#1\,\mathsc{#2}}}\else
{\mathrm{#1\,\mathsc{#2}}}\fi
\else\textup{#1\,{\mdseries\textsc{#2}}}%
\fi}

\newcommand{\hii}{\ION{H}{ii}}

\newcommand{\HII}{\ion{H}{ii}}

\jname{Xxxx. Xxx. Xxx. Xxx.}
\jvol{AA}
\jyear{YYYY}
\doi{10.1146/((please add article doi))}

\begin{document}

\markboth{S.F.S\'anchez}{Properties of SFGs}

\title{Spatially-Resolved Spectroscopic Properties of Low-Redshift Star-Forming Galaxies}

\author{Sebasti\'an F. S\'anchez$^1$ 
\affil{$^1$Instituto de Astronom\'ia, Universidad Nacional Aut\'onoma de  M\'exico, A.~P. 70-264, C.P. 04510, M\'exico, D.F., Mexico; email: sfsanchez@astro.unam.mx}}

\begin{abstract}
  I review here the spatially-resolved spectroscopic properties of
  low-redshift star-forming galaxies (and their retired
  counter-parts), using results from the most recent Integral Field
  Spectroscopy galaxy surveys. First, I briefly summarise the global
  spectroscopic properties of these galaxies, discussing the main
  ionization processes, and the global relations described between the
  star-formation rates, oxygen abundances, and average properties of
  their stellar populations (age and metallicity) with the stellar
  mass. Second, I present the local distribution of the ionizing
  processes, down to kiloparsec scales, and I show how the global
  scaling relations found between integrated parameters (like the
  star-formation main sequence, mass-metallicity relation and Schmidt-Kennicutt law)
  present local/resolved counter-parts, with the global
  ones being just integrated/average versions of the local ones. I discuss the local/resolved
  star-formation and chemical enrichment histories and their implication
  on the inside-out growth of galaxies. Third, I present
  the radial distributions of the surface densities of the
  properties explored globally, and how they depend on the integrated
  galaxy properties. Finally, I summarise all these results and discuss
  what we have learned from them regarding the evolution of galaxies.  
\end{abstract}

\maketitle

\tableofcontents


\section{INTRODUCTION}
\label{sec:intro}



The formation of stars from gas is one the physical processes that
contributes more deeply to shape our Universe. Without the ignition of
the thermonuclear reactions that power stars it would be impossible to
conceive our physical world. There would be basically no chemical
elements besides hydrogen and helium (and traces of lithium), no complex chemical compounds,
no rocky planets, and no life. Therefore, the study of the conditions
required to form stars is of key importance to understand our
existence. As individual entities the birth of stars is governed by
detailed physical laws, and therefore, given some particular
conditions of gas density, pressure and metallicity, their ignition is
triggered \citep[e.g.][]{bonnell98}.  However, when large stellar
populations are considered
(M$_*$$\sim$10$^4$M$_\odot$), star-formation is regulated by rather
statistical laws that correlate the main observed properties, like the
neutral gas density and the star-formation rate (SFR), averaged over
hundreds of parsecs or kiloparsecs scales across the galaxies
\citep{kennicutt1998}.  In this context it is relevant to understand
whether these laws are the same for all galaxy types or they depend
either on local (kilo-parsec scales or the scale of molecular clouds)
or general properties of the galaxies in which star-formation happens.

It is evident that all galaxies formed stars in the past and in
general the star-formation rate was higher in the past both in
individual galaxies \citep[e.g.][]{heavens04,speagle14} and averaged
over the entire population \citep[e.g.][]{madau14}. However, at low redshift and
at least up to
z$\sim$1 galaxies present a clear bimodality regarding their
SFR: (i) those that are more actively forming stars
(defined broadly as star-forming galaxies, SFGs), and (ii) those that
present little or no star-formation activity (defined broadly as
passive, retired or quenched, RGs)
\citep[e.g.][]{sta08}. \begin{marginnote}[] \entry{SFGs}{Star forming
    galaxies: those that are more actively forming stars.}
  \entry{RGs}{Retired galaxies: those that present little or no
    star-formation activity.}
\end{marginnote} These two groups are well separated in different diagrams that compare
either the current SFR or proxies of the age of the
stellar population with the integrated stellar mass or absolute
magnitude, like the SFR vs. Mass diagram
\citep[SFR-M$_*$, e.g.][]{renzini15}, the Color-Magnitude diagram
\citep[e.g.][]{bell04}, or the D4000-Mass diagram \citep[e.g.][ where
D4000 is an age tracer]{blanton09}. Even more, this bimodality is
strongly correlated with the morphological, structural and dynamical
properties of the galaxies
\citep[e.g.][]{blanton09,drory07,graham18}. In general, star-forming
galaxies are late-type, disk dominated \citep[e.g.][]{brin04}, while
retired galaxies are more early-type, bulge dominated
\citep[e.g.][]{drory07}. Even more, we know that the mass of
early-type galaxies at low redshift have grown at least a factor two
in the last
$\sim$8 Gyrs \citep[e.g.][]{bell06}, although they show little or no
significant star-formation during the same time period
\citep[e.g.][]{lopfer18,sanchez18b}. This implies that there should be
a transition of galaxies from the star-forming to the retired
galaxy population. This event may last only a few Gyr, and it involves a profound
morphological and structural tranformation of those galaxies, as well
as a dramatic ageing of their average stellar populations. Broadly
speaking, this process is known as quenching or rapid halting of the
star-formation activity \citep[e.g.][]{casado15,saint16}. Beside that
dramatic process, current star-forming galaxies have steadily
decreased their SFR along the same time period, showing slightly older
stellar populations on average, a smooth proccess generally known as
aging \citep[e.g.][]{casado15}.

\begin{marginnote}[]
\entry{ISM}{Interstellar Medium: gas and dust within a galaxy distributed among and beyond stars.}
\end{marginnote}
Our understanding of nearby galaxies and their evolution has
deeply improved since the advent of large imaging surveys,
complemented for a lower number of objects with single-fiber
spectroscopy \citep[e.g., Sloan Digital Sky Survey, SDSS, Galaxy and
Mass Assembly survey, GAMA,][, respectively]{york2000,gamma}. These
surveys provide us with photometry and multi-band imaging of millions
of galaxies, and spectroscopy of hundreds of thousands of them.  The
main results produced by these massive datasets were reviewed by
\citet{blanton09}, describing in detail the physical properties of
these galaxies, including their global distributions in luminosity,
stellar and atomic gas and their corresponding mass/luminosity
functions, for different environments.  Despite the huge amount of
information and the step forward introduced by these surveys, they had
a severe limitation: they do not provide resolved spectroscopic
information.  Galaxies have long been known to be spatially extended
objects, with observed properties that vary across their optical
extent \citep[e.g.~][]{hubble26,hubble36}. Many of these properties
vary systematically as a function of position relative to the galaxy
center, and radial gradients have been studied for decades in both the
gas and stellar population content
\citep[e.g.~][]{pagel81,rene89}. Therefore, to characterize galaxies
by single aperture spectra or spatially resolved multi-band photometry
imposes severe limitations in both the description of their properties
and the understanding of the evolution that shaped them.

This is particularly important in: (i) the study of the variation of
the physical properties of the Interstellar Medium (ISM); (ii) the
detailed understanding of the composition of the stellar populations
across the optical extension of galaxies; and (iii) for the analysis
of the kinematic properties and dynamical state of galaxies. To
address properly these issues, spatially resolved spectroscopic
information is required, covering a substantial fraction of the
optical extension of galaxies and for a large, well defined,
statistically significant sample, that cover the widest range of
galaxy populations in different properties (e.g., masses, colors,
morphologies...). The advent of wide-field and multiplexed Integral
Field Units (IFUs) in the last decade have allowed us to perform these
studies in an efficient way, making it possible to observe large
samples of galaxies, and leading to the first generation of Integral
Field Spectroscopy (IFS) galaxy surveys (IFS-GS).
\begin{marginnote}[]
\entry{IFU}{Integral Field Unit: a device attached to an spectrograph that allows to obtain several spectra simultaneously of different contiguous locations in the sky.}
\entry{IFS}{Integral Field Spectroscopy: technique that allows to obtain spectra using an IFU.}
\entry{IFS-GS}{Integral Field Spectroscopy Galaxy Survey: Large program to acquire IFS data over a well defined and statistically representative sample of galaxies.}
\end{marginnote}

Along this review, I will summarize our current knowledge of the spatially
resolved spectroscopic properties of low-redshift galaxies based
mostly on the results of more recent IFS-GS. I will focus on
star-forming galaxies, although I will also present the properties of their
retired counter-parts too, as a necessity to discuss our knowledge about
what triggers and halts star-formation. This review is therefore
focused on the cycle of birth, evolution and death of stars from a
statistical point of view. I will summarize the main results on the distributions of
stellar populations, properties of the ionized ISM,
and their interconnections. First, I will present the main global
properties of galaxies, focused on those aspects, to present later the
spatial distributions of the same properties. Finally I will present
the local scaling laws that we think rule the star-formation and
enrichment processes in galaxies. I do not address in this review the
morphological and photometric properties of galaxies, or any results
that could be derived using only photometric or integrated
spectroscopic data \citep[see][ for a review on the
topic]{blanton09}. The kinematic properties and the dynamical stage of
galaxies is not addressed neither \citep[see][ for a review on the
topic]{cappellari16}. For a more detailed exploration of the emission line
properties of galaxies I refer the reader to the recent review by \citet{kewley19}.

\section{DATA}
\label{sec:data}

As indicated before, the results shown in this review are extracted
from the most recent published analysis on IFS-GS data. Instead of
using published figures, along this review I recreate the results
based on a compilation of IFS datasets created by combining the data
from four different surveys: AMUSING++ \citep{galbany16b}, CALIFA
\citep{sanchez12}, MaNGA \citep{manga}, and SAMI \citep{sami}. It is
beyond the scope of this manuscript to review in detail the different
IFS-GS. Nevertheless, I present a summary of their properties in
Appendix \ref{sec:IFS}, describing the number of galaxies sampled by
each survey and the differences in spatial and spectral
resolution.
Despite the big
differences in the nature of these surveys, in particular in terms of
the covered area  for the sampled galaxies and the spatial and spectral
resolution, all of them provide with spatially resolved spectroscopic
information for statistical significant and large samples of galaxies
mostly located at $z\sim$0.01-0.06. All together the adopted
compilation comprises 8,234 galaxies. That is, so far, the largest
compilation of IFS data. A summary of the properties of this
compilation is included in Appendix \ref{sec:IFS}, and a full
description will be given elsewhere (S\'anchez et al., in prep.).
Along the review I will use this full sample to describe global
properties, when required. To present the spatial resolved or radial
properties I selected a subset of this sample comprising only those
galaxies/datacubes with stellar masses within 10$^6$ and 10$^{13}$
M$_\odot$, and a redshift range between 0.005$<z<$0.05, and fulfilling
the following criteria: (i) the IFS data cover at least 1.5 R$_e$ of the
galaxy; (ii) the galaxy is well resolved, i.e., the effective radius
(R$_e$) is larger than 5 arcsec (since most of the IFS-GS have a typical spatial resolution of FWHM$\sim$2.5 arcsec) ;
\begin{marginnote}[]
\entry{R$_e$}{Effective radius: galactocentric distance at which it is encircled half of the total the flux intensity in a certain photometry band of a galaxy.}
\end{marginnote}
(iii) it is not highly inclined (i.e., ellipticity, $\epsilon <$0.75);
(iv) the FoV of the IFS data covers at least 25 arcsec projected in
the sky (i.e., sampling at least 2.5 R$_e$). This sub-sample comprises
1,516 galaxies with the best compromise between spatial resolution and
sampled IFS data. To homogenize as much as possible the dataset I will
make use of the dataproducts provided by the same analysis pipeline,
{\sc Pipe3D} \citep{pipe3d,pipe3d_ii}.  This pipeline extract the main
spectroscopic properties of the galaxies, both integrated, radial and
spaxel-wise, including: (i) the composition and kinematic properties
of the stellar population, providing with the star-formation and
chemical enrichment histories, and main average/integrated properties
(stellar mass, age, [Z/H], dust attenuation), together with the
velocity and velocity dispersion; (ii) the main properties of a large
number of emission lines within the considered wavelength range (flux,
velocity, velocity dispersion and equivalent width); (iii) a set of
high order physical parameters, like the ISM oxygen abundance derived
using different calibrators, the dust attenuation, estimations of
molecular gas content, etc. Full details of the delivered products are
provided in some recent articles
\citep[e.g.][]{ibarra16,sanchez18,ibarra19}. {\sc Pipe3D} is just one
of the different pipeline/tools developed in the last years with the
goal of analysing IFS data \citep[e.g., PyCASSO, LZIFU, MaNGA
DAP,][]{amorim17,ho16,dap}. I adopted it just to homogenize the
dataproducts included along this review. In most cases this pipeline
produces the same or very similar results to the ones published in the
reviewed articles (based in many cases on the tools listed before or
similar ones). In case of significant differences or lack of consensus
I tried to highlight them as much as possible.




\section{Global properties of Galaxies}
\label{sec:global}

As indicated in the introduction, recent reviews have dealt with
the description of the current knowledge of the integrated/global properties
of galaxies in the near universe, mostly based on single aperture
spectroscopic data combined with broad-band imaging photometry \citep[e.g.][]{blanton09}.
In this section I summarise the main integrated properties and relations
that are relevant for the main scope of this review, i.e., the local/resolved
properties and relations.

\subsection{What ionizes gas in galaxies?} 
\label{sec:diag_gal}

\begin{figure}[h]
\includegraphics[trim=5 15 3 30,clip,width=5in]{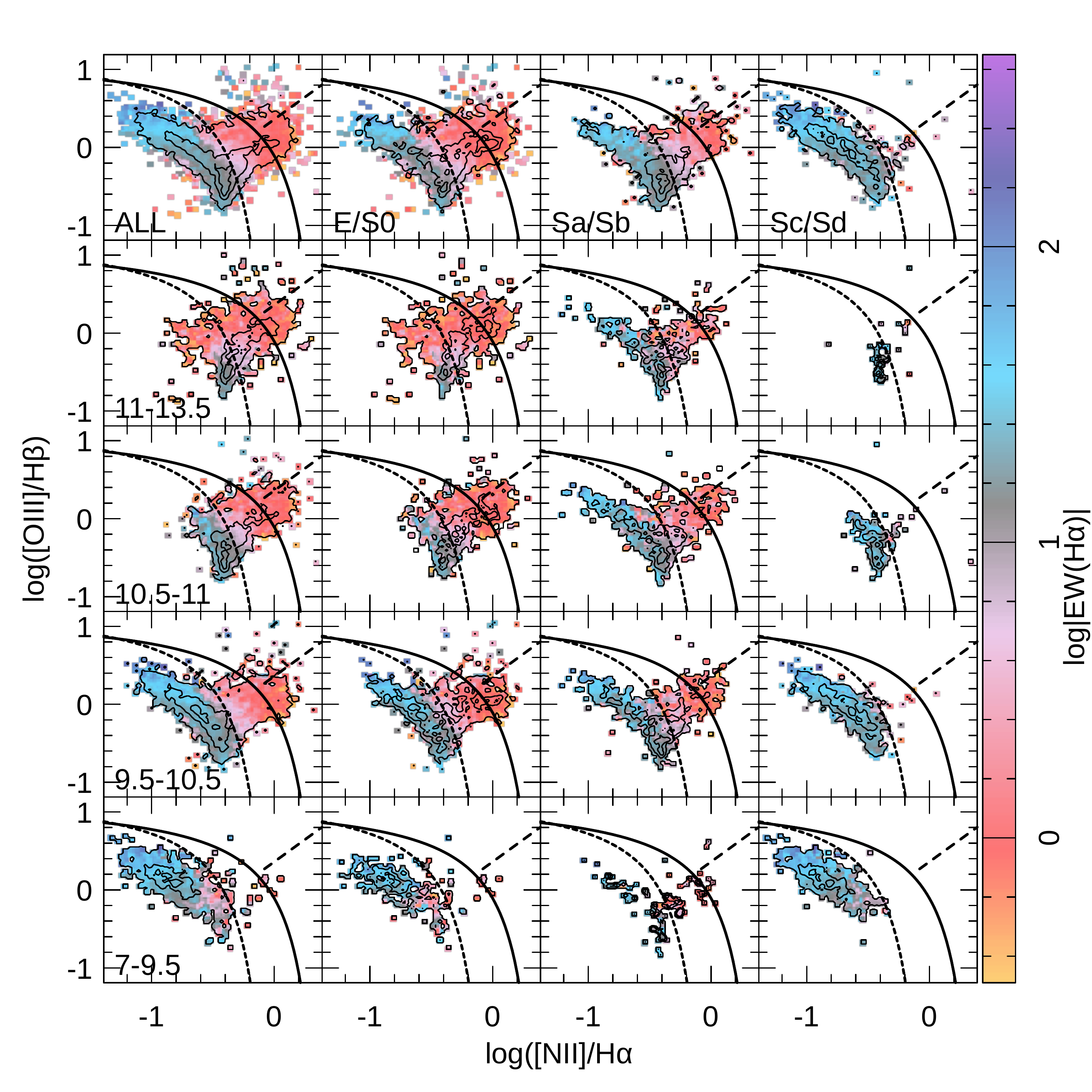}
\caption{Distribution along the classical [\ION{O}{iii}/H$\beta$] vs. [\ION{N}{ii}]/H$\alpha$ diagnostic diagram, where each of the 8,234 galaxies included in the adopted compilation contributes as a single point. Color coded is indicated the average equivalent with of H$\alpha$ across the field-of-view of each datacube. Contours represent the density of galaxies, with each contour including a 95\%, 50\% and 10\% of the points. In all panels, the solid-line represents the location of the \citet{kewley01} boundary line, with the separation between Seyferts and LINERs is indicated with a dashed-line. Finally, in the dotted line represents the location of the \citet{kauff03} demarcation line. Each panel represents a sub-sample of galaxies: the full sample is shown in the upper-left panel, and each panel shows a sub-sample of galaxies segregated by morphology (from left to right) and stellar mass (from top to bottom), with the covered ranges labeled in top- and left-most panels.}
\label{fig:BPT_mass}
\end{figure}

The ISM in galaxies is observed in the optical range mostly by the
emission lines produced either by recombination or by radiative
desexcitation of collisionally excited levels. To produce emission
lines, the gas should be first ionized by high energy photons (of at
least 13,6 eV) emitted either by stars (young and old) or active
galactic nuclei \citep[e.g.][]{osterbrock89}, or generated by the
energy dissipation in a shock front travelling through the ISM
\citep[e.g.][]{veilleux87,veilleux2005}. All the four different
sources of ionization may be present in a single galaxy: (i) regions
ionized by young OB stars residing in \ion{H}{ii} regions
\citep{strom39}, associated with star-forming areas (like spiral arms)
\citep[e.g.][]{baldwin81}; (ii) regions ionized by old stars,
post-AGBs or HOLMES \citep[e.g.][]{binette94,flor11}, associated with
retired regions in galaxies \citep[e.g.][]{sign13,gomes15};
\begin{marginnote}[]
\entry{post-AGB}{Final evolution period of low- and intermediate-mass stars is a rapid transition from the Asymptotic Giant Branch towards the white dwarf phase.}
\entry{HOLMES}{Hot low-mass evolved stars, mostly in the post-AGB or white dwarf phase.}
\end{marginnote}
(iii) AGN ionization, mostly concentrated in the central regions
\citep[e.g.][, although the extension depends on the AGN
luminosity]{husemann10,husemann14}; and (iv) shock ionization, either
driven by AGN or star-formation outflows
\citep[e.g.][]{heckman90,bland95,carlos16}, or driven by low density
gas \citep[e.g.][]{dopita96,kehrig12,cheung18}. Since the source of
ionization is always local, and not global, to describe the ionized
gas of a galaxy by average or integrated properties it is always
misleading. A galaxy is neither a star-forming galaxy, a retired galaxy, an AGN or
an outflow. A galaxy may present, locally, star-forming and/or
post-AGB ionization, and it may host an AGN and enhibits outflows which are able
to ionize the gas via shocks.  Therefore, the classification for
source of ionization in a galaxy should be local, and may require
spatially resolved information of the emission line
diagnostics. Moreover, as postulated by different authors
\citep[e.g.][]{sta08,cid-fernandes10,papa13,sanchez14,lacerda18}, it
may require some knowledge of the properties of the underlying stellar
population, and even the spatial distribution of the ionized gas
structure \citep[e.g.][]{carlos19}.  Based on all those results I
present a practical scheme to classify the ionization in Appendix
\ref{sec:class}.


Figure \ref{fig:BPT_mass} shows the BPT diagram \citep[][]{baldwin81}
for the global [\ION{O}{iii}/H$\beta$] and [\ION{N}{ii}]/H$\alpha$
line ratios, averaged along the entire field-of-view of the IFU data
for each galaxy. The plot is color coded by the average EW(H$\alpha$)
of each galaxy. I present the distribution for the full compilation of
galaxies adopted in this review, segregated by both mass and
morphology in the different panels. Thus, each galaxy contributes as a
single point in each panel, and it is characterized by a single pair
of the considered line ratios \citep[e.g.][]{kauff03}. In
addition I include the classical demarcation lines that are usually
adopted to classify the main ionizing processes
\citep{kewley01,kauff03}. Line ratios at the left-side of the diagram,
with high EW(H$\alpha$) are usually associated with ionization by
young stars, and therefore related to star-formation processes (\HII\
regions). On the contrary, the right-side of the diagram (above the
considered demarcation lines), is associated with hard ionization that
could be due to AGNs or shocks (high EWs), or ionization by old stars
(e.g., post-AGBs, with low EWs).

Although it is clear that ionization has a local nature, there are clear trends
between the global properties of galaxies (e.g., stellar mass,
morphology...) and the dominant ionizing source
\citep[e.g.][]{kauff03}. This is clearly appreciated in
Fig. \ref{fig:BPT_mass}: more massive/early-type galaxies show a
stronger contribution of ionization due to old stars, while less
massive/later-type galaxies have a stronger presence of ionization due
to young stars. This reflects the known evolution of galaxies, since
more massive/early-type galaxies formed stars earlier in the
cosmological time \citep[e.g.][]{perezgonzalez08,thomas10,rosa14}, and
therefore they have a larger fraction of old stars
\citep[e.g.][]{rosa16}.  Finally, AGNs are more frequently found in
massive galaxies, either early-type \citep[e.g.][]{kauff03} or early
spirals \citep[e.g.][]{schawinski+2010,sanchez18}. A more detailed
discussion on the ionization processes in galaxies and the use of
emission lines to explore their evolution was recently presented by
\citet{kewley19}.



\subsection{SFR-M$_*$ diagram and the SF-law}
\label{sec:SFMS}

\begin{figure}[h]
  \includegraphics[width=5in,trim=15 1 10 5,clip]{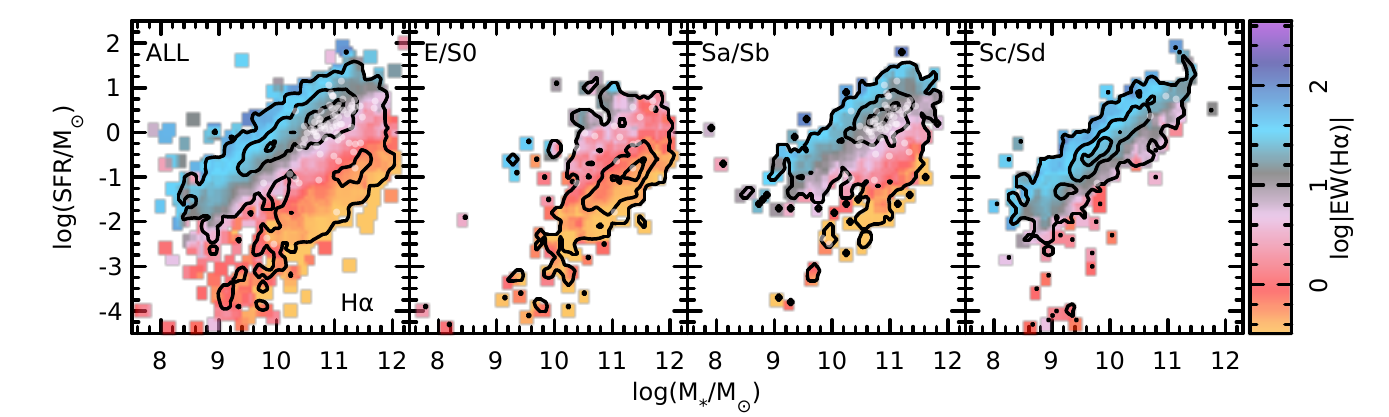}
\caption{SFR-M$_*$ diagram for all the galaxies included in the IFS compilation presented along this review (left panel), and in different sub-groups segregated by morphology, from more early-types to more late-types, from left to right. All panels represent the distribution adopting the SFR estimated based on the dust corrected integrated the H$\alpha$ luminosity following \citet{sanchez18}.  In each panel, the contours corresponds to the density of points, with each contour including a 95\%, 50\% and 10\% of the points. In all panels, the logarithm of the equivalent with of H$\alpha$ is color coded, following the same code previously shown in Fig. \ref{fig:BPT_mass}. Finally, the white circles in each panel, show the location of the optically selected AGNs \citep[as selected in ][too]{sanchez18}.}
\label{fig:SFMS}
\end{figure}

Star-forming galaxies (SFGs) are those in which the main ionizing
source is the vast presence of OB-stars, directly connected with
recent star-formation due to their short
life-time \citep[e.g.,][]{Pozzetti+2010}. On the contrary, retired
galaxies (RGs) are those with a lack of ionized gas or with gas
ionized by old stars, in which there is little \citep[e.g.][]{gomes16}
or no star-formation \citep[e.g.][]{sta08,cid-fernandes10}. In
general, there are no galaxy (or very few) where the ionization is
dominated by shocks induced by outflows. An interesting counter
example is the case the filamentary or biconical ionized gas detected
in weak AGNs and radio-galaxies \citep[already observed using narrow
band images in cD galaxies decades ago,
e.g. M87,][]{jarvis90}. \citet{kehrig12}, using CALIFA data, first
proposed that the ionization in these structures (and indeed across
all the optical extension of these galaxies) is dominated by
shocks. \citet{cheung18} proposed them as a new type of objects (named
Red Geysers), and their connection with radio-sources was revisited
recently by \citet{roy18}.  However, those galaxies comprises less
than a 4\%\ of the total elliptical ones (i.e., less than 1\% of the
total galaxy population), as shown by Lopez-Coba et al. in prep.

On the other hand, AGN ionization may dominate the
integrated properties of the emission lines in  galaxies, in particular at the central
regions. Thus, some galaxies are classified as AGNs, although
AGN hosts is a better term \citep[e.g.][]{kauff03}.
When the SFR derived using the dust-corrected H$\alpha$ luminosity of
a galaxy \citep[or any other SFR estimator like UV or FIR luminosities, e.g.,][]{catalan15} is plotted
along the stellar mass (M$_*$), the two populations of SFGs and RGs are
clearly distinguished \citep[and indeed well separated by the average value of the  EW(H$\alpha$), with 3-6 \AA\ being a clear boundary, e.g.][]{sta08}.
This is clearly appreciated in Figure \ref{fig:SFMS}, where I present
the SFR-M$_*$ diagram for the galaxies in the current compilation.

The distribution of SFGs is well characterized by a log-linear, tight
relation, observed between the SFR and M$_*$. This relation, known as
the star-formation main sequence (SFMS) of galaxies has been well
studied using both single-aperture and IFS data at low-redshift
\citep[e.g.][]{brin04,Daddi07,Elbaz07,Noeske07,gavazzi15,renzini15,catalan15,mariana16,sanchez18}. In
general, it has a scatter of $\sim$0.2-0.3 dex in SFR at any redshift,
with a slope slightly smaller than one at low-redshift, $\sim$0.8
dex/log(M$_*$). It presents a clear evolution in the zero-point that
increases with redshift \citep[e.g.][]{speagle14,rodriguez16}.  On the
other hand, RGs are distributed well below the SFMS, following either
a loose relation with the M$_*$ (or a cloud), broadly corresponding to
the location expected for an EW(H$\alpha$)$\sim$1\AA\
\citep[e.g.][]{sta08}. In between the two groups there is an area with
much lower number density of galaxies, known as the green valley (GV).
Galaxies in the GV (or GVGs) are considered in transit between SFGs
and RGs, and their limited number has been interpreted as a
consequence of a fast process transforming the former to the later
ones \citep[e.g.,][]{bell04,Faber+2007,Schiminovich+2007}.  More
recently, it was found that AGN hosts are also located in the GV of
the SFR-M$_*$ diagram
\citep[e.g.][]{torres-papaqui12,schawinski+2014,mariana16,sanchez18},
as appreciated in Fig. \ref{fig:SFMS}. Actually, their distribution is
spread from the SFMS towards the RGs cloud, covering mostly the GV,
and sharing many properties with GVGs \citep[e.g.][]{sanchez18}. This
result suggests that there is a connection (or co-evolution) between
the AGN activity and the quenching of star-formation in galaxies.




The key ingredient of star-formation is cold gas, from which stars are
formed \citep[see e.g.,][]{kennicutt12,krumholz+2012}. Indeed,
\citet{schmidt59} already suggested a relation between the
SFR and the interstellar gas volume density. This
relation was later observed as a relation between the surface
densities of both quantities, and it is known as the Schmidt-Kennicutt
or star-formation law \citep{kennicutt98}. This relation is maintained
at kpc-scales only for the molecular gas \citep[e.g.,][and references
therein]{Kennicutt07,bigiel08,leroy13}. It is beyond the scope of this
review to explore in detail the global star-formation law and the
detailed connection of cold gas with the star-formation
process. However, for the sake of understanding the local processes
described later, I should indicate that RGs present a clear deficit
of both atomic and molecular gas
\citep[e.g.][]{saint16,calette18}. Thus, the lack of cold gas seems to
be the primary reason for the limited SFR (if any).  How this deficit
of gas is connected to the presence of an AGN is a topic of study,
being both heating and removal the main explanations invoked through
the so-called negative feedback \citep[e.g.][]{silk97}. Indeed, there
are other possible sources of low gas content, related with gas interactions, harassment
and stripping, more related with externals processes \citep[e.g.][]{GASP}.

However, even for the same amount of cold gas, galaxies may present a
different star-formation efficiency, SFE=SFR/M$_{gas}$,
\citep[e.g.][]{saintonge2011} or different scaling factors between the
two parameters involved in the star-formation law
\citep[e.g.][]{elmegreen07}. Indeed, \citet{colombo18} show that the
SF is less efficient in more early-type galaxies beside the fact that
they have less amount of molecular gas. In summary, the lack of gas
is the primary driver for the halting of SF, but a secondary cause
involves a decline in the SFE.

\subsection{Gas phase Mass-Metallicity relation}
\label{sec:MZR}

\begin{figure}[h]
\includegraphics[width=5in,trim=15 1 10 5,clip]{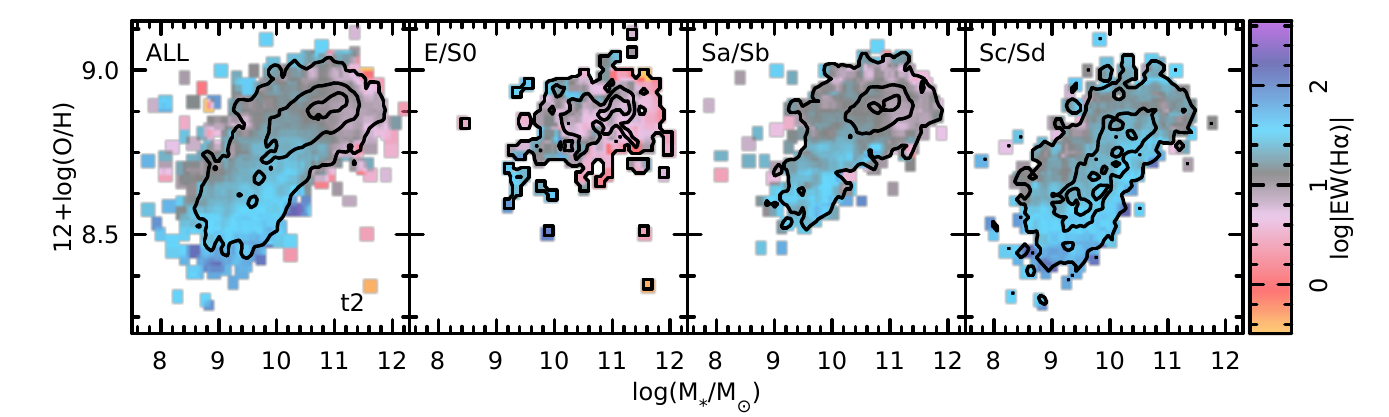}
\caption{Gas-phase mass-metallicity relation for the collection of IFS observed galaxies adopted along this review with abundances measured at the effective radius ($\sim$4600 galaxies). All panels correspond to the MZR derived using the $t2$ calibrator \citep[defined in][]{sanchez19}. The left-most panels show the distribution for all the galaxies, with similar distributions segregated by morphology shown in the different panels, from earlier types to later ones from left to right. Like in Fig. \ref{fig:SFMS}, the distributions are shown as density contours, with each contour including a 95\%, 50\% and 10\% of the points. Finally, the color indicates the mean value of the EW(H$\alpha$) for all galaxies contributing to each point in the distribution, following the same code previously shown in Fig. \ref{fig:BPT_mass}.}
\label{fig:M_OH_morph}
\end{figure}

Metals are produced as a result of the thermonuclear reactions that
make the stars shine, and during their super-novae explosion phase
(when it happens).  After their death \citep[and along their life-time
too, e.g.][]{yate12}, metals are expelled to the ISM, polluting it,
and enriching the next generations of stars. Therefore, the metal
content in the ISM is a tracer of the previous generations of stars,
modulated by the gas inflow, outflow and depletion processes. Among
the different metals, oxygen is particularly interesting, being the
most abundant one \citep[and the most frequent element after H and He,
e.g.,][]{peim07}. Being an $\alpha$-element, it is expelled mostly by
core-collapsed supernovae generated by the death of short-lived
massive stars, and therefore its relative enrichment is tightly
related to last star-formation events. As a consequence it is both a
tracer of the overall chemical evolution of galaxies, being a
by-product of SF, and a local proxy of the differential SF activity.

Averaged across the optical extension of a galaxy, the oxygen
abundance correlates with the integrated stellar mass. This relation,
known as the MZ-relation or MZR (although a more correct name should
be the M$_*$-O/H relation), connects the two main products of
star-formation integrated along the cosmic time. Known for decades
\citep[e.g.][]{vila92}, it was originally expressed also as a
luminosity-O/H relation \citep[e.g.][]{garnett:2002p339}.
\citet{tremonti04} made the first systematic exploration of this
relation using a statistically large and significant sample of
galaxies, showing that the two parameters exhibit a tight relation,
with a dispersion lower than $\sim$0.1 dex, covering more than four
orders of magnitude in M$_*$. More recent analysis reduces this
scatter to $\sim$0.05 dex \citep[e.g.][]{sanchez18}. Figure
\ref{fig:M_OH_morph} shows the distribution along the M$_*$-O/H
diagram for the full sample of galaxies adopted along this review,
using a particular calibrator. The shape of the MZR is clearly
appreciated. A full exploration of the differences reported when using
different calibrators was already discussed by \citet{kewley08}, and
more recently revisited by \citet{bb17}, \citet{sanchez17a} and
\citet{sanchez19}. The MZR presents an almost linear regime for
M$_*<$10$^{10}$M$_\odot$, flattening between
10$^{10}$-10$^{10.5}$M$_\odot$, and reaching a plateau for more
massive galaxies. The extension of the linear regime, i.e., the
location of the M$_*$ knee, and the actual asymptotic value depends
on the adopted calibrator or procedure used to measure the oxygen
abundance \citep[e.g.][]{bb17}, and the Initial Mass Function
\citep[IMF][]{salpeter55} adopted in the derivation of the
M$_*$. However, the shape is almost universal both in the nearby
universe \citep[e.g.][]{kewley08,sanchez17a,sanchez19}, and at
different redshifts \citep[e.g.][]{erb06,erb08,henry13,salim15}.

At low stellar masses the linear relation between the two parameters
is interpreted as a direct consequence of the star-formation history
(SFH) in galaxies, dominated at this regime by a consistent growth of
both parameters as integrals of the SFR (modulated by
the effective yield in the case of O/H). This regime of the relation
indicates that the evolution does not depart too much from the one
predicted by a close-box model \citep[e.g.][]{pily07}. The larger
differences from a close-box model are found in the asymptotic regime,
in which an increase in the stellar mass does not produce a
significant increase in the oxygen abundance. \citet{tremonti04}
already interpreted this flattening as a result of galactic outflows
that regulate the oxygen abundance (by removing metal rich gas from
galaxies). Different studies support this interpretation
\citep[e.g.][]{dave11,lilly13,belf16a,wein17}. An alternative
interpretation is that gas inflow can produce a similar shape for the
MZR, with the asymptotic value being a natural consequence of the
maximum amount of oxygen that can be produced by stars, i.e., the
yield.  \citet{pily07} already showed that the observed asymptotic
value is compatible with the yield predicted by a closed-box model for
a gas fraction f$_{gas}\sim$5-10\%, assuming a value of the oxygen
abundance derived using calibrators anchored to the direct
method. However, recent studies indicate that gas outflows are still
required to reproduce the observed shape \citep[e.g.][]{jkbb18}.

More controversial is the presence of a secondary relation between the
MZR and the SFR (once removed the primary relation between O/H and
M$_*$). \citet{elli08} first reported the existence of this secondary
relation, that latter on has been proposed as (i) a modification of
the stellar mass by a parameter that includes both this mass and the
SFR \citep[e.g.][known as the Fundamental Mass-Metallicity relation or
FMR]{mann10}, (ii) as a fundamental plane involving the three
parameters \citep[e.g.][known as the Mass-Metallicity-SFR Fundamental
Plane]{lara10a}, or (iii) as a dependence of the residuals of the
primarly relation with the SFR or the sSFR \citep[e.g.][]{salim14}. In
most of the cases it is described as a trend in which galaxies with
larger SFR at a fixed mass present a lower metallicity. Both galaxy
inflows and outflows have been proposed to explain the secondary
relation \citep[e.g.][]{salmeida19}.  The existence of this secondary
relation has been questioned by studies based on IFS data
\citep[e.g.][]{sanchez13,sanchez18,bb17,sanchez19}, contrary to most
of the previous ones based on single aperture spectroscopic surveys
(e.g., SDSS). In some cases, the existence or not of the FMR depends
on the interpretation of the data, as shown by the re-analysis of the
IFS data presented by \citet{salim14} and \citet{cresci19}.  It is
beyond the scope of this review to enter in this discussion, that has
been addressed in more detail recently by \citet{maio19}.

\subsection{Age and Metallicity distributions in galaxies}
\label{sec:M_ZH}

It is known that galaxies present a clear bimodal distribution in many
different properties that are likely physically connected. In
particular, they present a clear bimodality in the SFR-M$_*$ diagram
discussed in Sec. \ref{sec:SFMS}, which is directly connected with the
morphological segregation between early- and late-type galaxies
already introduced by \citet{hubble26}. This bimodality has been more
frequently explored using color-magnitude diagrams \citep[CMD,
e.g.][]{strat01,blanton03,bell03}, with early-type/RGs located along a
well defined region known as the {\it red-sequence}, and
late-type/SFGs distributed in the so called {\it blue-cloud} \citep[as
nicely reviewed by][]{blanton09}. A more physically motivated version
of the CMD is the Age-M$_*$ diagram, where galaxies clearly show a
sharp bimodal distribution
\citep[e.g.][]{gallazzi05,gallazzi08,sanchez18}.  This diagram shows
that early-type galaxies (E/S0) are strongly dominated by a very old
stellar population of $\sim$10 Gyr, with a narrow distribution of
ages. This is a direct consequence of a SFH dominated by a strong
burst at early cosmological times
\citep[e.g.][]{panter03,thomas05,rgb17}, frequently modelled with a
single burst of star-formation. On the contrary, late-type spirals
(Sc/Sd) contain a considerable fraction of old stars, of a similar or
slightly lower ages, but with a substantial amount of stars younger
than 1 Gyr, and a much wider range of stellar ages. Again, this
indicates that they present a smoother evolution
\citep[e.g.][]{lopfer16,rgb17,sanchez18b}, and a wider variety of SFHs
\citep[e.g.][]{ibarra16}. As expected, early-type spirals (Sa/Sb)
present a mixed behavior. Their older populations have similar ages to
those of the early-type galaxies. However, the contain a considerable
fraction of young stellar populations too, with ages similar to those
of the late-type spirals.

Contrary to the global stellar ages, stellar metallicity ($[Z/H]_*$)
does not show any clear evidence of a bimodal distribution, showing a
smooth relation with M$_*$
\citep[e.g.][]{gallazzi05,panter08,vale09,rosa14b,sanchez18}. Stellar
metallicity is a by-product of the star-formation process. However,
contrary to the gas phase abundance, which increases with each
generation of new massive stars formed (and death), modulated by
inflows and outflows, stellar metallicity traces the metal content in
the surviving stars (i.e., those of intermediate and low
mass). Therefore, its interpretation and the relation with the stellar
mass is less evident than that of the oxygen abundance. On the other
hand, the derivation of the stellar metallicity depends strongly on
data used (photometry/spectroscopy, wavelength range...) and the
method adopted to explore the stellar populations: full fossil
records, SFHs and Chemical Enrichment Histories (ChEHs) adopted or
not, comparison with single stellar populations, inclusion or not of
the [$\alpha$/Fe] relative abundances
\citep[e.g.][]{walcher11,conroy13}. Despite these caveats the lack of
bimodality indicates that the quenching process does not affect the
metal enrichment significantly, that, in any case, is frozen to the
last episode of SF.


\section{Resolved properties of Galaxies}
\label{sec:resolved}

Along the previous sections I have summarized the main properties of
galaxies derived from integrated or aperture limited spectroscopic
surveys.  In the current section I will review how these properties
are extended to local scales (i.e., $\sim$1 kpc), as uncovered by
more recent IFS Galaxy Surveys.


\begin{figure}[h]
  \includegraphics[width=5in, clip, trim=0 0 0 0]{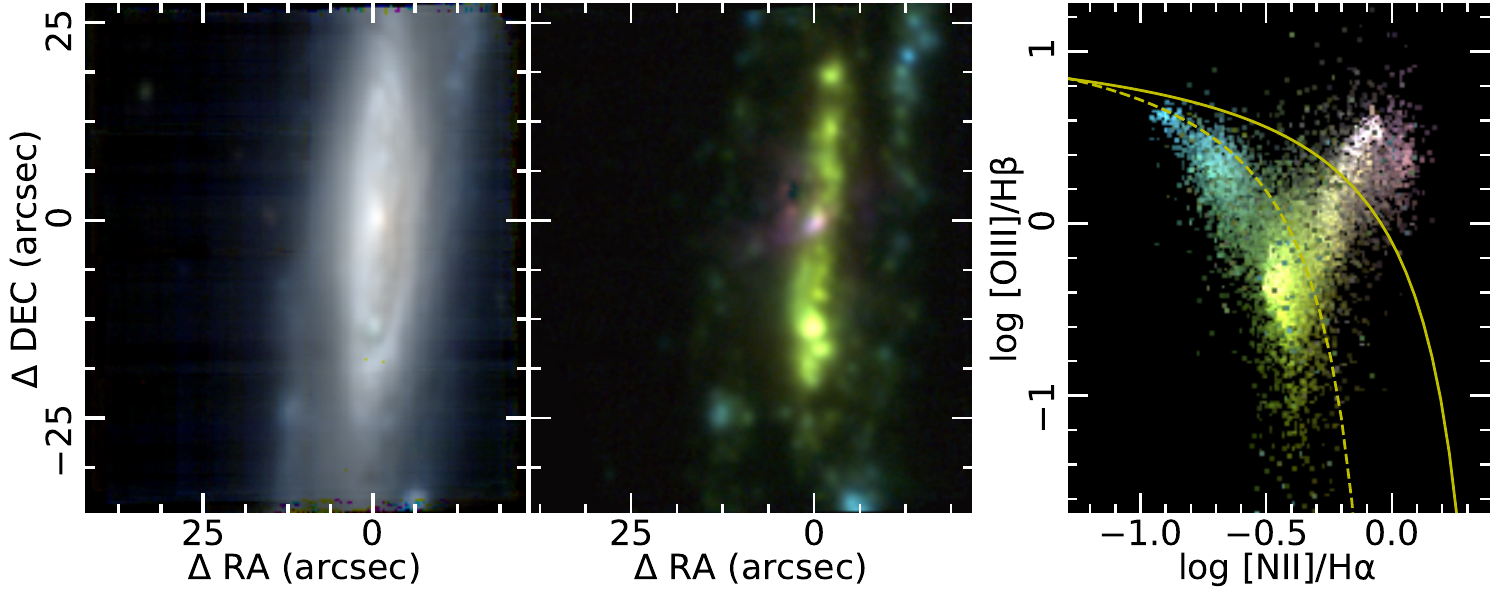}
\caption{Each panel shows, for galaxy IC1657 observed with MUSE
  (PI:R.Bacon): (i) the continuum image created using $g$ (blue),$r$
  (green) and $i$-band (red) images extracted from a MUSE datacube, by
  convolving the the individual spectra with each filter response
  (left-panel), (ii) the emission line image created using the
  [\ION{O}{iii}] (blue), H$\alpha$ (green) and [\ION{N}{ii}]  (red)
  emission line maps extracted from the same datacube using the {\sc
    Pipe3D} pipeline (central-panel), and (iii) the classical BPT
  diagnostic diagram involving the [\ION{O}{iii}]/H$\beta$ and
  [\ION{N}{ii}]/H$\alpha$ line ratios. Each point in this diagram
  corresponds to a single pixel (spaxel) in the other two maps,
  represented with the same color shown in the emission-line image
  (central panel). The solid- and dashed-lines represent the location
  of the \citet{kewley01} and \citet{kauff03} demarcation lines,
  respectively. {\it (credits. C. L\'opez-Cob\'a)}}
\label{fig:bpt_maps}
\end{figure}
\begin{marginnote}[]
  \entry{Ionization is a local process}{all sources of ionizing photons are local,
  not global. A galaxy does not have a single source of ionization.}
\end{marginnote}



\begin{figure}[h]
\includegraphics[width=5in, trim=5 15 1 5,clip]{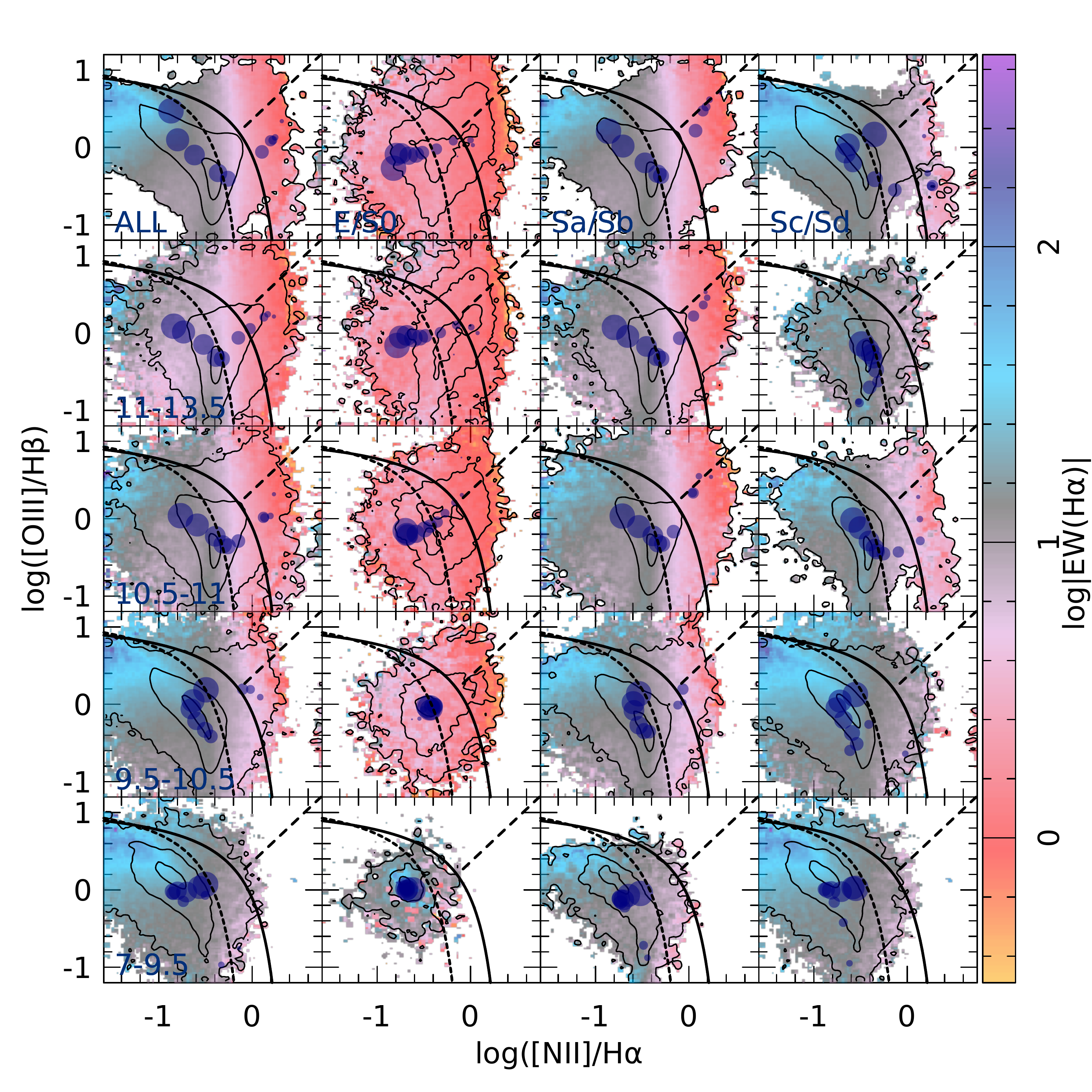}
\caption{Classical [\ION{O}{iii}/H$\beta$] vs. [\ION{N}{ii}]/H$\alpha$
  diagnostic diagram for the spatially resolved distribution of the
  ionized gas across the optical extension of the considered sample of
  galaxies. Color coded is the average of the logarithm of
  the equivalent with of H$\alpha$ at each location, and contours
  represent the density of spaxels, with each contour including a
  90\%, 50\% and 10\% of the points. Blue solid-circles represent the
  baricenter of the density distribution at different galactocentric
  distances, represented by the size of circles, ranging from 0.1 R$_e$
  (smallest circle) to 2.1 R$_e$ (largest circle). In all panels, the
  solid-line represents the location of the \citet{kewley01} boundary
  lines, with the separation between Seyferts and LINERs indicated
  with a dashed-line. Finally, in the dotted-line represents the
  location of the \citet{kauff03} demarcation line. Each panel
  represents a sub-sample of galaxies: the full sample is shown in the
  upper-left panel, and each panel shows a sub-sample of galaxies
  segregated by morphology (from left to right) and stellar mass (from
  top to bottom), with the covered ranges labeled in top- and
  left-most panels.}
\label{fig:BPT_mass_res}
\end{figure}

\subsection{Ionized gas across the optical extension of galaxies}
\label{sec:diag_pos}

As indicated in Sec. \ref{sec:diag_gal} the ionization in galaxies is
produced by local processes, that may be different in different
locations within each galaxy. Averaged across this extension, the
properties of the ionized gas reveal either the dominant physical
process or a mixture of all of them \citep[e.g.][]{kauff03}. This is
illustrated in Figure \ref{fig:bpt_maps} where two composite images
are shown, one constructed based on the $g$ (blue), $r$ (green) and
$i$-band (red) continuum images and the other one based on the
[\ION{O}{iii}] (blue), H$\alpha$ (green) and [\ION{N}{ii}] (red)
emission line maps, extracted from the MUSE observations of the galaxy
IC1657 corresponding to the AMUSING project \citep[PI: J. Anderson,
e.g.][]{laura18}. The right-most panel shows the classical BPT
diagnostic diagram (already discussed in Sec. \ref{sec:diag_gal} and
shown in Fig. \ref{fig:BPT_mass}). In this case each dot in the BPT
diagram corresponds to a pixel in the RGB emission-line image, and it
is represented by the same color shown in that image (central
panel). This allows us to easily identify the location within a galaxy
at which different ionization happens. A clear bi-conical ionized gas
structure is seen in the center of this galaxy, highlighted as pink colors, indicating strong
[\ion{N}{ii}]/H$\alpha$ and [\ion{O}{iii}]/H$\beta$ line ratios (as
seen in the BPT diagram). This structure is a clear signature of shock
ionization induced most probably by an outflow driven by a strong
nuclear star-formation event \citep[e.g.][L\'opez-Coba et al., in
prep.]{carlos16,carlos19}. In addition to this structure, it is easy to
identify, in the emission-line image, the star-forming regions across
the inclined disk, with a clear gradient in the line ratios, which
are recovered by the green colours towards the center, and bluish colors in the outer
part. This change in the line ratios is frequently interpreted as a
signature of the oxygen abundance gradient observed in disk galaxies
\citep[e.g.][]{sear71,vila92,sanchez14}.
\begin{marginnote}[]
\entry{Ionization is a local process}{integrated/average line ratios do not provide with the full information about the ionization conditions in galaxies}
\end{marginnote}
This object is used to illustrate that different ionization sources
could be present at different locations within each galaxy.  In
general all sources of ionization described in
Sec. \ref{sec:diag_gal} maybe be present simultaneously (or not)
in different galaxies.


Despite the different ionization sources that may be present in a
single galaxy, there are patterns within galaxy types, similar to
those described for the integrated/average properties. Figure
\ref{fig:BPT_mass_res} shows the distribution of individual ionized
regions in a subset of the galaxy sample explored in this review
within the classical BPT diagnostic diagram (i.e., the same adopted in
Fig. \ref{fig:BPT_mass}, for integrated quantities).  The position of
individual regions depends on the mass and morphology of the host
galaxy; regions belonging to more massive and early type galaxies
populate the right-hand side of the diagram, with lower values for the
EW(H$\alpha$). On the other hand, for late and less massive galaxies
more ionized regions populate the left-hand side of the diagram with
larger values of the EW(H$\alpha$). However, contrary to what is seen
for the integrated properties of galaxies (shown in
Fig. \ref{fig:BPT_mass}), for the spatially resolved regions a wider
range of parameters is covered in most of the diagrams. This reflects
the fact that the average properties do not represent the full range
of ionization conditions found across the optical extension of
individual galaxies (as shown in Fig. \ref{fig:bpt_maps}).

The distribution of ionized regions does not depend only on the
properties of the hosting galaxy, but also on the location within the
galaxy. The baricenter
of the density distributions along the diagram for different
galactocentric distances in included in Fig. \ref{fig:BPT_mass_res} as
blue solid-circles. For doing so I estimate
the same density distribution shown as contours, but in consecutive
radial bins centred at consecutive galactocentric distances between
0.1R$_e$ and 2.1R$_e$, and a width of 0.2R$_e$ (following the
position angle and ellipticity of each individual galaxy). The size
of the circles increases with the galactocentric distance of the considered ring.
For most
galaxy types and stellar masses there is a clear trend with the radial
distance.  The ionized regions in the center of galaxies (R$<$R$_e$)
are mostly located in the upper-right of the diagram. On
the other hand, the outer regions are located either at the upper-left (or left) (for most of late-type galaxies, apart from the lowest mass
ones) or at the bottom-left (for early-type ones). However,
there are substantial differences for different galaxy types and
stellar masses. Early-type galaxies lack the left-side branch
corresponding to SF regions in general, and the trend goes just from
the upper-right to the bottom-center. Finally, as lower is the mass,
as narrower seems to be the covered range of parameters, in particular for the
[\ION{N}{ii}]/H$\alpha$ ratio, and the trend is even reverse for Sc/Sd galaxies.

The explanation for these distributions has been discussed in
different articles. For the center of most massive and early type
galaxies, the location in the diagram and the low observed values for
the EW(H$\alpha$) indicate that the dominant ionization is due to
ionizing old stars, post-AGBs or hot low-mass evolved stars
\citep[HOLMES][]{binette94,sta08,cid-fernandes10}. Recent results have
shown that this ionization, observed as a diffuse extended ionization
in the central regions of early-type galaxies by \citet{sarzi10} and
\citet{gomes16}, is ubiquitous in galaxies of different morphology at
the location of old-stellar populations
\citep{sign13,papa13,sanchez14}. This result was recently confirmed by
\citet{belf17} and \citet{lacerda18}. The observed trend from the
upper-right towards the lower left of the diagram from the center to
the outer regions could be explained by a change in the average age of
the ionizing population, in agreement with the predictions by models
\citep[e.g.][]{mori16}. This result highlights the connection between
the observed properties of the ionized gas and those of the stellar
populations, showing that in early-type galaxies there should be a
negative gradient in the average age of the stellar population, as
indeed it is observed \citep[e.g.][]{rosa14}. Finally, for a few
early-type galaxies, in particular the less massive ones, there is some
marginal star-formation activity, as shown by \citet{gomes16}. This
star-formation could be a residue of a former activity that declines
due to the natural dimming or ageing of the disk
\citep[e.g.][]{casado15}, or a rejuvenation due to the capture of
pristine gas or a gas-poor low-mass galaxy by the early-type one \citep[e.g.][]{bernd11}.

The connection with the properties of the stellar populations is even
more clear for the late-type galaxies. In those galaxies there is
some ionization due to old stars in the central regions for early-type
spirals, i.e., spirals with bulge (Sa/Sb) at R$<$R$_e$. This is more
clearly appreciated in the average distribution of these galaxies and
for most massive ones (upper panels of third column in
Fig. \ref{fig:BPT_mass_res}). For the later-type spirals (Sc/Sd),
galaxies without a prominent bulge, the location of the central
ionized regions in the diagram is not well defined. This is in agreement
with the results by \citet{sign13}, as indicated before,
since bulges are dominated by older stellar populations. However, the
dominant ionization in these galaxies is star-formation, and indeed
the corresponding line ratios are found in the regions usually associated
with this ionization, i.e., below the \cite{kewley01} curve with
large values of the EW(H$\alpha$). This is where most of the ionized regions
are found, in particular for galactocentric distances larger than 0.5
R$_e$. In these galaxies, and in particular in those with a stellar
mass larger than 10$^{9.5}$M$_\odot$, there is a trend from the
bottom-middle towards the upper-left area of the diagram as we move
further away in the disk (from 0.5 to 1.5 R$_e$). This trend can be
easily explained as a consequence of the well known negative gradient
in the gas-phase metallicity in spiral galaxies, that has been found
to be very similar for all galaxies in this mass regime
\citep{sanchez14,laura16}. For the less massive late-type galaxies
($<$10$^{9.5}$M$_\odot$), there is a less clear or inverted trend,
reflecting that in this regime the abundance gradient is less
prominent, flat, or even inverted \citep{belf17,larim17}. In summary,
the observed distribution reflects the connection between the
ionization conditions in star-forming regions and the overall
evolution of the underlying stellar population \citep[as described
by][]{sanchez15}. As a consequence of this connection, the location of
an \ION{H}{ii} region in a diagnostic diagram is mostly determined by
the stellar-mass and morphology of its host galaxy and its
galactocentric distance.  Indeed, this is the reason why we can use
\ION{H}{ii} regions to trace the chemical evolution of spiral
galaxies, including our own galaxy
\citep[e.g.][]{molla05,laura18,esteban18,carigi19}.

These results show that the ionization through the full optical
extension of the general population of galaxies is dominated by
stellar processes (either young or old). However, in some particular
galaxies (e.g., AGNs, galactic outflows hosts, mergers), there are
other ionization processes that can clearly over-shine the effect of
these stellar ionizing sources. This is particularly true for certain
regions within galaxies (i.e., central regions), but it is also
evident when the average/integrated ionization properties are explored
(as appreciated in Fig. \ref{fig:BPT_mass} where AGNs are identified
as blue points above the Kewley curve). However, the frequency of
those processes is so low (at least in the nearby universe) and in
some cases the spatial extension is so confined to certain regions in
galaxies, that they have little influence in the average spatial
distributions shown in Fig. \ref{fig:BPT_mass_res}. In other words,
they are blurred by the overwhelming and ubiquitous presence of
stellar ionization processes.

\subsection{Local Age bimodality: $\Sigma_{*}$-age bimodality}
\label{sec:local_Age_bi}

\begin{figure}[h]
\includegraphics[width=5in, clip, trim=15 10 50 20]{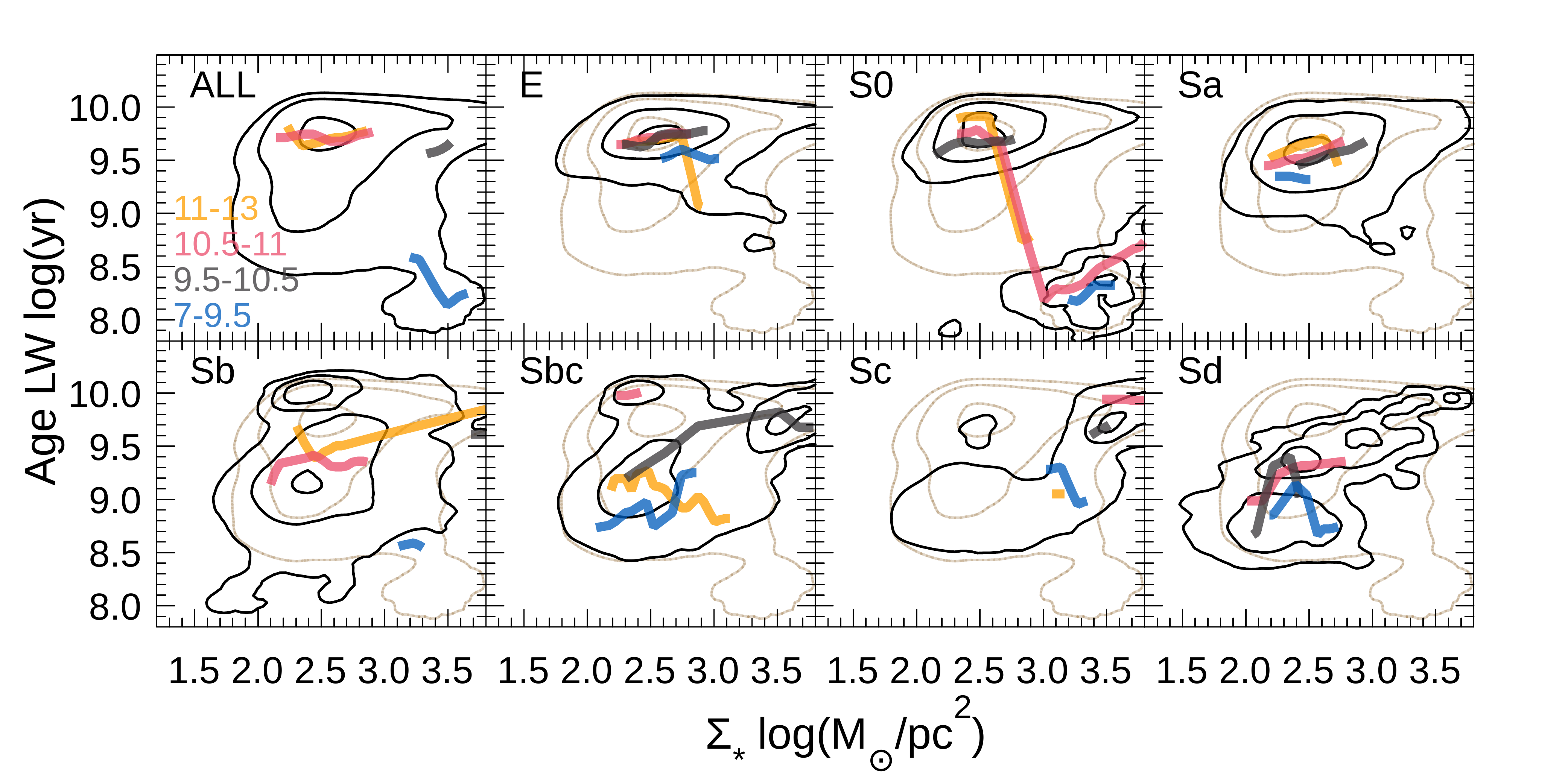}
\caption{Distribution of the luminosity-weighted ages of the stellar populations along the stellar mass surface densities shown as density contours (encircling 85\%, 50\% and 15\% of the regions), in solid-black lines. Each panel shows the distribution for all galaxies (top-left), and segregated by morphology (from earlier galaxies to later ones, from top-left to bottom-right). In addition, it is shown the trace of peak densities for different stellar mass bins as color solid-lines, with each color representing a M$_*$ bin as indicated in the label (only densities above a 50\% are considered in this derivation). Finally, the light-brown solid contours in the panels for each different morphology reproduce the distribution of the full population (shown in the top-left panel)}
\label{fig:local_Mass_Age}
\end{figure}

I show in Section \ref{sec:M_ZH} that galaxies present a clear
bimodal distribution along the M$_*$-Age diagram. As indicated in that
section this diagram is a more physical-driven version of the
color-magnitude diagram, used for decades in astronomy to select
star-forming and non-starforming galaxies
\citep[e.g.][]{mcintosh+2014}. \citet{rosa14} first showed the
distribution along the $\Sigma_{*}$-Age diagram for spatially resolved
kiloparsec-scale regions of different galaxies, describing a clear trend towards older
populations being located in denser regions of more massive galaxies
(and in the central regions). More recently, \citet{zibetti17}
demonstrated that this distribution is bi-modal: the distribution of
individual regions within galaxies in the $\Sigma_{*}$-Age diagram
present two clear density peaks clustered around old ($\sim$9 Gyr) and
young ($\sim$1-4 Gyr) ages, respectively, mostly related with retired
regions (dominant in early-type galaxies), and star-forming regions
(more frequent in late-type galaxies). In late-type galaxies (in
particular in early-spirals) the central regions, associated with
bulges, are found around the old-age peak, while their disks and
spirals arms are located around the young-age one.  This is a clear
indication of an internal/local age bimodality. This result is 
found as a statistical effect when exploring the full sample,
and also in individual galaxies.
\begin{marginnote}[]
  \entry{Local bimodality}{areas within galaxies present a local bimodality in they $\Sigma_*$-age distribution.}
\end{marginnote}

%

Figure \ref{fig:local_Mass_Age} reproduces the results by
\citet{zibetti17}, showing the distribution of individual spaxels in
the $\Sigma_{*}$-Age diagram for the compilation explored in this
review. I show the distribution for all galaxy types (top-left) and
segregated by morphology from earlier to later types (from top-left to
bottom-right). In addition, I show the density peak along $\Sigma_{*}$ for different
stellar masses (colored solid-lines). The bimodality uncovered by \citet{zibetti17} is
clearly seen, showing a strong morphological dependence (with
younger regions located in later type galaxies), and a trend with
the stellar masses (with younger regions located in less massive
galaxies). Intermediate spirals (Sb/Sbc) show a clear bimodal
distribution, highlighting the connection between the two populations
with the bulge (old) and the disk (young). Like in the case of the
M$_*$-Age (and CM-diagram), the regions with old stellar populations
follow a well-defined region, called old-sequence
(resembling the red-sequence in a CM-diagram), with very similar ages
($\sim$6 Gyrs) for a wide range of $\Sigma_{*}$ values. On the other
hand, young stellar populations cover a considerable range of values,
from a few hundred of Myrs to a few Gyrs within the same $\Sigma_{*}$
range. The bimodality implies an abrupt/short-lived transition,
leading to a local quenching of the star-formation or sharp change in
the SFHs from the inside-out \citep[e.g.][]{lopfer18,belfiore17a}.
Moreover, this distribution indicates that all regions with old-stellar
population have very similar SFHs (due to the limited range
in ages). However, regions with young stellar populations may
present a large variety of SFHs (due to the spread of ages).
This was indeed described in \citet{ibarra16} and
\citet{rgb18}.

\subsection{Local SFMS: $\Sigma_{*}-\Sigma_{SFR}$.}
\label{sec:local_SFMS}

\begin{figure}[h]
\includegraphics[width=5in, clip, trim=1 10 52 20]{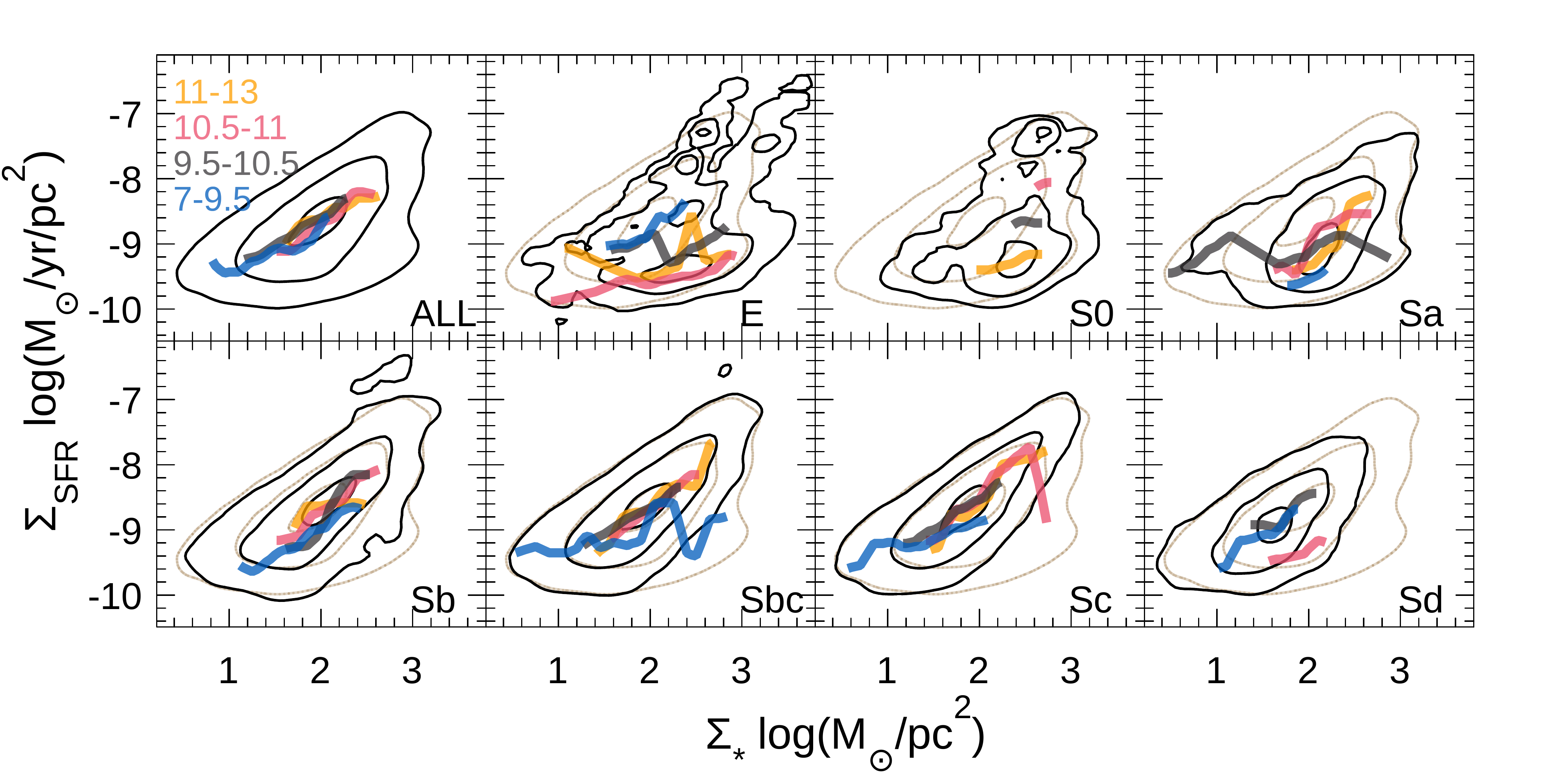}
\caption{Distribution of the star-formation rate surface densities along the stellar mass surface densities shown as density contours following the nomenclature of Fig. \ref{fig:local_Mass_Age}}
\label{fig:local_SFMS}
\end{figure}

I have shown in Sec. \ref{sec:SFMS} that galaxies also present a bimodal
distribution in the SFR-M$_*$ diagram, with SFGs following a tight
correlation (the SFMS) between the two parameters ($\sigma\sim$0.2-0.3
dex, once removed the main trend). In 2013 two almost simultaneous studies, \citet{sanchez13} and
\citet{wuyts13}, analysing completely different samples (at $z\sim$0
and $z\sim$1, respectively) and using different techniques (IFS and
narrow-band HST imaging), showed that star-forming regions present a
very similar relation between the SFR surface density ($\Sigma_{SFR}$)
and stellar mass surface density ($\Sigma_{*}$), a relation that holds
at kilo-parsec scales. This relation is now known as the resolved SFMS
(or rSFMS), and has been confirmed by many different authors using
mostly IFS-GS mainly for galaxies in the nearby universe
\citep[e.g.][]{mariana16,rosa16,maragkoudakis16,abdurro17,hsieh17,pan18,ellison18,medling18,erroz19,mariana19}. They
all found that this relation is as tight as the global one, with a
dispersion between $\sim$0.2-0.3 dex (once removed the main trend), and a slope slightly below
one. To our knowledge there is no formal published comparison between
the high-z rSFMS and the low-z ones. However, comparing the best
fitted values reported by \citet{wuyts13} adopting a log-linear
relation with those reported by the most recent analyses at low-z, it
seems that the rSFMS relation present a weak evolution in the slope
(from $\alpha_{z\sim1}\sim$0.95 to $\alpha_{z\sim0}\sim$0.78-0.94,
respectively) and a very strong evolution in the zero-point (from 
$\Sigma_{SFR,2}$ $_{z\sim1}$$=$10$^{-6.8}$
M$_\odot$yr$^{-1}$pc$^{-2}$ to 
$\Sigma_{SFR,2}$ $_{z\sim0}$$=$10$^{-8.5}$ M$_\odot$yr$^{-1}$pc$^{-2}$, with
$\Sigma_{SFR,2}$ being the SFR surface density at $\Sigma_{*}$
$=$10$^2$ M$_\odot$pc$^{-2}$). This evolution is similar to the one
reported by the global SFMS \citep[e.g.][]{speagle14,rodriguez16}, and
it is broadly reproduced by the most recent analysis of
hydrodynamical/cosmological simulations \citep[e.g.][]{tray19}.

Figure \ref{fig:local_SFMS} shows the distribution along the
$\Sigma_{SFR}$-$\Sigma_{*}$ diagram for the individual spaxels in my
compilation, segregated by both morphology and mass. In this
particular case I adopted the $\Sigma_{SFR}$ derived from the
H$\alpha$ emission irrespectively of the ionizing source, following
\citet{mariana16}, \citet{sanchez18} and \citet{mariana19}. This
way H$\alpha$ traces either the SFR for star-forming areas (SFAs) in
each galaxy, or just an upper-limit to the SFR for non
star-forming/retired areas (RAs). The most recent results show that
there is a bimodality in the distribution of areas in this diagram
\citep[e.g.][]{hsieh17,mariana19}, with SFAs tracing the described
rSFMS relation and RAs located in a cloud well below that
relation. Depending on the S/N cut applied to the H$\alpha$ detection
(or any other tracer of the SFR), this second cloud is more or less
evident. In our particular compilation this cloud is clearly seen only
for earlier type galaxies (E, S0 and less clear for Sa), and only as a
bump in the distribution for all galaxy types. As reported by the same
authors, the RAs cloud is dominated by regions in early-type galaxies,
while the rSFMS is dominated by regions in late-type ones, with a
clear morphological evolution \citep[as found by][]{rosa16}.  For the
bulk population of galaxies, irrespectively of their morphology, the
density peak in the diagram traces the rSFMS at any mass. Thus, the
mass dependence is weaker than the morphological one. Like in the case
of the $\Sigma_{*}$-Age bimodality, this is a consequence of the local
evolution, with SFAs associated with regions in the disk and RAs
associated with regions in the bulge (dominated by quenched regions).

It is worth noticing that the applied cuts in the S/N and the detection limits for properties
involved in resolved relations have to be treated with
care. \citet{mariana19} explored the effects of detection limits in
the shape of the rSFMS and in the artificial generation of broken
distributions resembling two-linear regimes \citep[e.g.][]{erroz19}.

Besides the increase of the fraction of RAs, some authors
\citep[e.g.][]{rosa16,mariana19}, reported a change in the rSFMS
itself with the morphology of the host galaxy. Later-spirals (Sc/Sd)
present slightly larger $\Sigma_{SFR}$ than earlier-spirals (Sa/Sb)
for a fixed stellar mass.  This is somehow appreciated in
Fig. \ref{fig:local_SFMS}, with the rSFMS of Sa (Sc) galaxies being
slightly below (above) that of the full population. This result may
indicate that the SFH of SFAs in earlier-spirals present a sharper
decline with the cosmological time (shorter times scales) than in
later-ones, having a lower $\Sigma_{SFR}$ at $z\sim0$ than in the past. Thus, they
present a faster evolution or ageing that the SFAs of purely disk-dominated galaxies. More recently, Ellison et al. (submitted) has
demonstrated that the dispersion across the rSFMS is due to changes in
the local star-formation efficiency (SFE). This agrees with the
described morphological trend, since the SFE does indeed change with
galaxy type both globally and locally \citep[e.g.][]{colombo18}. Most
probably there are dynamical effects, like the stabilisation induced
by the bulge proposed by \citet{martig09}, that precludes the
star-formation in earlier-spirals. However, the SFE does not explain
why a region is retired, that seems to be more directly connected with
a lack of gas (with the conditions required to form stars), and being
driven by global properties, in particular by the presence of a
massive central black-hole, as demonstrated by Bluck et al. (in
prep.).  In other words, SF is governed by local processes
(auto-regulation/local feedback) while quenching is driven by global
ones \citep[in agreement with the results presented by][]{rosa14}


\subsection{Local Mass-Metallicity relations}
\label{sec:local_MZ}

\begin{figure}[h]
\includegraphics[width=5in, clip, trim=15 10 50 20]{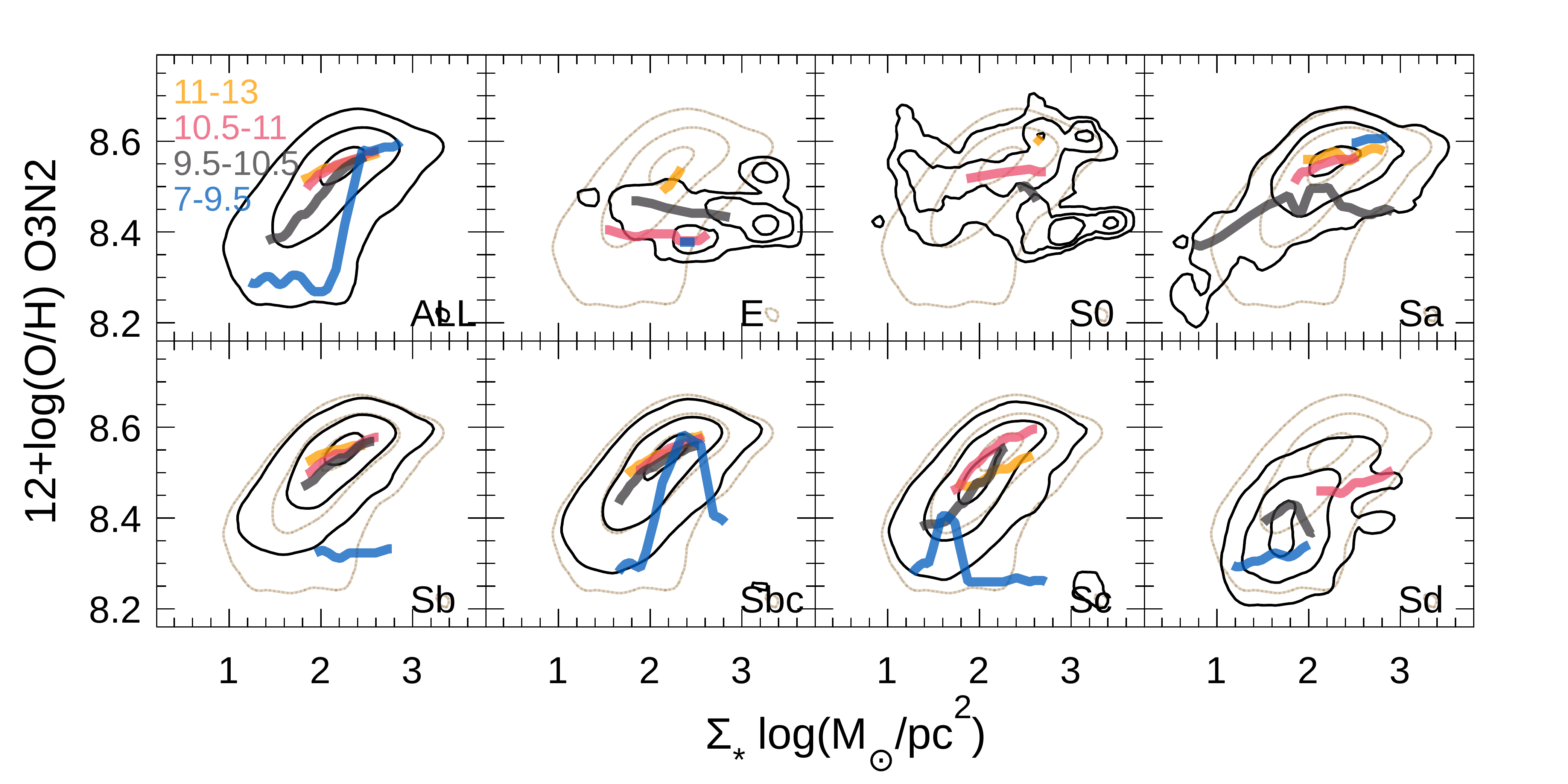}
\caption{Distribution of the gas-phase oxygen abundances along the stellar mass surface densities
shown as density contours following the nomenclature of Fig.\ref{fig:local_Mass_Age}. The O/H was derived based on the M13 calibrator \citep{marino13}}
\label{fig:local_Mass_OH}
\end{figure}

In the previous sections I showed that the stellar populations present
a local bimodality in the Age-M$_*$ and SFR-M$_*$ diagrams, with RAs
located in a tight {\it old-sequence} in the first diagram and SFAs
following a local rSFMS, in the latter one. These local/resolved
counter-parts of global relations and properties have been explored
for other parameters, like the MZR. \citet{ryder95} already showed
that local/spatially-resolved oxygen abundance presents a correlation
with the surface-brightness in the optical. This relation suggests the
existence of a local/resolved MZR (rMZR hereafter). Like in the case
of the rSFMS, this relation was proposed in two independent studies by
\citet{moran12} and \citet{rosales12}, using different techniques. In
the first case classical slit-spectroscopy was used to recover the
oxygen abundance along the aperture for a sample of star-forming
galaxies in combination with SDSS photometry to estimate
$\Sigma_{*}$. The second study explored the relation between oxygen
abundance, $\Sigma_{*}$ and EW(H$\alpha$) for the SFAs (HII regions
and clusters) detected using IFS \citep{rosales-ortega10}. Both
studies show that there is a relation between $\Sigma_{*}$ and O/H
with a shape that resembles the global $M_*$-O/H relation. It presents
two regimes, one, just below $\Sigma_{*}<$10$^2$ M$_\odot$pc$^{-2}$,
with an almost log-linear increase of the oxygen abundance with
$\Sigma_{*}$, followed by a flattening to an asymptotic value at
$\Sigma_{*}\sim$ 10$^3$ M$_\odot$pc$^{-2}$.  Indeed, \citet{rosales12}
explicitly show that the global MZR relation can be recovered from the
rMZR integrating the local one along the optical extension of each
galaxy. Subsequent studies confirmed the existence of this local
relation using larger samples of galaxies and/or SFAs or IFS data with
better spatial resolution \citep[e.g.][]{sanchez13,jkbb16,erroz19}. In
some cases two-linear relations are proposed for that relation rather
than a smooth transition between the two regimes
\citep[e.g.][]{erroz19}.

Figure \ref{fig:local_Mass_OH} shows the distribution in the
$\Sigma_{*}$-O/H diagram for the individual spaxels of the explored
data-set, once more segregated by morphology and stellar mass. In this
particular case only the SFAs have been selected to derive the oxygen
abundance, following the criteria outlined in the Appendix
\ref{sec:class}. For the remaining regions the adopted oxygen
abundance calibrators may not be valid, as they are anchored to
measurements derived for \ION{H}{ii} regions or photoionization models
created to describe these SF ionized structures. Like in the case of
the global MZR the shape of the relation is basically preserved
irrespectively of the calibrator adopted to estimate the O/H or the
assumptions made to derive the stellar mass density. However, the
absolute scale of the oxygen abundance changes with the calibrator
\citep[in agreement with the expectations of the results found by
][]{kewley08}. Contrary to what it is found for the age bimodality and
the rSFMS, the distribution of SFAs in the $\Sigma_{*}$-O/H does not
seem to present clear deviations neither with the stellar mass (for
M$_*>$10$^{9.5}$M$_\odot$) nor with the morphology (for spiral
galaxies) besides the wide range of values covered by each group.  In
other words, all SFAs seems to be distributed along the same rMZR
relation, with more massive galaxies covering the upper-right of the
diagram and low-mass (and later-types, e.g., Sd) covering the
lower-left range. Clear deviations from this global trend are
appreciated for low mass galaxies (M$_*<$10$^{9.5}$M$_\odot$), and
maybe the few star-forming regions in elliptical galaxies (E and
S0). In the first case the statistic is poor (low number of
galaxies). However the result could be interpreted as a consequence of
an outside-in chemical enrichment, that I will discuss in the upcoming
sessions. On the other hand, for the ellipticals and S0, maybe the
few galaxies with detected star-formation \citep{gomes16} are the
consequence of the capture of metal-poor galaxies, rather than an
effect of disk dimming. Indeed, in the case of S0 the distribution
shows two peaks, one following the main rMZR trend, and another one
more similar to the distribution appreciated for pure
ellipticals. Unfortunately, even for the large compilation included in
here the statistic is too low for SFAs in early-type galaxies.


Recent studies have explored whether the rMZR presents a possible
secondary relation with the SFR ($\Sigma_{SFR}$), following the
concept of the FMR. This secondary relation is commonly interpreted as
the consequence of gas inflows and outflows (as discussed in
Sec. \ref{sec:MZR}). The results of these explorations are not fully
conclusive. Some authors claim that there is no secondary relation
\citep[e.g.][]{jkbb16,bb17}, based on the analysis of the possible
dependence of the residuals of the rMZR with the sSFR and
$\Sigma_{SFR}$. Other authors claim that there are secondary trends
between the residuals of the radial distributions of the
$\Sigma_{SFR}$ and O/H \citep[e.g.][]{laura19}, that at low-mass are
negative and at high-mass positive.  This may be interpreted as
evidence for a local/resolved FMR \citep[or rFMR,][]{salmeida19}.
However, radial migration and the local effect of spiral arms may
produce similar effects \citep[e.g.][]{laura16b,vogt17} without
involving the same physics claimed to explain the FMR (i.e., a strong
effect of outflows and inflows). Finally, there are authors that have
reported the existence of individual regions in galaxies with
anomalous metallicities (AMR), that deviate from the rMZR
\citep[e.g.][]{hwang18}. Like in the case of \citet{laura19}, they
consider that gas accretion is the main cause of these deviations. The
AMRs are located mostly in interacting systems and in the direction of
the close companion, which indeed suggests a non secular origin,
different than the concept of the FMR or rFMR (aimed to describe the
quiescent evolution of star-forming galaxies in general).  So far,
with the current compilation of data, the results by \citet{jkbb16}
are reproduced, showing no significant correlation between the
residual of the rMZR with the $\Sigma_{SFR}$ (i.e., no
general/fundamental secondary relation with the SFR is appreciated).


\begin{figure}[h]
\includegraphics[width=5in, clip, trim=15 10 50 20]{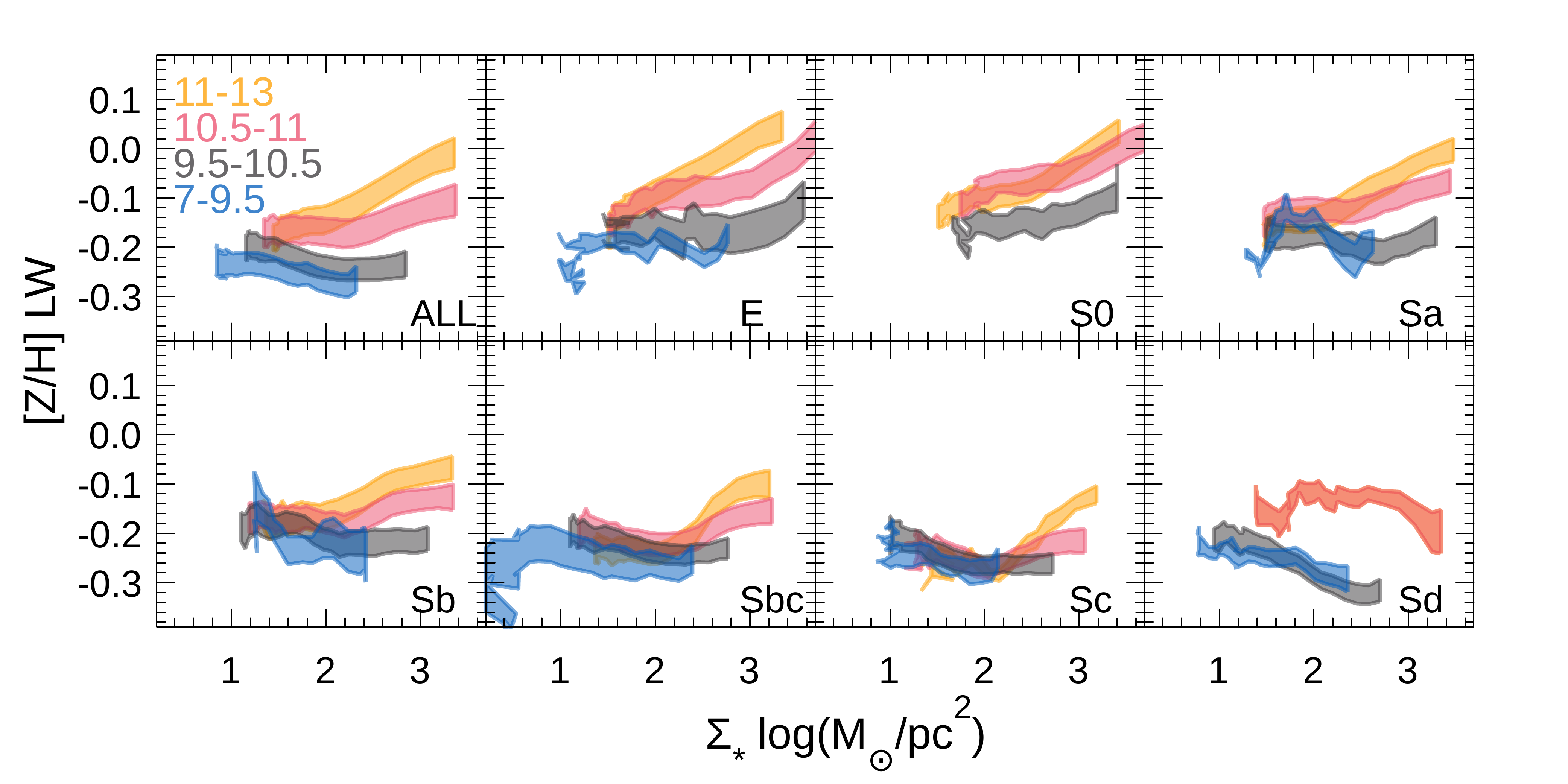}
\caption{Distribution of the luminosity-weighted metallicities of the stellar populations along the stellar mass surface densities derived from the azimuthal averaged radial distributions
described in Sec. \ref{sec:gradients} (Fig. \ref{fig:RAD_ZH_LW} and Fig. \ref{fig:RAD_Sigma_M}),
  for the galaxies considered in this review segregated by stellar mass and morphology.
 The top-left panel includes the distributions for all the galaxies with different colors indicating different mass ranges and the remaining panels include the same distributions for different morphologies, from earlier (top-left) to later types (bottom right).}
\label{fig:local_Mass_ZH}
\end{figure}

The global and local relations of the oxygen abundance with both M$_*$
and $\Sigma_*$ are a consequence of how $\alpha$ elements (O-like) are
produced (i.e., expelled to the ISM by type-II SN, the direct
by-product of SF). On the other hand, non-$\alpha$ elements (Fe-like),
are introduced to the ISM by different mechanisms, involving not only
the production of new stars but the death of intermediate mass ones
(by the production of type Ia SNe). Indeed, the [$\alpha$/Fe] ratio is
strongly related to the shape of the star-formation histories in
galaxies \citep[e.g.][]{delarosa11}.  As the SFH changes for galaxies
of different mass \citep[e.g.][]{panter03}, and morphologies
\citep[e.g.][]{lopfer18}, and from the inside-out
\citep[e.g.][]{eperez13,ibarra16}, the dependence of the stellar
metallicity with the stellar mass is less obvious than that of the
oxygen abundance. The existence of a global stellar mass-metallicity
or M$_*$-[Z/H] relation in galaxies, summarized in
Sec. \ref{sec:M_ZH}, highlights the relation between the shape of the
SFH and the stellar mass density (and therefore, the integrated stellar
mass too).

A local counter-part of this global relation has not been broadly
explored. To our knowledge it was presented in \citet{rosa14b}, where they presented the distribution along the
$\Sigma_*$-[Z/H] diagram for the azimuthal averaged mass-weighted
stellar metallicities of 300 galaxies extracted from the CALIFA
survey. I present in Fig. \ref{fig:local_Mass_ZH} a similar
distribution for the collection of data adopted here. Following
\citet{rosa14b}, in order to minimize the scatter introduced by the
large uncertainties in the derivation of [Z/H] in individual spectra,
I first derive azimuthal average values in bins of 0.1 Re for each
galaxy (i.e., I create radial distributions of [Z/H], that I will
discuss later, in Sec. \ref{sec:grad_age}).  Broadly speaking the
results are consistent with those presented by \citet{rosa14b}, with
the local [Z/H] increasing with $\Sigma_*$, in particular for the same
stellar mass regime (M$_*>$10$^{9.5}$M$_\odot$, once considered the
differences in the adopted IMFs). This agrees with the existence of a
negative stellar metallicity gradient in most massive galaxies
\citep[e.g.][]{rosa14,oyarzun19}.  However, contrary to what happen
with the rMZR, the local stellar metallicity-$\Sigma_*$ relation is
not the same for all galaxies. It depends strongly on galaxy mass and
morphology. In particular, for low-mass and late-type galaxies the
distribution of [Z/H] along $\Sigma_*$ is almost flat or even negative. This
indeed agrees with a flattening of the stellar metallicity gradient
observed in this kind of galaxies \citep{rosa14}. In addition, the
observed differences between the behaviour of the gas phase (oxygen)
and stellar metallicities across galaxies implies a change in the SFHs
and therefore a differential chemical enrichment history for $\alpha$
and non-$\alpha$ elements.


%
%

\subsection{Local star-formation law}
\label{sec:local_SK}

\begin{figure}[h]
\includegraphics[width=5in, clip, trim=0 5 50 20]{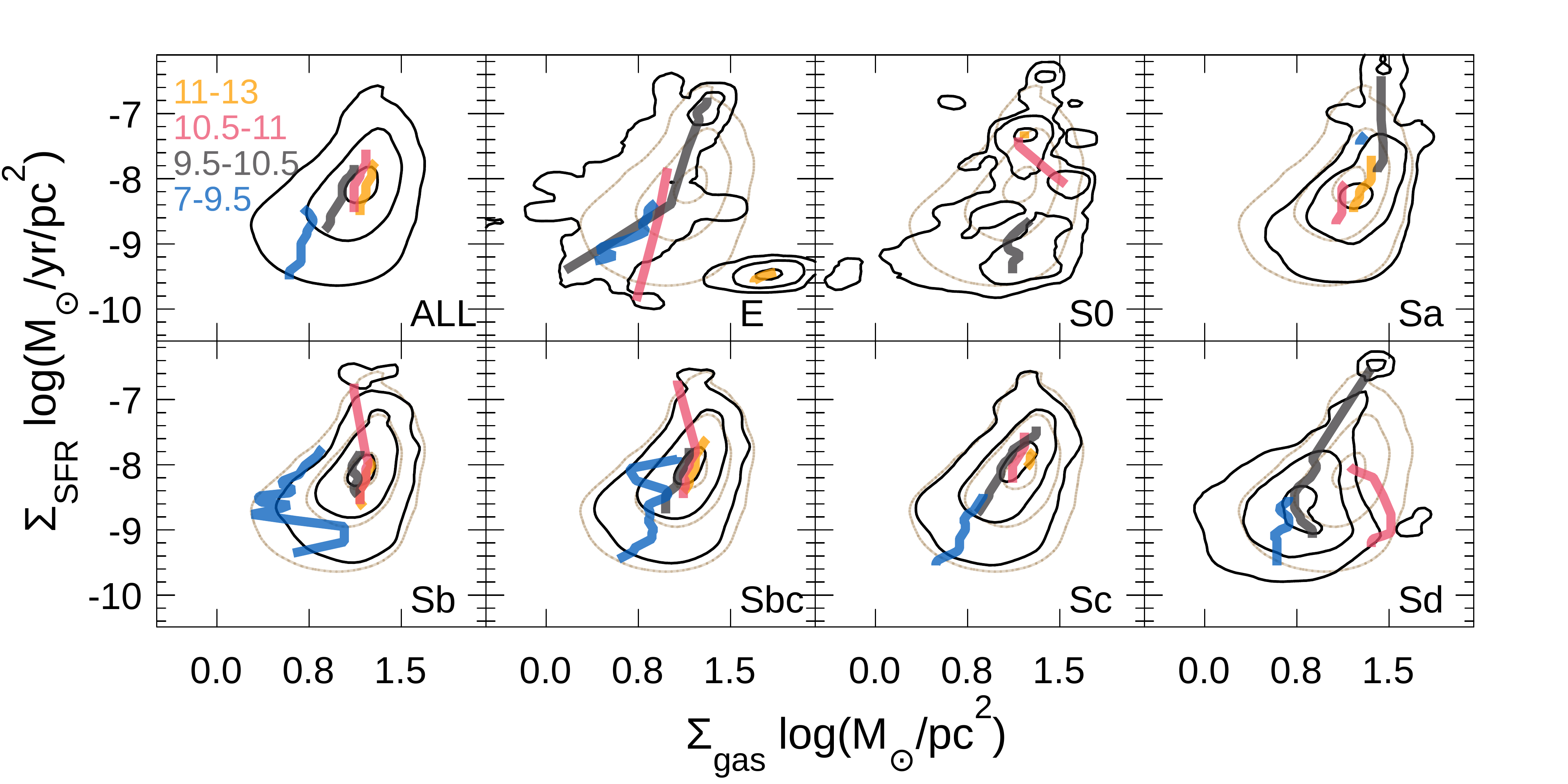}
\caption{Distribution of the star-formation rate surface densities along the gas mass surface densities derived from the azimuthal averaged radial distributions shown in Fig. \ref{fig:RAD_OH_O3N2} and Fig. \ref{fig:RAD_Sigma_M}
  for the galaxies in this review segregated by stellar mass and morphology. I adopted the same nomenclature and symbols used in Fig. \ref{fig:local_Mass_Age}.}
\label{fig:local_SK}
\end{figure}

The birth of new stars comprises two different processes. First, the
diffuse atomic gas is transformed into high dense molecular
gas. Second, stars are born due to the dynamical collapse of
self-gravitating portions of those molecular clouds. \citet{schmidt59}
proposed that the SFR should depend on the mass of gas included in a
volume, just assuming that all stars are formed locally from the
original available gas, and that there was no inter-exchange (of stars
and gas) with the near environment (i.e., $SFR=M_{gas}^n$). However,
the power-scaling factor ($n$) between both quantities was not
defined.  \citet{kennicutt89} proposed that this scaling factor could
be derived through the gas free-fall time ($\tau_{ff}$), following the
equation $\rho_{SFR} \propto \rho_{gas}/\tau_{ff}$, by assuming that
all gas collapsed within this time is transformed into
stars. Considering that $\tau_{ff}\propto \rho_{gas}^{-0.5} $, they
conjectured that the basic star-formation law should follow a relation
$\rho_{SFR} \propto \rho_{gas}^{1.5}$.  Empirically,
\citet{kennicutt89} and \citet{kennicutt98} demonstrated that indeed
the surface density of the SFR follows a relation with the gas surface
density similar to proposed one. This relation was derived originally
galaxy wide, and it is known as the SK-law (Schmidt-Kennicutt law) or
star-formation law.

As discussed in previous sections SF, and in particular
the collapse of molecular clouds, is a local process that happen at
much lower scales than a galaxy size. Spatially-resolved observations have shown that
indeed the SK-law is verified only for molecular gas
\citep{wong02,Kennicutt07}, a result confirmed for integrated
quantities too \citep[e.g.][]{reyes19}. At kpc-scales it shows an
almost linear relation \citep[e.g.][]{bigiel08,leroy08}. Therefore,
the depletion time ($\tau_{dep} = \Sigma_{gas}/\Sigma_{SFR}$) or
star-formation efficiency (SFE$= \Sigma_{SFR}/\Sigma_{gas}$) are almost
constant at these scales \citep{bolatto08,hughes10}, with a value of
$\sim$2 Gyr \citep[e.g.][]{bigiel08,rahman12,leroy13}. However, these
studies are biased towards low-mass, late type galaxies. It is still
not known if this result holds in more massive, or more early-type
galaxies. The particular conversion factors to derive the molecular
gas and the adopted IMF may play a role in these results. It is known
that there are clear deviations from the SK-law (thus variations in
$\tau_{dep}$) at sub-kpc scales \citep{schruba11,kruij14,kruij19}, in
extreme starburst galaxies \citep[e.g.][]{daddi10,genzel10}, in the
center of galaxies \citep{leroy13,jogee05,utomo17}, in regions near
strong spiral arms or with different local dynamical states
\citep{schru19}, and in early type galaxies \citep{davies2014}. The
average $\tau_{dep}$ within a galaxy presents a clear dependence with
sSFR and M$_*$ \citep{saintonge17}. All these results clearly indicate
that the combination of the optical spectroscopic information provided
by IFS-GS, for a wide variety of galaxies, with estimations of the
cold gas content at similar spatial scales is crucial to understand
these processes \citep[e.g.][Ellison et al., in prep.]{bolatto17}. It
is beyond the scope of this review to make a detailed exploration of the
cold gas content in galaxies (atomic or molecular), or to explore the
star-forming law at different scales \citep[see][for detailed reviews
on the topic]{kennicutt12,bolatto13}. In this section I just summarize
the most recent results based on the combination of IFS data with
molecular gas at kiloparsec resolutions.

So far the number of galaxies sampled by both IFS-GS and spatial
resolved molecular gas observations is rather small. Pioneering
studies cover mostly early-type galaxies observed by the SAURON and
Atlas3D surveys \citep[e.g. $\sim$40 galaxies,][]{alatalo12}. More
recent explorations range from a handful of objects
\citep[e.g.][]{belli17,stark18}, or a few tens \citep[like the
$\sim$47 galaxies in the ALMAQUEST collection,][Ellison et al. in
prep.]{li17} to a well defined sample of over one hundred galaxies
\citep[EDGE-CALIFA sample,][]{bolatto17}. There are a few on-going
surveys, like PHANGS \citep[][]{roso19}, but they are focused on large
galaxies in the Local Universe that can hardly be a representative
sample of the $z\sim0$ population due to the cosmic variance and the
limited sampled volume.

To highlight the urgent need of complementary explorations of the
molecular gas on a large sample of galaxies already covered by IFS-GS,
I present, in Figure \ref{fig:local_SK} the spatially-resolved
distribution in the $\Sigma_{SFR}-\Sigma_{gas}$ diagram for the sample
of galaxies used in this review. Lacking of spatially resolved
information for the molecular gas I adopted the dust-to-gas conversion
proposed by Barrera-Ballesteros et al.(in prep., using data from the
EDGE-CALIFA survey), already used in \citet{galbany17} and
\citet{sanchez18}. It involves an additional scatter of $\sim$0.3 dex
to the relation, but it is the best proxy I can adopt so far. Despite
these limitations, there is a clear relation between the two
parameters, with a slope near one ($\alpha$=1.12$\pm$0.09), and an
average $\tau_{dep} \sim$1.7 Gyr, in agreement with published
results. The dispersion of the relation, $\sim$0.1 dex, is clearly
below the expected individual errors, which validates the approach
adopted.  When this relation is explored using estimations of the
molecular gas based on direct CO observations the scatter is as low as
$\sim$0.05 dex \citep[e.g.][Lin et al.in prep., Ellison et al. in
prep. ]{bolatto17}. Indeed, it is possible that the rSFMS (and
the SFMS) are a direct consequence of this relation through a tight
$\Sigma_{gas}-\Sigma_{*}$ (M$_{gas}-$M$_{*}$) relation \citep[shown by
Lin et al. in prep. and ][respectively]{calette18}.


Earlier-type galaxies present shallower relations
($\alpha\sim$0.8-0.9) with a larger dispersion, and larger
$\tau_{dep}$ ($\sim$10 Gyr) than late types, that present steeper
slopes ($\alpha\sim$1.2-2.7), smaller dispersion, and lower
$\tau_{dep}$ ($\sim$0.5 Gyr). Both results agree with the most recent
explorations by \citet{colombo18}, \citet{utomo17}, and
\citet{li17} using direct estimations of the molecular gas derived
from CO observations, and broadly the model presented by
\citet{belfiore19}. These results indicate that the ageing and
quenching of the star-formation is driven not only by the lack of
molecular gas \citep[e.g.][]{saint16}, but also by a decline in the
SFE \citep[e.g.][]{sanchez18} (i.e., increase in $\tau_{dep}$).  The
observed differences imply that most probably there is no unique
$\tau_{dep}$ (or SFE) value or single SK-law for all galaxies and
areas within them (e.g., Ellison et al. in prep.), contrary to the
early evidence. In other words, $\tau_{dep}$ (SFE) may depend on
additional parameters whose inclusion in the star-formation law would
reduce the scatter in the $\Sigma_{SFR}\propto\Sigma_{gas}$ relation
\citep[e.g.][ Bluck et al. in prep.]{dey19}. 

Following this reasoning it is possible to reconcile the observed
differences for high-redshift starburst galaxies and nearby disks, and
even between the observed differences in the central and outer regions
of galaxies. The processes involved would be: (i) the inclusion of the
orbital or dynamical time ($\tau_{dyn}$), that connects the
gravitational instability (due to bar or spiral arms) with a bursting
in the SFE \citep{silk97,elme97}; (ii) the gas velocity dispersion as
a measurement of the local pressure that increases due to the SF
feedback \citep{silk97}; (iii) the stellar velocity dispersion, in
particular in bulges, that stabilize the molecular clouds preventing
the SF \citep{martig09}; (iv) the local gravitational potential traced
by $\Sigma_*$ (that would connect the SK-law with the SFMS)
\citep[e.g.][]{saintonge2011} ; (v) the gas metallicity that
facilitates the cooling \citep[e.g.][]{casado15}; or (vi) a combination
of all of them \citep[e.g.][]{dey19}.  In summary, the current results
indicate that the star-formation law follows the SK-law only at a
first order.

Furthermore, considering the SK-law, the SF quenching (either global
or local) may be related to a deficit of cold gas. As indicated
before, there is a global (and local) M$_{gas}$-M$_*$
($\Sigma_{gas}-\Sigma_{*}$) relation for SFGs (and SF regions). This
relation is not verified for early-type galaxies, that present a
deficit of neutral (at least molecular) gas with respect to their
late-type galaxies counterparts of the same M$_*$ \citep[e.g.][and
references there in]{calette18}. This deficit is considered as the
primary cause of the global decline of the star-formation rate
observed in these objects \citep[e.g.][]{saintonge2011}. The same
deficit is appreciated at local scales, where retired regions present
lower $\Sigma_{gas}$ than star-forming ones at the same $\Sigma_*$.
With the explored compilation of data I confirm this trend.

%
%

The origin of this deficit is still not clear. Several authors
consider that it is originated by AGN feedback, either by removing gas
due to galactic winds or heating it preventing to collapse and form
stars \citep[e.g.][]{sanders96,hopkins+2009}. However, the evidence
connecting AGN and quenching is still circumstantial: (i) it
seems to be the only process energetic enough to produce this effect
\citep[e.g.][]{croton06,henriques19}; (ii) their hosts are located in the green
valley between SFGs and RGs
\citep[e.g.][]{kauff03,sanchez04,schawinski+2014}, as indicated in
Sec. \ref{sec:SFMS}, sharing many of their properties with GV galaxies
\citep[e.g.][, Lacerda in prep.]{sanchez18}. However, there are still problems in this
scenario, like the lack of direct evidence \citep[although
see][]{fabian12} and the mismatch of the time scales between quenching
($\sim$1-2 Gyr) \citep[e.g.][]{sanchez18b} and the length of an AGN
phase \citep[$\sim$ 1-2 Myr, e.g.][]{urry95}. Other authors
indicate that the presence of the bulge itself could stabilize the
atomic gas preventing it to collapse forming molecular clouds or
decreasing the SFE \citep[e.g.][]{martig09}. In this case, a decline
in the general SFE (and towards the center) could contribute to the
halt of the star-formation (or gives support to this halt, as
suggested by Bluck et al. in prep).  An additional complication is
that the presence of an AGN is directly linked with the presence of a
bulge through the \citet{magorrian98} relation
\citep[e.g.][]{kormendy+2013}. Therefore, a clean distinction between the two
processes is difficult.

\subsection{Spatially resolved star-formation histories}
\label{sec:local_CMF}

\begin{figure}[h]
\includegraphics[width=5in, clip, trim=12 0 20 20]{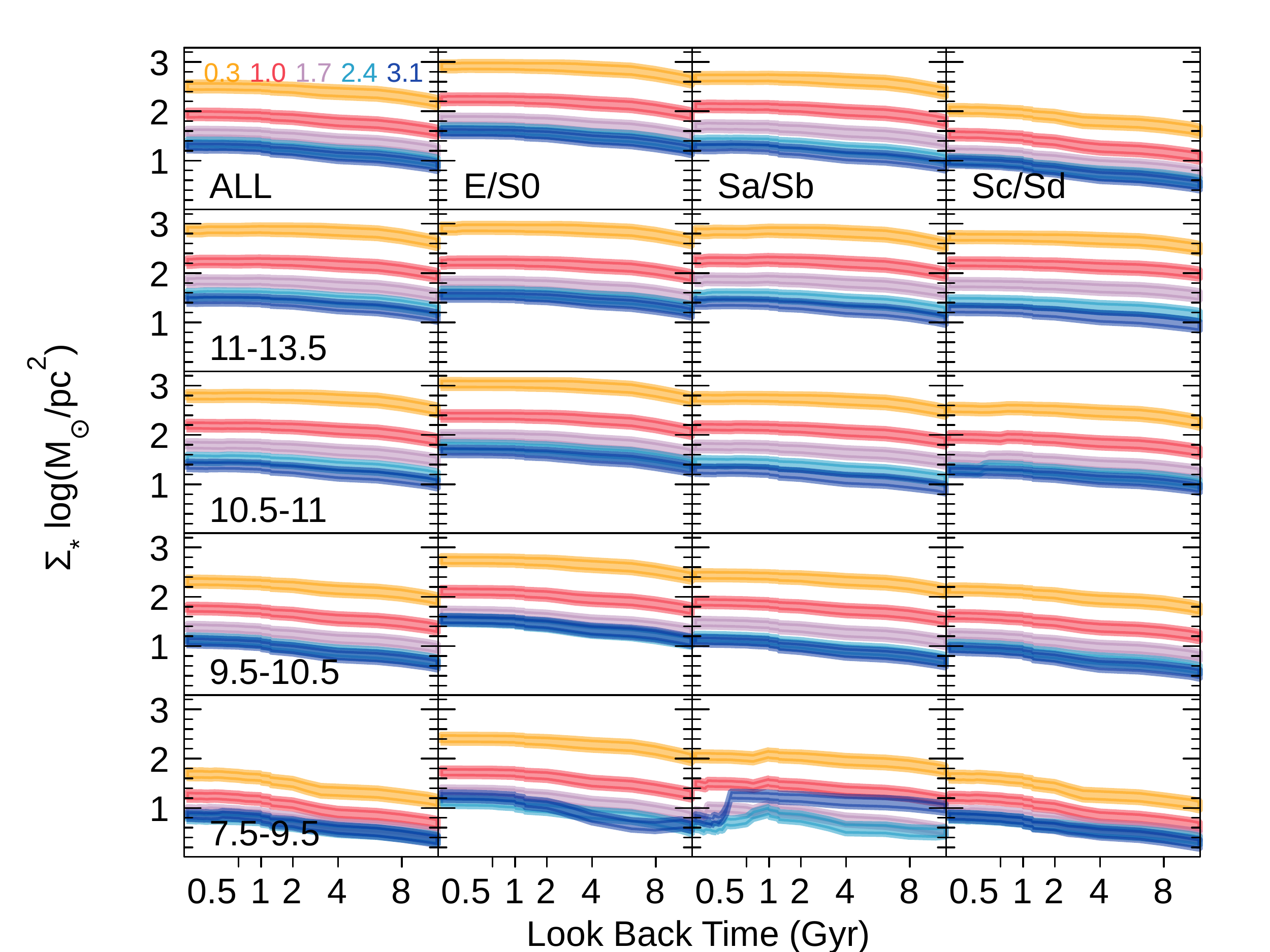}
\caption{Distribution of the cumulative stellar mass surface densities
  at different galactocentric distances (labelled with different
  colors) along the look-back time for all galaxies with spatially-resolved information (top-left panel), and for galaxies of different
  morphology (panels from left to right) and/or stellar masses (panels
  from top to bottom).}
\label{fig:local_CMF}
\end{figure}



The analysis adopted in the different publications reviewed in the
previous sections exploring the properties of the stellar populations
relies on the assumption that the observed spectra of galaxies (and
regions within galaxies) preserve the imprints of their cosmological
evolution, and that these imprints can be used to trace back that
evolution. This method is known as the fossil record, and has been
reviewed in detail by other authors
\citep[e.g.][]{walcher11,conroy13}. In particular, \citet{walcher11}
explore how the main properties of the stellar populations (luminosity
and mass weighted ages and metallicities, dust attenuation and masses)
can be derived from the decomposition of the stellar continuum (and
its limitations).

In principle, from this analysis, it is possible to recover
star-formation and chemical-enrichment histories. The fossil record
methods have two main implementations. One adopts a particular shape for the SFHs and ChEHs
\citep[e.g.,][]{gallazzi05,thomas10,Bitsakis+2016,zibetti17}, and
compare the observed spectra with the predicted ones tuning/fitting
certain parameters that govern the functional forms considered: e.g.,
the decay time of an exponential SFH ($\tau$), or the time delay for
the ignition of SF in the Universe ($t_{delay}$). A second approach
consists in a non-parametric exploration of the SFHs (and ChEHs),
based on their decomposition in single bursts of SF. Each burst
produces a single-stellar population (SSP), a set of stars born all at
a certain time (and therefore with the same age as they evolve) from
the same gas with a certain metal content
\citep[e.g.,][]{panter07,vale09,eperez13,ibarra16,rgb17}. The first
approach is unable to reproduce in detail the full spectral features
of the observed spectra, and it is frequently used to compare with
broad (or medium) band photometry or with particular stellar features
\citep[like stellar indices, e.g.][Zibetti et al. in
prep.]{gallazzi05}. However, it provides with SFHs (and ChEHs) that
are more easily interpreted. The later approach is
able to reproduce the details of the observed spectra, however, the
interpretation of the results is frequently less obvious.  It is
technically complex, and prone to large uncertainties
\citep[e.g.,][]{cid-fernandes14,pipe3d}. The results depend strongly on
the selected templates of SSPs or the stellar libraries
\citep[e.g.,][]{rosa14}, and on how the errors propagate when
non-linear components are considered, like the dust attenuation or the
stellar kinematics \citep[][]{cid-fernandes13,cid-fernandes14}.
Finally it relies on the assumption that the SSPs are eigenvectors,
which is mathematically not right. Therefore, it may present different
degenerancies, like the age-metallicity or the metallicity-velocity
dispersion one \citep[e.g.][]{patri11}. Further issues, involving the
IMF assumed, stellar evolution isochrones, stellar template adopted,
dust extinction curve, and so on are discussed in detail in
\citet{walcher11}. I should clarify that most of these caveats
apply to the parametric methods too.

Using the fossil record method it was possible: (i) to confirm the
downsizing in galaxies
\citep[e.g.,][]{perezgonzalez08,thomas05,thomas10}; (ii) to explore
cosmic evolution of the SFR density in the Universe
\citep[e.g.,][]{panter07,lopfer18,sanchez18b} reproducing the
distribution derived from direct observations
\citep[e.g.][]{madau98,madau14}; and (iii) to predict the global
chemical enrichment history of galaxies \citep[e.g.,][]{asari07}. Most
of those results were based on single aperture spectroscopic data
and/or integrated multi-band photometry, and therefore, they trace the
evolution of galaxies as a whole. The advent of IFS-GS has allowed the
exploration of the spatially resolved SFHs of galaxies. The pioneering
study by \citet{eperez13}, using a limited sample extracted from the
CALIFA survey, first demonstrates that the downsizing is spatially
preserved: (i) the central and outer regions of more massive galaxies
grow faster than those of less massive ones; (ii) the inner regions of
galaxies more massive than $\gtrsim$10$^{9.5-10}$M$_\odot$ assemble
their mass faster than the outer regions, following an inside-out
growth; (iii) for less massive galaxies there is a possible transition
from the inside-out towards the outside-in. Recent updates of these
results, using the full CALIFA sample, including UV-photometry and/or
adopting a different inversion method were presented in
\citet{rosa17}, \citet{rgb17} and \citet{lopfer18}. These results were
broadly confirmed with the exploration by \citet{ibarra16} of a much
larger sample of galaxies extracted from the MaNGA survey. The main
difference reported was that the outside-in regime at low mass was not
fully confirmed (a regime not well covered by the original CALIFA
sample). In this later work the diversity of spatially resolved SFHs
(rSFHs, hereafter) was studied, finding that more massive galaxies
present very similar rSFHs at the different spatial regimes explored.
However, as the integrated mass decreases, galaxies present a much
large variety of rSFHs (what implies a larger dispersion in their
average properties, like age and metallicity). However, on average,
they follow an outside-in rSFHs, in agreement with the distributions
found in the $\Sigma_*$-O/H and $\Sigma_*$-[Z/H] diagrams
(Sec. \ref{sec:local_MZ}).  The validity and limitations of this
method and the results reported were explored through the analysis of
post-processed cosmological hydrodynamics simulations, where the rSFH
are well known \citep{ibarra19}.

I reproduce these results in Fig. \ref{fig:local_CMF}, which shows
the average evolution of the stellar mass density at different
look-back times and for different radial bins, $\Sigma_{*}$(r,t) for
the sample of galaxies explored here.  Different panels
show the distributions segregated by mass and morphology.  Following
\citet{eperez13} and \citet{ibarra16}, the radial bins are selected
normalized to the current ($z=0$) observed effective radius. The
figure shows that more massive galaxies indeed present
larger values of $\Sigma_{*}$(r,t) than less massive ones at any radius
and cosmological time (i.e., the local/resolved
downsizing). This segregation is also seen for different
morphologies, with earlier type galaxies having larger
$\Sigma_{*}$(r,t) than later types, not only at any $r$ and $t$, but
also at any stellar mass bin. Thus, the local downsizing does not
depend  only on the mass, but also on the morphology.  This result was
published by \citet{rgb17}, who shows that the local downsizing and
inside-out growth has a trend with $\Sigma_{*,cen}$ (the stellar mass
density in the center, $\sim$1kpc, of galaxies). Indeed, their
results suggest that the diversity in the rSFHs found by
\citet{ibarra16} for low-mass galaxies has a clear dependence with
this parameter.

\citet{rgb17} reported two additional results: (i) the scale-length of
galaxies in mass is smaller (on average) than the corresponding
parameter in light (characterized by r$_{50}$). This result,
already shown in \citet{rosa14}, indicates that galaxies grow faster
in mass than in light. Although I do not show a similar plot, this
result is clearly reproduced with the sample explored in this
review; (ii) the local downsizing (understood as the speed of mass
assembly) with morphology maybe be broken for E/S0, which seem to form
their outer regions at a slower rate than Sa. A possible explanation is that
in early-type galaxies captures/minor mergers affect
the mass-assembling history in the outer regions, resembling a change
in the local downsizing, as proposed by \citet{oyarzun19}.  Further
explorations, and in particular deep comparisons with mock/simulated
galaxies \citep[following][]{ibarra19} are required to fully
understand all these results.

\subsection{Radial gradients of galaxy properties}
\label{sec:gradients}

It is clear that the properties of the stellar populations and ionized gas
change along the optical extensions of galaxies, showing patterns and
relations that reassemble well-known ones derived for integrated
properties. In many cases the explored physical parameter
($\Sigma_{SFR}$, Age, [Z/H], O/H, or $\Sigma_{gas}$), present clear
dependencies with $\Sigma_{*}$. As this parameter present a clear
radial trend in galaxies \citep[e.g.][]{rosa14}, it is expected that many of the analyzed parameters present radial trends too
\citep[e.g.][]{jkbb16}. Beside this radial trends, due to the
different sub-structures observed in (late-type) galaxies, like spiral
arms and bars, azimuthal or local deviations from the
pure radial distribution are expected \citep[e.g.][]{laura16b}. In this section I
review the most recent results regarding the radial distribution of
the spectroscopic properties of galaxies and their relation with the
observed patterns described in previous sections.

Along this review all radial distributions are shown with distances
normalized to the effective radius of each galaxy (R$_e$, already used in
Fig.
\ref{fig:BPT_mass_res}). 
I adopt this scheme instead of representing them in physical scales
since this way it is possible to compare with galaxies covering a wide
range of sizes. This method was introduced by \citet{sanchez12b} and
nowadays is commonly used
\citep[e.g.][]{rosa14,rosa15,rosa16,laura16,belf17,laura18}, since it
does uncover patterns not clearly seen when using physical sizes
\citep[e.g.][]{eperez13,ibarra16}. Moreover, the effective radius has
been proved to be really effective, with most of the physical
parameters sampled at this distance being representative of the
average values across the optical extension of galaxies
\citep[e.g.][]{mous06,sanchez13,rosa14}.

\subsubsection{Stellar mass density gradients}
\label{sec:grad_Mass}

\begin{figure}[h]
\includegraphics[width=5in, clip, trim=0 15 50 20]{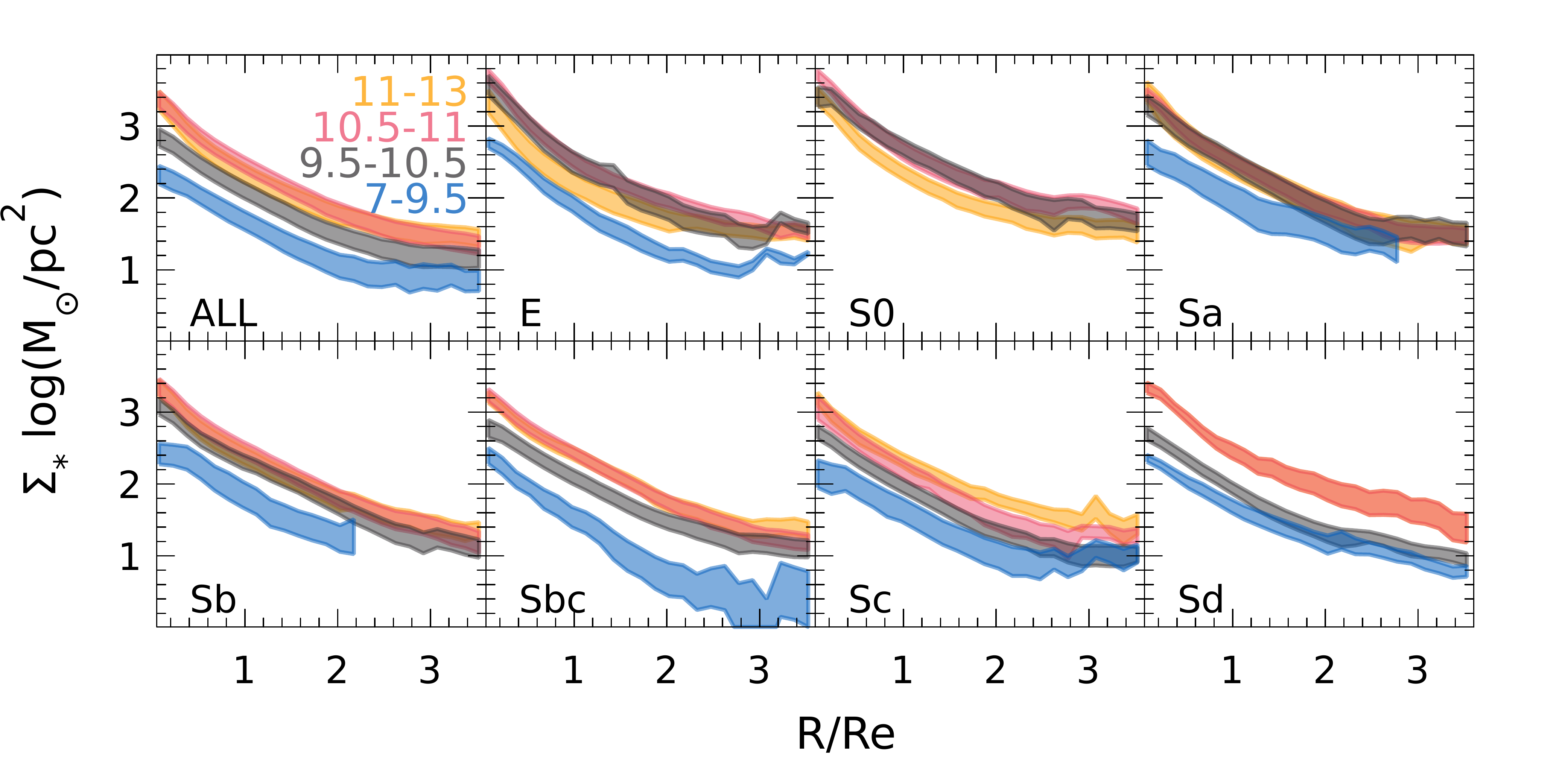}
\caption{Azimuthal-averaged radial distributions of the stellar mass
  surface density for the galaxies in this review segregated by
  stellar mass and morphology (as indicated in the inset). The
  top-left panel includes the radial distributions for all the
  galaxies with different colors indicating different mass ranges and
  the remaining panels include the same distributions for different
  morphologies, from earlier (top-left) to later types (bottom
  right).}
\label{fig:RAD_Sigma_M}
\end{figure}

It is well known that the radial distribution of the azimuthal
averaged surface brightness profiles (SBP) present an almost monotonic
decrease for almost all galaxy types of any stellar mass. In the case
of disk-dominated spiral-like galaxies this profile is well
characterized, as a first order, by an exponential function of the
radius \citep{free70} or a double exponential profile
\citep[e.g.][]{courteau96}. On the other hand, the SBP of early-type
galaxies (or the bulge of late-type ones) is better represented by a
sharper profile, well described by an exponential of the 1/4th power
of the radius \citep{deVauc59}, or a more general profile known as the
Sersic profile \citep{sersic68}, characterized by the power of the
radius known as the Sersic index ($n_s$). In general, early-type
spirals (Sa/Sb) require a combination of a single or a double
exponential plus one of the functional forms described to characterize
the bulge. The description of other structural components, like a bar,
requires the inclusion of further functional forms \citep[e.g.][and
references therein]{jairo17}. Irregular or interacting galaxies show
much complex radial profiles and in many cases their characterization
requires a full analysis of the two-dimensinal distribution using more
complicated functional forms \citep[e.g.][]{peng10,jairo17}.

The radial decline of the SBP is nowadays accepted to be a consequence
of an almost monotonic decline in the stellar mass density
($\Sigma_{*}$), via a conversion through the M/L-ratio
\citep[e.g.][]{bakos2009}. In general, the M/L-ratio presents a
log-linear relation with optical colors, a simple consequence of the
color-age relation in stellar populations
\citep[e.g.][]{bell01,zibetti09}. As galaxies contain older stellar
populations in their central regions \citep[e.g.][]{rosa14}, the
M/L-ratio enhances the radial decline of $\Sigma_{*}$ with respect to
that of the SBP. For this reason the R$_e$ in light is larger, in
general, than the one derived in mass (as indicated in the previous
section). The advent of IFS-GS has allowed to explore the spatially
resolved distribution of the stellar population content, providing
with more accurate M/L-ratios at kilo-parsec scales
\citep[e.g.][]{rgb18}. This improves the derivations based on colors
\citep[e.g.][]{bell01}, adopted in previous studies
\citep[e.g.][]{bakos2009}. These new datasets facilitate the
systematic exploration of the radial distributions of $\Sigma_{*}$ in
galaxies of different morphology and stellar mass.

Figure \ref{fig:RAD_Sigma_M} shows the radial distribution of
$\Sigma_*$ for the current adopted compilation. As stated beforehand
all galaxies, irrespective of their morphology and mass, present a
decline of $\Sigma_{*}$ with radius. The central ($\Sigma_{*,cen}$)
and effective ($\Sigma_{*,Re}$) stellar mass densities increases with
the integrated mass and morphology (as they become earlier) for
galaxies later than S0s. \citet{rosa14} already shown that Elliptical
galaxies, are more compact, with a steeper radial decrease of
$\Sigma_{*}$, in particular more massive ones. Note that when the
radial distribution of $\Sigma_{*}$ is not normalized by $R_e$, the
picture changes, since there is a correlation between M$_*$ and R$_e$
\citep[e.g.][]{cappellari16,vandesande19}.

\citet{rosa14} showed that galaxies, in general, present two regimes
in the radial distribution of $\Sigma_{*}$, one at R$<$0.5$R_e$,
steeper (with a slope ranging between $\alpha\sim-$0.5 dex/Re and
$\sim-$1.5 dex/Re), and another one at R$>$0.5$R_e$ (with an almost
constant slope of $\alpha\sim-$0.5 dex/Re). The first one corresponds
(broadly) to the bulge, while the later one corresponds to the disk
(for late-type galaxies). Finally, the radial profile is steeper
($\sim-$1.5 dex/Re) for more massive galaxies and shallower for less massive
ones ($\sim-$0.5 dex/Re). All these trends are qualitatively
appreciated in Fig. \ref{fig:RAD_Sigma_M}. They agree with the
explorations of the rSFHs reviewed in previous sections: 
the stellar mass assembles faster and with stronger intensities (higher SFRs)
in the inner regions of more massive/early-type galaxies than in the
outer-regions of less massive/late-type ones.

\subsubsection{Stellar age and metallity gradients}
\label{sec:grad_age}

\begin{figure}[h]
\includegraphics[width=5in, clip, trim=15 15 50 20]{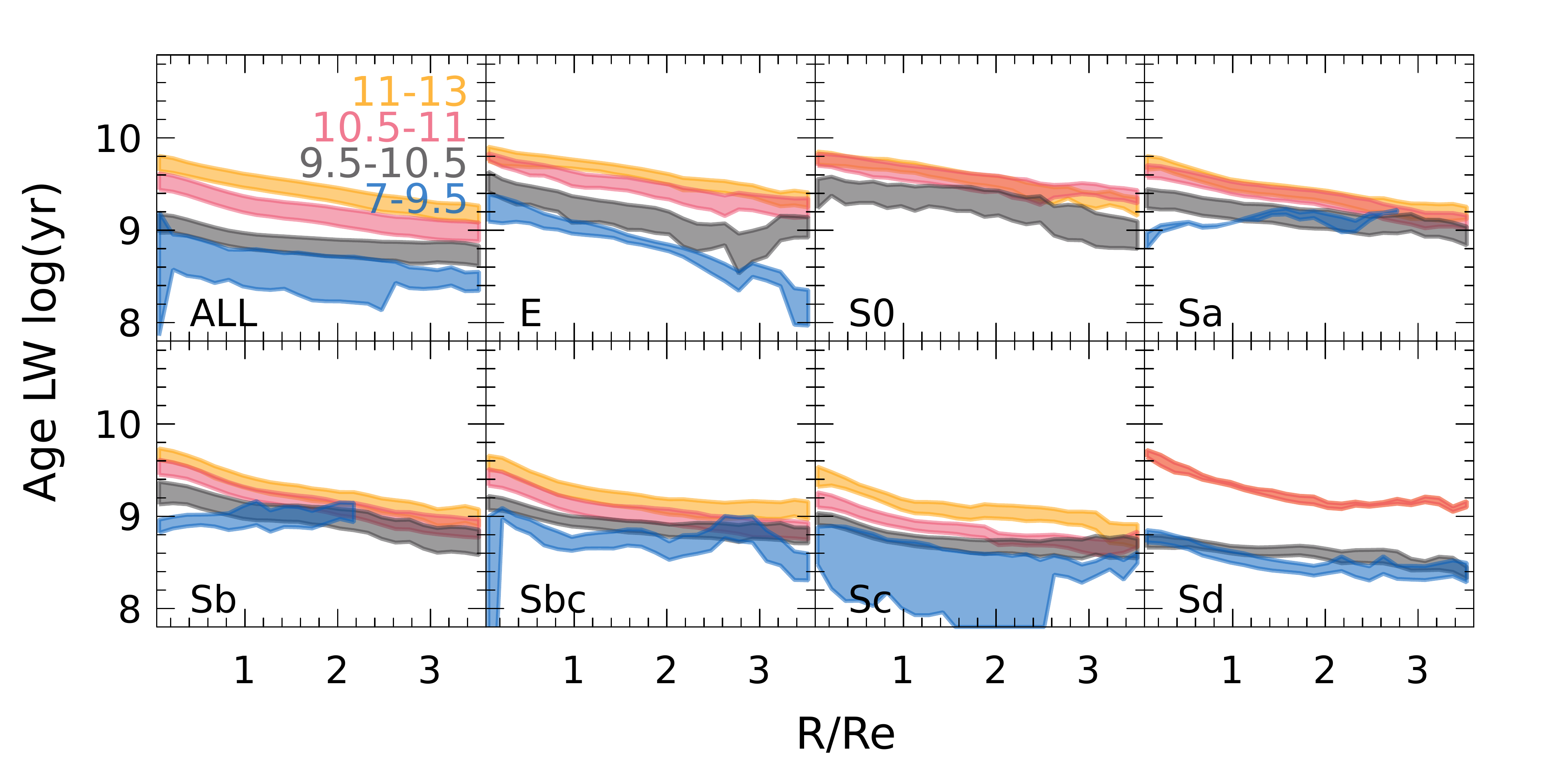}
\caption{Azimuthal-averaged radial distributions of the luminosity-weighted stellar ages for the galaxies in this review segregated by stellar mass and morphology; colors and symbols as in Fig. \ref{fig:RAD_Sigma_M}.}
\label{fig:RAD_Age_LW}
\end{figure}


Pioneering works on the radial gradients of colors and stellar indices
were already performed four decades ago \citep[e.g.][]{faber1977}.
Radial changes in both properties were interpreted as a physical
variation in the average ages (and metallicities) of the stellar
populations.  The combination of better quality long-slit spectroscopy
with broad-band imaging, focused on the study of elliptical galaxies
\citep[e.g.][]{pele90,gonza94}, shows that they present a shallow
gradient in age and a clear negative gradient in metallicity. These
early results were revisited based on the analysis of larger samples
of galaxies, using multi-broad band photometry, long-slit and
even IFS data
\citep[e.g.][]{mehlert03,patri07,rawle08,coccato10,kuntschner10,spolaor2010,koleva2011,mcdermid15,rosa14,rosa15,godd15,ruiz16,oyarzun19}. All
these studies agreed on reporting a negative gradient in the stellar
metallicity, [Z/H], but the consensus was less clear for the
presence of a negative age gradient. \citet{godd15} reported a flat or
even a positive gradient in the stellar ages of early-type galaxies,
explained by an "outside-in" scenario, based on data extracted from
the primary sample of the MaNGA survey (reaching 1.5 R$_e$). Similar
positive age gradients are sometimes described from analysis based on
broad-band colors \citep[e.g.][]{tortora10}. In spectroscopic based
studies clear positive age gradients were found before only in the
very central regions \citep[$\sim$Re/8][]{kunt15,mcdermid15}. At
larger scales, different authors have found contradictory
results. They either found a wide range of gradients, from negative to
positive \citep[e.g.][]{spolaor2010,koleva2011} or shallow negative
gradients or absence of a gradient
\citep[e.g.][]{patri07,rawle2010}. However, when the spectroscopic
data reach the outer regions of galaxies (R$>$2 R$_e$), most of the
studies report a negative gradient in the stellar ages supporting an
"inside-out" scenario \citep[e.g.][]{rosa14,cheng15,rosa15,zheng16}.

Figure \ref{fig:RAD_Age_LW} shows the radial gradients of the
luminosity-weighted ages derived using our own analysis based on our
compilation of IFS observations. For early-type galaxies of any mass
there is a clear negative gradient, in agreement with most of the
published results, although I should stress that there is no full
consensus in this regard. The lack of consensus is deeply ingrained in
the procedure adopted to derive the stellar ages. \citet{rosa14} and
\citet{rosa15} used a full spectral fitting based on the {\sc
  starlight} code \citep{starlight}. This code, like {\sc Pipe3D} implemented
in this review, uses the full optical spectral range and does not
removes the shape of the spectral continuum (i.e., features like D4000
are important in that derivation), contrary to the one adopted by
\citet{godd15}. Removing the shape has strong implications, since the
dust attenuation is no longer recovered \citep[although not removing
it requires a high quality spectrophotometric
calibration][]{walcher11}.  Recent results show that the polynomial
functions adopted in some methods to describe and correct the shape
are indeed very similar to the expected correction by a dust
attenuation law \citep{dap}. Finally, I should indicate that the use
of different wavelength ranges \citep[e.g., Atlas3D covers just the
wavelength range between 4800-5380\AA,][]{cappellari10}, or the
inclusion or exclusion of the alpha-enhancement in the models could
contribute significantly to the reported differences.

\begin{figure}[h]
\includegraphics[width=5in, clip, trim=15 15 50 20]{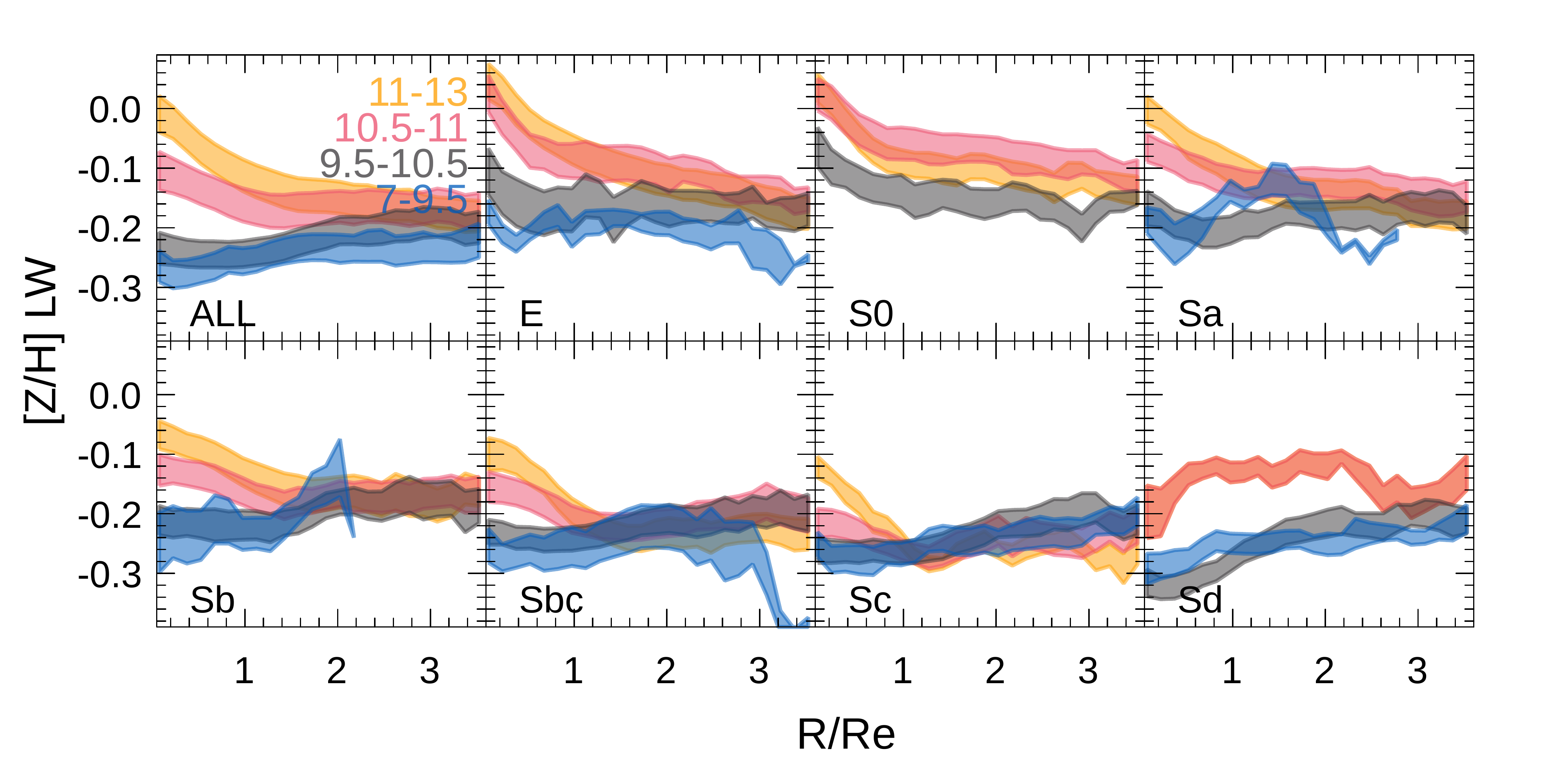}
\caption{Azimuthal-averaged radial distributions of the luminosity-weighted stellar metallicities ([Z/H]) for the galaxies in this review segregated by stellar mass and morphology; colors and symbols as in Fig. \ref{fig:RAD_Sigma_M}.}
\label{fig:RAD_ZH_LW}
\end{figure}

The number of similar studies for late-type galaxies is limited.
This is mostly due to the difficulty of removing the emission lines that
contaminate the spectra, in particularly important ranges, like the
regions covered by the Balmer absorption lines \citep[sensitive mostly
to the age, e.g., ][]{worthey94}. Different fitting techniques have
addressed this issue in different ways
\citep[e.g.][]{sarzi06,sanchez-blazquez14,pipe3d}, removing efficiently this
contamination. In this regard the consensus is that the averaged ages
of the stellar populations present a negative gradient for all
late-type galaxies more massive than 10$^{9.5}$M$_\odot$, with a
somehow similar slope \citep{patri14b,rosa15,godd15,ruiz17}. The stellar metallicity
presents a negative gradient only for earlier-type spirals
(Sa,Sb,Sbc) and more massive ones (M$_*>$10$^{9.5}$M$_\odot$),
with a slope that becomes shallower for low-mass and late-type
galaxies. This is illustrated in Figure \ref{fig:RAD_ZH_LW}, which
shows the radial gradients of the luminosity weighted stellar
metallicities derived for our data compilation. Again, these
results agree with the scenario reviewed in Sec. \ref{sec:local_CMF},
with an "inside-out" growth for more massive and
earlier-type galaxies, and an "outside-in" growth, for the less
massive and later-type ones. 

However, we should be cautious for the
low-mass/later-type galaxies. For galaxies with strong star-formation
with respect to their mass (i.e., the low mass SFGs, mostly Sc and
Sd, with high sSFR) that are more metal poor ones (following the
MZR), the estimation of age and metallicity is highly
degenerated. The degenerancy between these two parameters is well
known and widely studied \citep[e.g.][]{walcher11}. However, in this
range it is far more complicated: (i) the presence of strong emission
lines on top of the Balmer absorptions in combination with the fact
that the spectral features are narrower due to the combination of the
type of stars and the low velocity dispersion, affects the accuracy in
the removal of the contamination by the emission lines; (ii) the
absorptions by metallic elements are weaker than in the case of more
massive galaxies, making it more difficult to obtain a precise
metallicity; (iii) this, in combination with the limited range of values for
the 4000\AA\ break when the light of the stellar population is dominated
by young stars, affects the accuracy and precision in the derivation
of both ages and metallicities.

\subsubsection{Oxygen abundance gradients}
\label{sec:grad_OH}

Before the stellar population analysis had enough accuracy to explore
the metallicity gradients in SFGs, for decades, the study of the
chemical enrichment was based in the analysis of the gas-phase
abundance. In particular \ion{H}{ii} regions have been considered as
good tracers of the chemical composition in these galaxies.  It was
already in the seventies of the past century that the presence of a
negative abundance gradient across the discs of nearby galaxies
had been uncovered \citep{sear71,comte74}, even in the Milky Way
\citep[MW,][]{peim78}. This gradient has been confirmed by different
observations
\citep[e.g.][]{matt89,vila92,Martin:1994p1602,zaritsky94,esteban18},
and some deviations from this single negative gradient have been
appreciated \citep{belley92,pepe96,roy97}. Like in the case of other
gradients explored in this review, the use of the effective radius as
a scale-length has helped to uncover the patterns and differences
between galaxy types. \citet{diaz89}, and later \citet{vila92},
demonstrated that using a scale-length anchored to the galaxy
properties (like R$_e$ and R$_{25}$), instead of a pure physical
radial distance, reduces the scatter in the distribution of slopes of
the abundance gradients.
\begin{marginnote}[]
\entry{R$_{25}$}{radius of the elliptical isophote of a galaxy at which it is reached a surface-brightness of 25 mag arcsec$^2$ in the V-band.}
\end{marginnote}
Following this philosophy, with the advent of
large IFS observations, \citet{sanchez12} and \citet{sanchez14},
showed that all star-forming galaxies more massive than
M$_*>$10$^{9.5}$M$_\odot$, present a characteristic oxygen abundance
gradient of $\sim -$0.1 dex/R$_e$ when computed within the range
between 0.5-2.0 R$_e$ (i.e., the disk region).  Contrary to early
results \citep[e.g.][]{zaritsky94} the slope of the gradient in this
region does not seem to be strongly affected by other properties of
the galaxy, like morphology or the presence of a bar. This result was
confirmed also for the stellar metallicity \citep{patri14b}.
\begin{marginnote}[]
\entry{Characteristic gradient}{oxygen abundance presents similar slopes in their gradients for disk galaxies with M$_*>$10$^{9.5}$M$_\odot$.}
\end{marginnote}

\begin{figure}[h]
\includegraphics[width=5in, clip, trim=15 15 50 20]{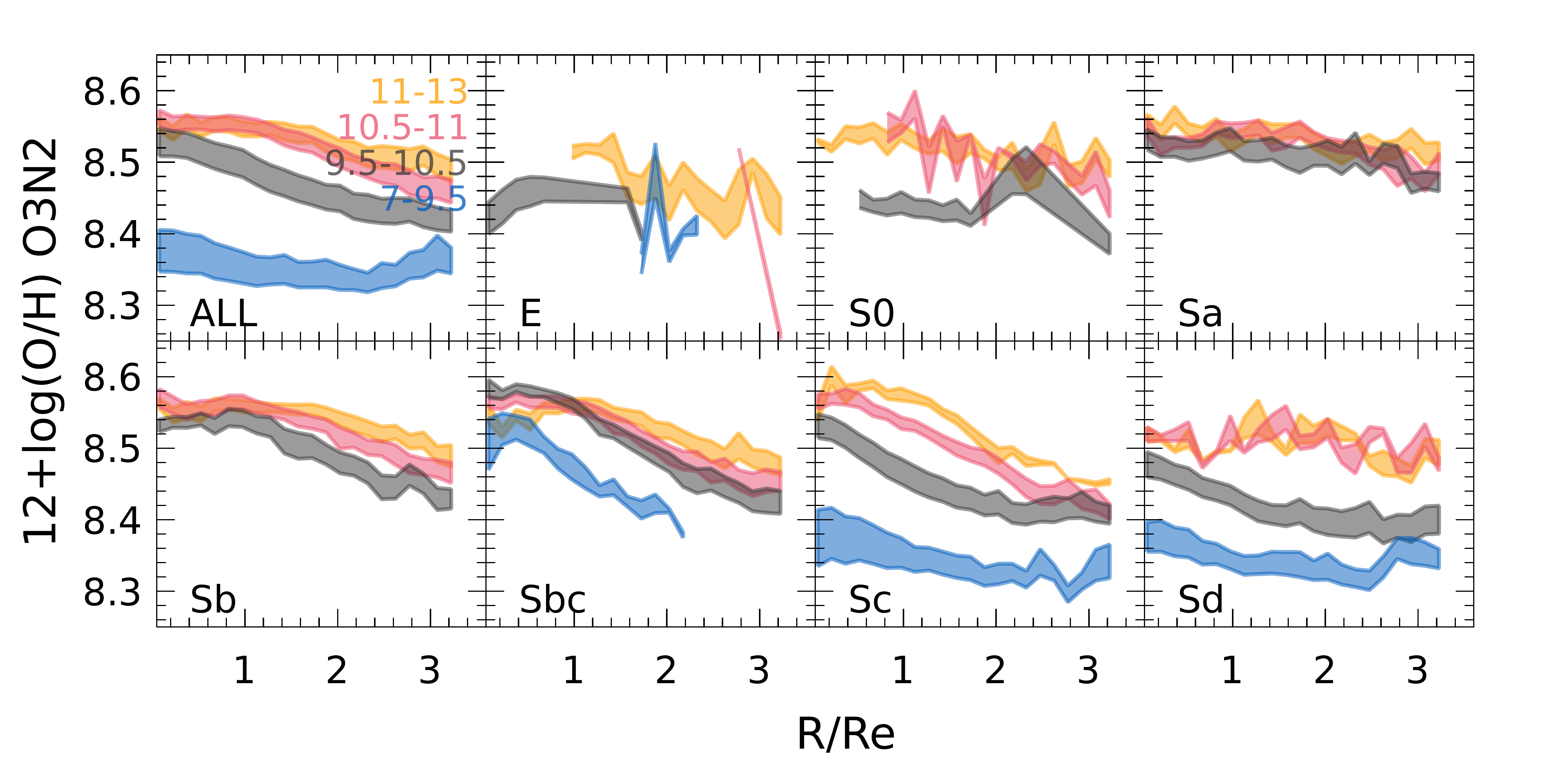}
\caption{Azimuthal-averaged radial distributions of the gas-phase oxygen abundances derived using the O3N2 calibrator for the galaxies in this review segregated by stellar mass and morphology; colors and symbols as in Fig. \ref{fig:RAD_Sigma_M}.}
\label{fig:RAD_OH_O3N2}
\end{figure}

Besides the presence of a common/characteristic gradient, for some
galaxies there is an apparent flatenning or drop of the abundance in
the central regions (R$<$0.5 R$_e$), and a flatenning in the outer
regions (R$>$2 R$_e$) for all the galaxies analyzed. These results
were confirmed using larger IFS datasets
\citep[e.g.][]{laura16,zinc16,epm17,belf17,poet18} or observations
with exquisite spatial resolution \citep[i.e., MUSE
data,][]{laura18}. Some works described a continuous change in the
abundance gradient from almost flat for low-mass galaxies
M$_*<$10$^{9}$M$_\odot$ towards the described characteristic slope,
reached at M$_*\sim$10$^{9}$M$_\odot$ \citep{bresolin15,belf17}. In
other cases no drop or flatenning is appreciated at any scale, and a
single slope for the abundance gradient seem to reproduce the data
\citep{pilyugin14}. The actual selection of the star-forming regions
used to derive the gradient may affect somehow the results. For
example, the inner drop is not detected when the SF/\ion{H}{ii}
regions in the intermediate area between the Kauffmann and Kewley
curves are excluded \citep[e.g.][]{zinc16}. I should stress here that
selecting just regions below the Kauffmann (or Kewley) curve may
exclude star-forming regions with large forbidden line ratios (i.e.,
nitrogen enhanced), like the ones described in the inner region of
galaxies by \citet{kennicutt89}, \citet{ho14} and
\citet{sanchez12}. The nature of those regions is still not clear, but
they are easily identified as ionized nebulae in narrow-band images or
emission line maps based on IFS data.

Figure \ref{fig:RAD_OH_O3N2} shows the radial profiles of the oxygen
abundance for the sample analyzed in this review, segregated by mass
and morphology. These profiles reproduce the results reported in
the literature well, in particular those by \citet{laura18}.  This article
presented the most detailed analysis so far on the shape of abundance
gradient. It used the super-high resolution data obtained by
MUSE, on a sample of more than 100 galaxies, confirming that the slope
of the abundance gradient in the disk regime is indeed very similar
for all galaxies when the distribution is normalized to the effective
radius \citep[as suggested by][]{sanchez14}. Regarding the shape, it
may present either a single slope, or the combination of a monotonic
decrease with either an inner drop and an outer
flatenning. \citet{laura16} and \citet{belf17} already showed that the
inner drop is ubiquitous in massive galaxies
(M$_*>$10$^{10}$M$_\odot$), and not present in low-mass ones
(M$_*<$10$^{9.5}$M$_\odot$). This is seen
Fig. \ref{fig:RAD_OH_O3N2}, in particular for Sb to Sc galaxies. The
effect of radial migration towards the Lindblad resonances, or the
freezing of the chemical enrichment associated with the quenching of the
star-formation in the inner (bulge dominated) regions are plausible
explanations for this deviation from the monotonic decrease. On the
other hand, the outer flatenning does not seem to have a clear
pattern. It does not be directly connected with the shape of the surface
brightness profile of disk galaxies \citep[e.g.][]{marino16}. A change
in the efficiency of the star-formation in the outer regions of the
disk \citep[e.g.][]{thil07}, or outer radial migrations, are possible
explanations to this change in the shape of the abundance
gradient \citep[e.g.][]{bresolin17}. \citet{laura18} present the concept of a chemical enrichment
scale-length in a galaxy (R$_{O/H}$), defined as the distance
at which the oxygen abundance decreases by $\sim$0.1 dex. They
show that this parameter presents a linear and almost one-to-one
correlation with the effective radius.  This relation was already
outlined by \citet{bresolin15}, when they showed the relation between
the slope (in physical scales for the galactocentric disances) and the
scale-length of the disk. As a consequence, when normalized by
this scale-length the inner drop and the outer flatenning happens
virtually at 0.5R$_{O/H}$ and 2.0 R$_{O/H}$ for all galaxies, with a
very narrow scatter.

Some recent results have questioned the existence of a characteristic
abundance gradient. \citet{belf17}, using data from the MaNGA survey,
showed that for some O/H calibrators \citep{maio08} there is a strong
dependence of the gradient with the stellar mass, with low-mass
galaxies (M$_*<$10$^{9.5}$M$_\odot$) presenting an almost flat
abundance gradient. However, for other calibrators \citep{pettini04}
the distribution is consistent with a single slope for most of the
galaxies. \citet{poet18}, using data from the SAMI survey, reported a
weak trend with the mass only when using an R23-based oxygen abundance
calibrator \citep[like the one proposed by][]{maio08}. For other
calibrators, they found an almost constant slope. Moreover, as
indicated before, the analysis by \citet{laura18}, using MUSE data
(with much better spatial resolution), does not provide any evidence
of a dependence of the slope with M$_*$. In the regime of low-mass
galaxies (down to M$_*<$10$^{8.5}$M$_\odot$), the most recent
explorations have not found any dependence with M$_*$ either
\citep{bresolin19}. Thus, the possible dependence with the stellar mass
has been reported only for low-mass galaxies, for particular calibrators,
and using IFS data of low spatial resolution.

The existence of a general rMZR, discussed in Sec. \ref{sec:local_MZ},
can explain the observed oxygen abundance gradients in galaxies simply
considering the existence of an inverse radial gradient in the $\Sigma_{*}$,
(Sec. \ref{sec:grad_Mass}). Indeed, \citet{jkbb16} was able to
reproduce not only the negative abundance gradient but also the
characteristic slope of the gradient in disk galaxies
($\sim -$0.1 dex/Re). The small scatter in the rMZR
relation ($\sim$0.05-0.07 dex) and the abundance gradients strongly
supports the idea that oxygen chemical enrichment is dominated by
local processes, being tightly related to the local star-formation
history, without little effects of migrations or mixing beyond few
kiloparsecs. This supports the idea that the radial-mixing scale
\citep[introduced by][, based on the scatter of the abundance
gradient]{censushii}, the typical distance of oxygen abundance mixing
in a galaxy, is rather small ($\sim$0.3-0.4 r$_e$).

In addition to the exploration of the radial gradient of the oxygen abundance,
recent studies have been focused on the exploration of possible azimuthal
variations in the oxygen abundance. These variations could be induced by the
nature of spiral arms as possible density waves, that may produce a
differential star-formation history \citep[e.g.][]{grand16,2018NatAs.tmp..159P} with
respect to the rest of the disk, and introduce an azimuthal variation
in the stellar or gas metallicity. Bars could also induce radial
motions that produce azimuthal variations in these physical parameters
\citep[e.g.][]{atha92,min12,dimat13}. Pioneering observations have
found that the abundance gradient shows subtle differences when the
galaxy is divided in quadrants \citep{rosales11}. More recent studies
have found significant azimuthal variations in the oxygen abundance in
a handful of galaxies
\citep[e.g.][]{li13,censushii,zinc16,laura16b,ho17,vogt17,ho18}. However, when making
a statistical analysis there are very little differences, even when
considering barred and unbarred galaxies \citep{laura16,zinc16}. An
alternative interpretation is that the observed azimuthal variations
are induced by the presence of anomalous abundance regions \citep{hwang18}, more
related to gas inflow than to secular processes \citep{laura19}. In
any case, azimuthal variations may be present in some 
galaxies induced by the combination of very particular dynamical
processes or even external effects. However, it seems to have a subtle effect
in the radial abundance gradient.


While there are several studies of the oxygen abundance gradients in
nearby star-forming galaxies, the systematic studies of nitrogen
abundances (N/H) or nitrogen-to-oxygen abundances (N/O) are less
numerous. In general, these studies have found a
steady monotonic decrease of the N/O ratio from the inner to the outer
regions \citep{epm14,belf17}. There are no significant deviations
from this pattern described in the literature. In general it is
accepted that most of the Nitrogen observed in galaxies in the nearby
universe is due to secondary production, at least for galaxies more massive than
M$_*>$10$^{9.5}$M$_\odot$. Thus the N/O gradient follows the
oxygen abundance one in general \citep[e.g.][]{belf17}.

\subsubsection{Star-formation rate gradients}
\label{sec:grad_SFR}

\begin{figure}[h]
\includegraphics[width=5in, clip, trim=0 15 50 20]{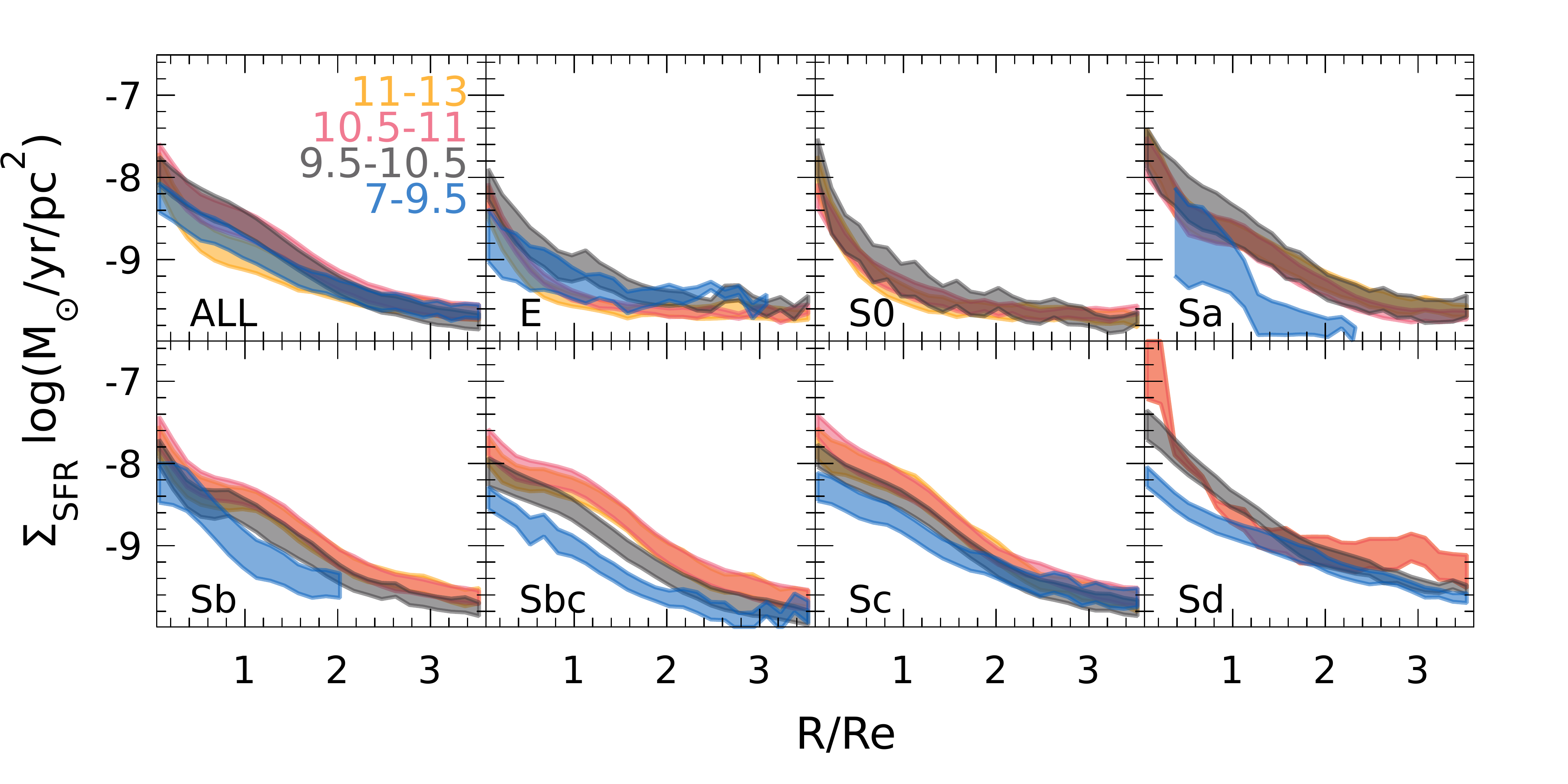}
\caption{Azimuthal-averaged radial distributions of the surface density of the SFR derived from the H$\alpha$ intensity for the sample of galaxies, segregated by stellar mass and morphology; colors and symbols as in Fig. \ref{fig:RAD_Sigma_M}.}
\label{fig:RAD_SFR}
\end{figure}

The presence of a monotonic decrease of the gas phase oxygen and
nitrogen abundances in star-forming (disk mostly) galaxies, described
in the previous section, has been interpreted as a direct evidence of
an inside-out formation of these objects \citep{matt89,bois99}. Under
this scenario the gradient is formed as a consequence of the increased
timescales of the gas infall with galactocentric distance. The
differential amount of gas supply in the inner regions compared to
that of the outer ones implies a differential SFR,
assuming a constant or similar depletion time across the disk
\citep[i.e., a common S-K law][]{schmidt59,kennicutt89}. The
subsequent differential chemical enrichment naturally explains the
oxygen (and nitrogen) abundance gradients. Most chemical abundance
codes aimed to reproduce the described gradients, in particular those
in our Galaxy, are based on this scheme on a first order
\citep[e.g.][]{molla05,tissera:2013aa,carigi19}. In summary, this
scenario predicts and requires the existence of a monotonic decrease
of the star-formation with the galactocentric distance.

Figure \ref{fig:RAD_SFR} shows the radial distribution of the
SFR density ($\Sigma_{\rm SFR}$). There is a clear peak in the
central regions with a decrease towards the outer ones, broadly in
agreement with the proposed inside-out scenario. \citet{rosa16} was
one of the first studies exploring the radial dependence of the
star-formation with both the stellar mass and morphology of
galaxies. They found results qualitatively similar to the ones
presented in Fig. \ref{fig:RAD_SFR}. Besides the radial decline in the
$\Sigma_{SFR}$ for all galaxies, early-type galaxies (E/S0) present
lower values of star-formation than the ones seen in late-type
counterparts at any stellar mass, for any galactocentric
distance. Thus, in these galaxies the decline of the star-formation
happens at every radial distance. However, the maximum decline is found
in the central regions. As already indicated in previous sections, the star-formation
activity in early-type galaxies is a somewhat recent topic (see
Sec. \ref{sec:diag_pos}), being explored using IFS by a very few
number of studies \citep[e.g.][]{gomes16}. In general, these galaxies
host less number of star-forming areas, and those star-forming areas
are forming stars at a lower rate with respect to those in late-type
galaxies \citep[e.g.][]{mariana19}.


\begin{figure}[h]
\includegraphics[width=5in, clip, trim=0 15 50 20]{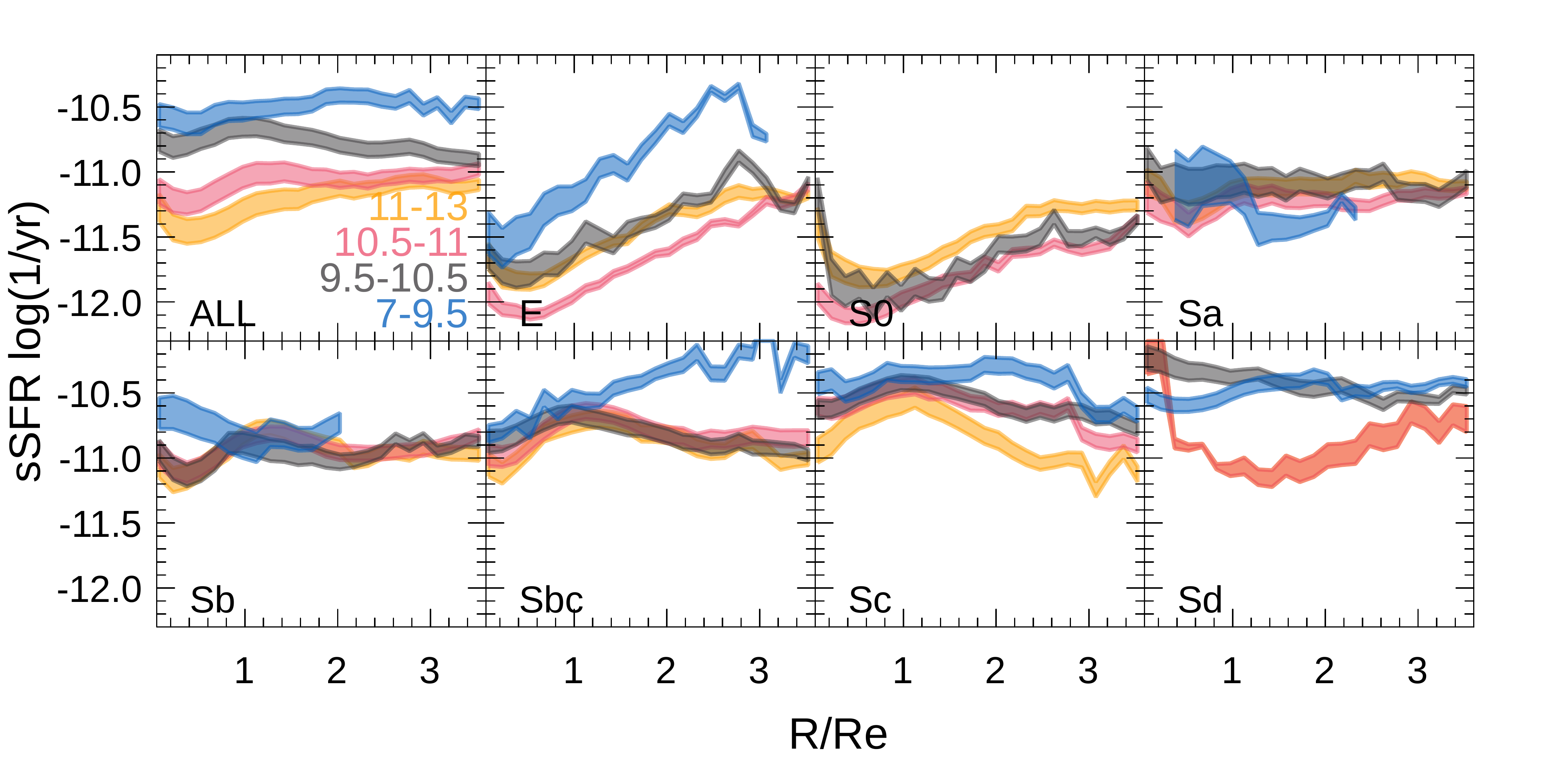}
\caption{Azimuthal-averaged radial distributions of the specific SFR for the galaxies segregated by stellar mass and morphology; colors and symbols as in Fig. \ref{fig:RAD_Sigma_M}.}
\label{fig:RAD_sSFR}
\end{figure}

For late-type galaxies (LTGs), the most common morphological type among SFGs,
$\Sigma_{\rm SFR}$ is 0.5-1 dex stronger than that of early-types
at any galactocentric distance. Among them, there is a clear variation
with the stellar mass and the morphology: more massive and later
LTGs present stronger SFRs. However, once normalized by
the stellar mass density ($\Sigma_{*}$), the differences are less
evident. Figure \ref{fig:RAD_sSFR} shows the radial distribution of
the spatially-resolved specific SFRs
($=\frac{\Sigma_{SFR}}{\Sigma_{*}}$). For LTGs (from Sa to
Sc), the sSFR presents a very similar value of about
$\sim 10^{-11}$yr$^{-1}$, with a possible variation of ~0.2 dex
between earlier and later spirals \citep[in agreement
with][]{rosa16,belfiore17b,sanchez18}. The sSFR has units of the
inverse of time, and indeed it inverse could be interpreted as the
amount of time required to form the current stellar mass at the
current SFR \citep[e.g.][]{rosa16}. In purity, this is
true if the current observed M$_*$ is corrected by the mass locked
in dead stars along cosmological times. However, that
correction has a maximum of $\sim$30\%, for the adopted IMF
\citep[e.g.][]{court13}. Therefore, it does not affect significantly
the order of magnitude of the estimation. This result indicates that at the
current $\Sigma_{\rm SFR}$ galaxies would require hundreds of Gyr to
form their current $\Sigma_*$ (being particularly true for the
early-type galaxies). Thus, the SFR in the past must have been
considerably higher than today, at any galactocentric distance. This
is in agreement with the observations of the evolution of the SFR
along the time based on integrated quantities
\citep[e.g.][]{speagle14}, or by the exploration of the cosmic
star-formation density in the universe
\citep{madau14,driver17,sanchez18b}.

The low dispersion of the sSFR at any galactocentric distance for
late-type (mostly SF) galaxies was interpreted by \citet{rosa16} as a
direct consequence of the existence of the rSFMS. Just comparing the
radial distributions of the two involved quantities, $\Sigma_*$
(Fig. \ref{fig:RAD_Sigma_M}) and $\Sigma_{\rm SFR}$
(Fig. \ref{fig:RAD_SFR}), it is clear that a rSFMS should
hold. Contrary to late-type galaxies, early-type (E/S0) ones exhibit a
clear decrease of the sSFR towards the inner regions, at any mass,
despite the peak observed in $\Sigma_{SFR}$ for these galaxies.  This
has been interpreted by different authors as a clear evidence of an
inside-out quenching
\citep{rosa16,belfiore17b,li17,ellison18,sanchez18}.  that is
evidently related to the presence or influence of a bulge in the
halting of the SF activity, either directly \citep[e.g.][]{martig09},
or by hosting an AGN. In terms of their distribution along the
$\Sigma_{\rm SFR}-\Sigma_{*}$, early type galaxies, dominated by
retired areas, are located well below the loci of the rSFMS
\citep[][and Fig. \ref{fig:local_SFMS}]{mariana19}.

\subsubsection{Gas mass density gradients}
\label{sec:grad_Mgas}

\begin{figure}[h]
\includegraphics[width=5in, clip, trim=0 15 50 20]{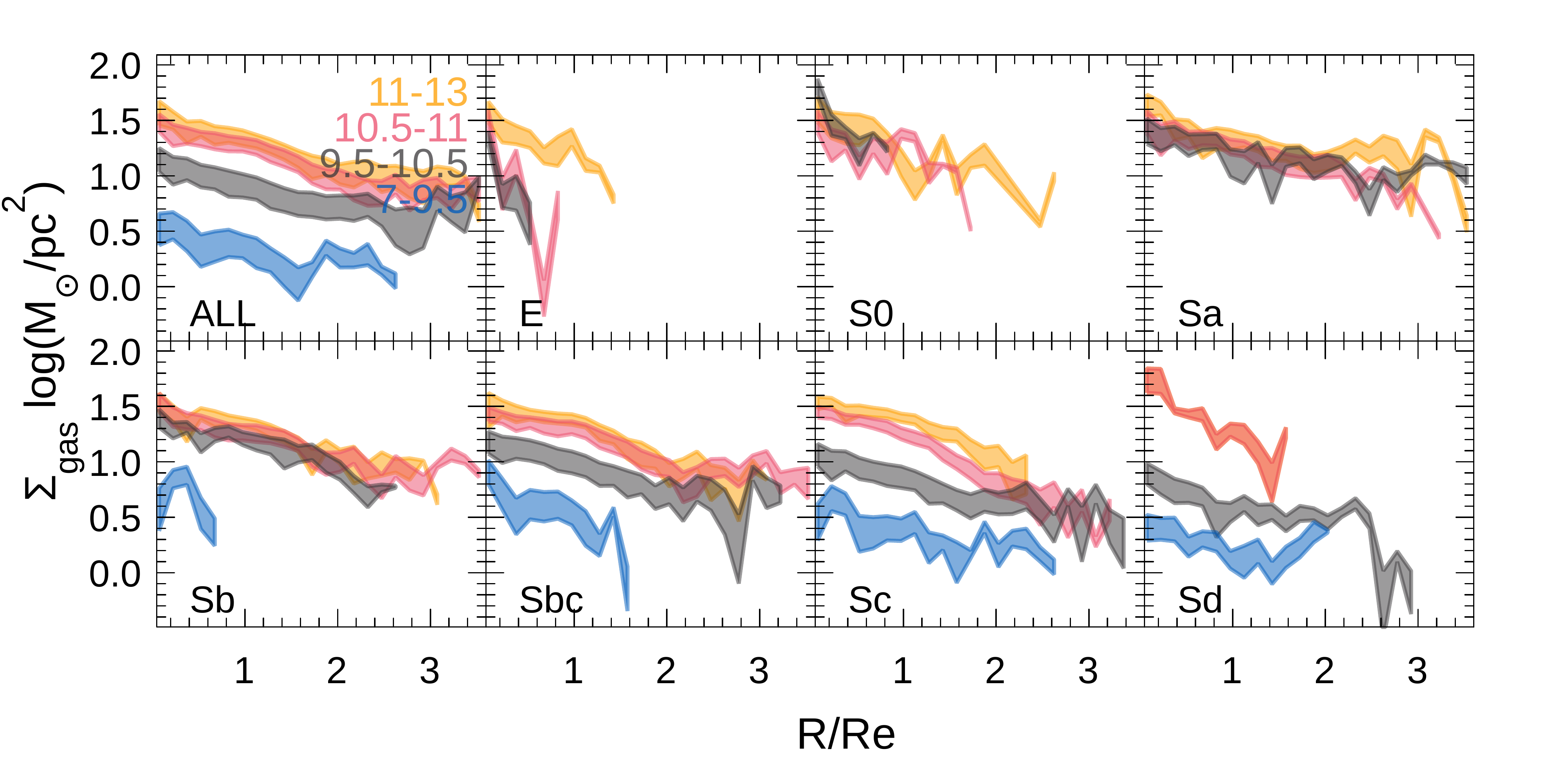}
\caption{Azimuthal-averaged radial distributions of the surface density of the gas mass estimated based on the dust-to-gas ratio for the galaxies in this review segregated by stellar mass and morphology; colors and symbols as in Fig. \ref{fig:RAD_Sigma_M}.}
\label{fig:RAD_Mgas}
\end{figure}

The exploration of the radial distribution of $\Sigma_{gas}$ and its
connection with $\Sigma_{\rm SFR}$ (through the SFE) are giving us
clues of how the SF is maintained (or enhanced) and the reason for its
quenching. In Sec. \ref{sec:local_SK} I described the few cases of
galaxies covered by IFS-GS and spatially-resolved molecular gas
observations.  Regarding the radial distribution, the molecular gas
content peaks softly in the inner regions \citep{bolatto17,li17}, with
a more or less monotonic decrease from the central regions outwards
\citep[e.g.][Fig. 4]{colombo18}, with a scale-length (effective or
half-light radius) that matches well that of the stellar
light \citep[e.g.][, Fig. 13]{bolatto17}, confirming the results by
previous studies \citep{regan01,davies13}. Figure \ref{fig:RAD_Mgas}
shows the radial distribution of the molecular gas surface density
derived using the dust-to-gas proxy described before
(Sec. \ref{sec:local_SK}) for the well-resolved galaxies in our IFS
compilation. The general radial decline
reported in the literature is clearly reproduced.

I segregate the radial distributions by stellar mass and morphology,
following the scheme of previous figures. In general, the gas surface
density strongly depends on the total stellar mass for late-type
galaxies \citep[e.g.][]{saintonge2011}. However, for early-type
galaxies (E,S0 and Sa), there is no clear trend with the mass, in
agreement with the results by \citet{young11}. In addition, for the
same mass galaxies of different morphology present similar radial
gradients of $\Sigma_{gas}$, as reported by
\citet{colombo18}. Finally, the molecular gas fraction decreases
for earlier and more massive galaxies. This decline is stronger in the
inner regions, as it is evident when comparing the distributions shown
in Fig. \ref{fig:RAD_Sigma_M} and Fig. \ref{fig:RAD_Mgas}.  This seems
to indicate that the observed decline in the sSFR from the inside-out
(Fig. \ref{fig:RAD_sSFR}) is due primarily to a relative lack of
molecular gas. However, a change in the SFE also may play a role.



\begin{figure}[h]
\includegraphics[width=5in, clip, trim=0 15 50 20]{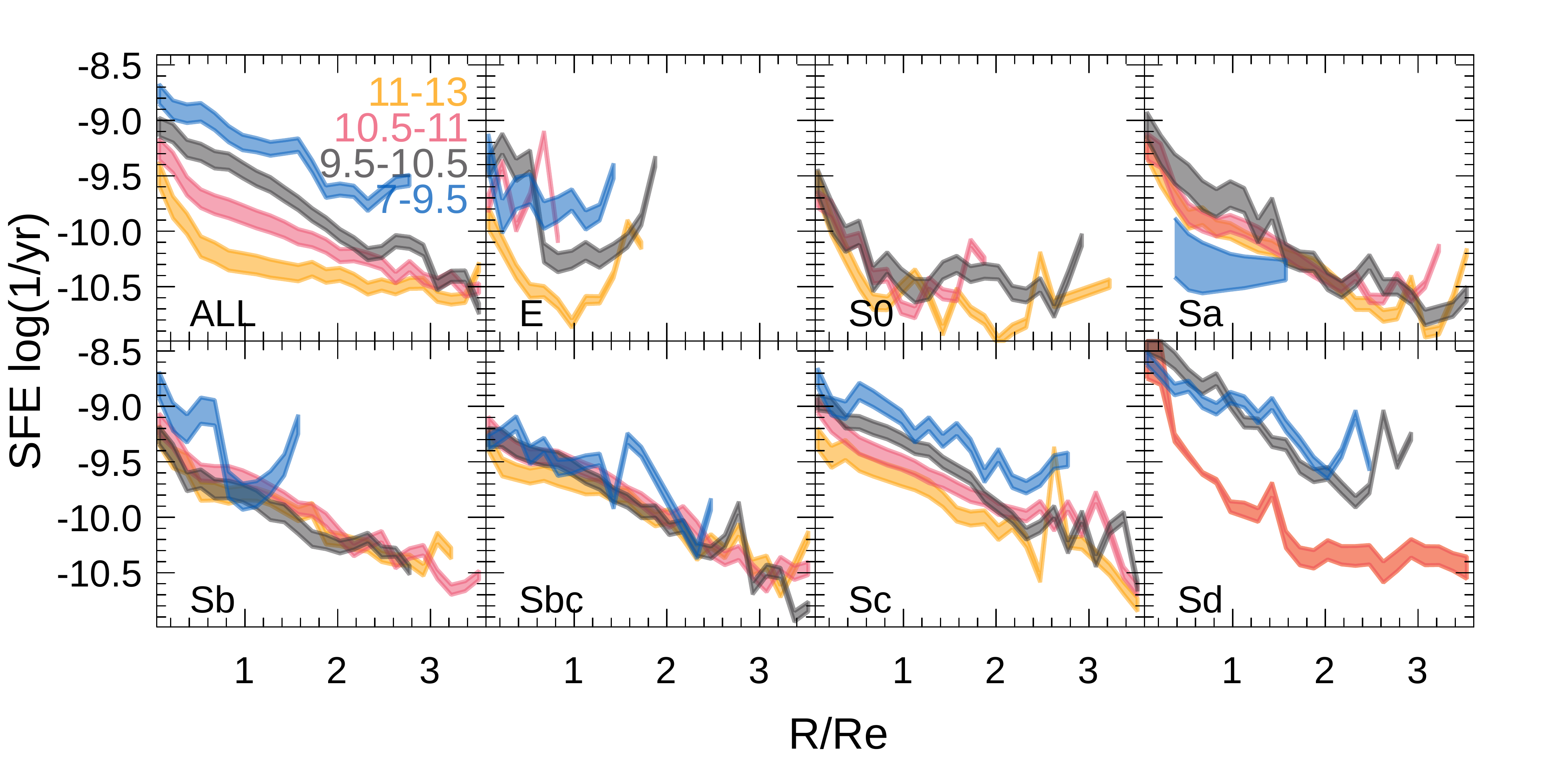}
\caption{Azimuthal-averaged radial distributions of the star-formation efficiency for the galaxies considered in this review segregated by stellar mass and morphology; colors and symbols as in Fig. \ref{fig:RAD_Sigma_M}.}
\label{fig:RAD_SFE}
\end{figure}

\citet{utomo17} found that, on average, the radial distribution of the
depletion time $\tau_d$ (the inverse of the SFE) is rather flat when
analyzed spaxel by spaxel at a kpc-scale. However, this quantity
presents a clear radial trend, when radial averages are considered, as
shown by \citet{colombo18}. In the inner regions $\tau_d$ has lower
values, of the order of $\sim$1 Gyr (SFE$\sim$10$^{-9}$yr$^{-1}$), and
up to $\sim$10-20 Gyr (SFE$\sim$10$^{-10.5}$yr$^{-1}$) in the outer
ones, with a clear segregation by mass and morphology (e.g., Fig. 4 of
their article). Despite the different proxies adopted to derive the molecular
gas mass (dust attenuation vs. CO observations), similar results are
recovered for our dataset. Figure \ref{fig:RAD_SFE}
shows the radial distribution of the SFE segregated by
stellar mass and morphology. The monotonic decrease from the center
outwards is clearly shown, with more massive galaxies presenting lower
SFE (larger $\tau_d$) than less massive ones. When segregating by
morphology the picture is less clear. While for some morphological
types (e.g., E or Sc) there is still a segregation by mass, for other
ones (e.g., S0 or Sbc) its seems that all galaxies have a similar SFE
distribution respective of their mass, without a clear
pattern. However, when comparing galaxies of different morphology and
the same stellar mass the trend is clear: in general, later galaxies
present larger SFE (shorter $\tau_d$) than earlier ones, at any
galactocentric distance. Recent results (Ellison et al., in prep.),
indicates that this change in the SFE is one of the causes of the dispersion
in the rSFMS, that can be observed as a segregation by morphology
\citep[e.g.][]{mariana19}. If $\tau_d$ is interpreted as the time
required to consume the current molecular gas mass with the current SFR,
then this time is shorter in the inner regions of less massive and
later galaxies and in in the outer regions of more massive and earlier
galaxies.

I should note that the reported distributions may change in the
presence of bars, that are known to induce radial movements of gas \citep[e.g.][]{shlos89,boone07,knapen19},
and in particular they can produce an increase of the molecular gas
and even the SFE in the inner regions of galaxies (in contrast with
what is found in here).

\section{What we have learned?}
\label{sec:end}

Along this review I summarised the most recent results in the
characterisation of the spatially-resolved spectroscopic properties of
galaxies in the nearby universe ($z<$0.1). For doing so, I compile the
largest possible sample of galaxy data observed using IFS, and
replicate the published results (to a certain extend).

In summary, recent results have found that the evolution of galaxies
is governed by local properties, but influenced somehow by global
ones. The duty-cycle of gas-to-stars, the chemical enrichment, and the
quenching and halting of both follow local/spatially resolved
scaling/evolutionary laws that are verified at kiloparsec scales in
galaxies. Among them, the SK-law, rSFMS and rMZR relations, verified
for star-forming regions in galaxies, are followed by regions of
different galaxies that to some extent follow the same resolved SFHs
and ChEHs.  These evolutionary sequences depend both on the
galactocentric distance, the stellar mass, and morphology, indicating
that most probably they actually depend on the local strength of the
gravitational potential and the dynamical/orbital structure. The
central regions of more massive and earlier-type galaxies evolve
faster, showing sharper SFHs, initiated by a steady and rapid increase
of the SFR, that last less than 1 Gyr, reaching a peak followed by a
sharp decay. The outer regions of less massive and
later-type galaxies evolve in a slower fashion, with smoother SFHs,
initiated with a slow increase of the SFR, reaching its peak of SF at
much recent cosmological times (in some cases, they are still in the
rising phase of the SF). This, together with a narrow range of
depletion times (within a factor $\sim$2) across galaxy types,
galactocentric distances, and final stellar masses, explain the
observed local relations.

The SFHs shapes the ChEHs, that in general is governed by local
processes, with recycling, inflow, and outflow happening more at local
than at global scales. The SFHs and ChEHs shape or are shaped by the
scaling relations indicated before, that evolves with cosmological
times (in the case of the rSFMS and rMZR) or are rather constant
(SK-law). Above the kiloparsec scale, integrated/averaged through the
optical extension of galaxies, the local/resolved relations are
observed as global ones (SFMS, MZR). In some cases, depending on the
fraction of regions actually forming stars within a galaxy, the global
relations may present shallower relations than the local ones (e.g.,
SFMS vs. rSFMS). However, not all global relations are induced by
local ones. This is the case of the stellar MZR, that relates the
average [Z/H] and the stellar mass. There is no parallel
$\Sigma_*$-[Z/H], since the stellar metallicity depends both on the
amount of generated stars and the way that these stars are formed:
i.e, both the strength and shape of the SFH. Moreover, a certain time is
required before intermediate-mass stars pollute the ISM with non-alpha
(Fe-like) elements, and therefore, different SFHs leading to the same
$\Sigma_*$ imply different [Z/H] values.

The described relations shape the radial gradients observed in many
different quantities, as a pure consequence of the radial
gradient of $\Sigma_{*}$, that is formed in early cosmological times. It
seems that the early evolution is dominated by the trigger of SF
happening at the same time across galaxies, consuming the gas
reservoir that is larger in the central regions by a pure effect of
the shape of the gravitational potential. The ignition of the SF,
following a SK-law shaped by the free-fall time of collapsing
molecular clouds, automatically creates a radial gradient in
$\Sigma_{*}$. Due to the local relations indicated before and the
differential evolution of the stellar populations (formed faster and
earlier in the central regions of more massive galaxies), the observed
properties of galaxies present clear radial gradients with central
regions having older and more metal-rich stellar populations, higher rates
of SF, and more metal rich gas. In general, the evolution of galaxies
is governed by an inside-out growth during most of their life-time.

This evolution is verified for galaxies with stellar mass
M$_*>$10$^{9.5}$M$_\odot$.  Below that mass most of the regions at any galactocentric
distance are still in the rising phase of the SFHs, and therefore
their chemical enrichment is slightly different, presenting flatter
oxygen abundance gradients and even inverse stellar metallicity
gradients. If this evolution is scaled towards higher-redshifts for
more massive galaxies, it may indicate that the very early phase of
the evolution of galaxies had an outside-in growth phase, that  is
shorter in massive galaxies, and has left almost no trace in the
current fossil record. It is maybe possible that the outer regions of
massive galaxies are still in this evolutionary phase, which may explain the
flattening of some observed gradients in external regions of galaxies.
I speculate here that the transition happens in the regime in which
the infalling gas is of the same order as the original reservoir, which keeps
the SFHs in a continuous (but smooth) rising phase.

The lack of (molecular) gas is the main driver for halting the SF
activity, triggering a quenching that also evolves from the
inside-out. In other words, the central regions of more massive
galaxies, that have followed the sharper SFHs, exhaust the cold gas
earlier than the outer regions of less massive ones. In addition,
there is a decline in the star-formation efficiency,
related to dynamical effects (i.e., a shift from a pure SK-law). It
may well be that AGN feedback is the main driver for the absence of
the SF fuel, either due to physical removal or by heating beyond the
cooling time-scales. However, the dynamical effects introduced by the
bulge dynamics seem to maintain the effects of the quenching for
longer periods.

One of the main outcomes of large explorations of galaxy properties
using IFS is that most galaxies present many different ionization
processes at different locations. Thus, we should consider and study
ionization as a local process. A fraction of galaxies may present AGN
activity that could ionize (mostly) the central regions. But this does
not preclude them for presenting star-formation activity in the disk,
which produces a different kind of ionization. Central outflows could
be observed in galaxies hosting an AGN or galaxies with strong nuclear
SF activity. Even more, it is now clear that LINER-like emission is not
dominated by weak-AGNs, as commonly assumed for decades (although
ionization by weak-AGNs may also be present in some galaxies).  The
most probable reason for this kind of ionization, observed as diffuse
ionized gas, is the presence of sufficiently old stellar populations
(older than 1.6 Gyr, on average). However, in the case of
radio galaxies, shock ionization may be a significant contribution to
the ionization (i.e.  the so called Red Geysers). But they represent
less than a 1\% of the total number of galaxies.
Finally, in a very few cases, collimated synchrotron radiation
associated with radio-jets could be sufficiently energetic to ionize
the ISM (e.g., 3C120, M87). Alltogether, the description of the
ionization stage of a galaxy using only integrated properties (e.g.,
line ratios) is fundamentally wrong. It has a pure statistical
meaning, and only if there is a single ionization process that clearly
dominates over the remaining ones. A particular care should be taken
in the interpretation of the classical diagnostic diagrams, which make
little sense unless they are combined with the properties of the
underlying continuum (stellar population mostly). If not, they may
lead to fundamental mistakes.

Finally, despite recent advances, there are caveats to the emerging
picture: (i) we have a limited knowledge of the atomic and molecular
gas content at the same spatial scales explored using the most recent
IFS-GS (thus, covering a wide range of integrated properties as
stellar masses and/or morphologies). Despite the individual efforts on
a handful of targets \citep[e.g.][]{li17}, or pioneering explorations
in large samples \citep[e.g.][]{davies13,bolatto17}, there is still a
need for a comprehensive exploration of the HI and H2 distribution at
a kiloparsec scale for samples of galaxies already observed using
IFS. Current instruments, like the VLA could easily make such
exploration for the atomic gas for those samples at low redshifts
(AMUSING++, CALIFA, and the lower redshift targets of the MaNGA and
SAMI surveys). For the molecular gas, ALMA would be ideal, in
particular considering that the field-of-view of this instrument match
pretty well that of MUSE and the optical extension of the galaxies of
some of the quoted IFS-GS (e.g., CALIFA, whose diameter selection
match with the FoV of ALMA). In other cases, like MaNGA or SAMI, the
FoV of ALMA is considerably larger and therefore there is no perfect
match of the covered regime between both observations. The lack of
this dataset limits our knowledge of the gas-to-stars duty-cycle and
our understanding of the feedback and quenching processes; (ii) we
still do not know the real nature of the scaling relations. So far, we
know that they are verified at kiloparsec scale, and we also know that
they break at lower scales. Star-formation and chemical enrichment are
very local processes and at the scale of an individual \ION{H}{ii}
region these scaling relations are clearly not verified. This
indicates that they have a statistical nature, and are only verified
at certain volumes and time-scales: (a) volumes in which the
integrated molecular and stellar masses are big enough to average the
physical processes; (b) time-scales sufficiently long to compensate
the instant star-formation with the metal mixing and stellar feedback,
in different locations within the area (or volume) considered. Thus,
they are the result of physical processes that indeed do not verify
those scaling relations. IFS-GS exploring galaxies at much smaller
physical scales \citep[like PHANGS][]{roso19} may tackle this problem
in the near future. Following this reasoning, the derived SFHs and
ChEHs also have a statistical nature, reflecting the envelop evolution
of more pseudo-stochastic processes. In other words, the SFHs at lower
scales do not present the described patterns, being dominated by peaks
of star-formation that reflect the single events observed as
individual \ION{H}{ii} regions; (iii) we need to extend our
explorations to different wavelength ranges to improve our
understanding of the properties of the stellar populations and the
ionized gas. In particular, the combination of UV and FIR data with
the information provided by IFS seems to be a promising tool to
improve our estimations of the resolved SFHs and ChEHs
\citep[following][]{lopfer18}. So far we are limited by the coarse
spatial resolution of the data in these wavelength ranges; (iv) we
lack of a detailed knowledge of the AGN feedback and its connection
with the bulge growth to the extend that we could explain the
quenching using simple causal connections. We do not know whether the
effect is purely mechanical (gas removal) or thermal (gas heating), or
a combination of both and to which extent. We do not know which
frequency is required to sustain the quenching or if it is a single
event process, sustained by other processes (like bulge-induced
dynamical effects). Actually, we do not know which is the real
mechanism that makes bulges grow in time in sufficient detail; (v) we
need to extend our current IFS explorations to larger samples at
higher redshifts in order to study the evolution of the observed
quantities and to determine whether our inferences based on the fossil
records are compatible with observations.  This would require new
instruments still not foreseen or those installed in new facilities
(like the IFS at the James Webb Space Telescope); finally, (vi) an
effort should be made to develop detailed hydrodynamical and N-body
simulations, with more precise recipes of the feedback, to a
resolution good enough to compare with the current and future
dataset. Little has been done to reproduce simultaneously the
spatially-resolved and integrated scaling relations observed in
galaxies using simulations to the extend that we can understand better
the physical processes \citep[following e.g.,][]{tray19}.


In summary, the most recent results based (mainly) on IFS-GS data and
their combination with spatial resolved information of the cold gas
content, have improved our understanding of the interconnections
between star-formation, chemical enrichment and stellar evolution in
general. New patterns and scaling relations have been uncovered, in
particular at kiloparsec scales. In addition, previously known ones
have been reinterpreted, mostly global scaling relations. However,
there is still a vast number of unsolved problems that would require new
data, new techniques, and a deeper inter-exchange and cross-talk
between observational and theoretical studies. As frequently happens
in any branch of knowledge, new answers have lead to new questions that
will require further explorations in the upcoming years.



\begin{summary}[SUMMARY POINTS]
\begin{enumerate}
\item Summary point 1. Evolution of galaxies is governed by local
  properties but affected by global ones. Most scaling
  relations (like SK-law, MZR and SFMS) have local/resolved
  counterparts verified at kiloparsec scales, from which the global
  ones are integrated versions.
\item Summary point 2. Not all global relations are verified at local scales for all galaxy types and at any stellar masses (e.g., stellar MZR).
\item Summary point 3. Local scaling relations (and the lack of them) are a consequence of the narrow range of depletion times and the similarities of the local SFHs at fixed galactocentric distances of galaxies of similar final stellar mass, together with a chemical enrichment dominated by local processes.
\item Summary point 4. Radial gradients are a consequence of the local relations and the initial radial distribution of gas following the gravitational potential.
\item Summary point 5. Deviations from local relations (and gradients) are either (i) associated with dynamical and mixing processes, together with local exchange of gas (inflows, outflows, fountains), or (ii) described by diffrences in $t_{dep}$ or the local SFHs among galaxy types.
\item Summary point 6. Ionization is a local process, that may be driven by different physical processes, and it cannot be clearly understood using purely integrated quantities.
\item Summary point 7. The dominant ionization in a certain location in a galaxy is mostly dictated by the properties of the underlying stellar population (that are connected with those of the ionized gas).
  \end{enumerate}
\end{summary}

\begin{issues}[FUTURE ISSUES]
\begin{enumerate}
\item Future issue 1. How the stellar and ionized gas properties connect with those of the cold gas (atomic and molecular) at kiloparsec scales? We require  HI and H2 explorations at similar physical scales of current IFS-GS on large and statistically significant samples of galaxies already covered by those surveys.
\item Future issue 2. At which scales local relations are still verified? What this tell us about their nature? Are those statistical laws? IFS-GS expand to a smaller physical scales.
\item Future issue 3. IFS-GS should expand towards larger samples and at higher redshifts to constrain even more the evolutionary paths already uncovered by recent surveys. Covering lower mass ranges will be also fundamental.
\item Future issue 4. We need to refine our methods to recover the SFHs and ChEHs. Spatially-resolved spectroscopic information should be complemented with multiwavelength information at similar scales (that would require new UV and FIR explorers).
\item Future issue 5. Data at higher spectral resolution, deeper, and on wider wavelength ranges are required to use fainter emission lines to obtain precise information on the ionization conditions and the chemical composition of the gas.
\item Future issue 6. Which is the real quenching mechanism? Mergers? AGNs? stabilization? The involved mechanisms, in particular AGN feedback, should be understood in much more detail to be constrained better observationally.
\item Future issue 7. Larger efforts in detailed hydrodynamical and N-body simulations at kiloparsec and sub-kiloparsec scales are required to understand in detail the local/resolved scaling relations.
\end{enumerate}
\end{issues}

\appendix

\section{Integral Field Spectroscopy Galaxy Surveys}
\label{sec:IFS}

This review is focused in the recent results on the spatial resolved
properties of galaxies derived using mostly Integral field
spectroscopy (IFS).  IFS is the technique that allows us to obtain
simultaneously several spectra (within a defined field-of-view, FoV)
of a quasi-continuous region in the sky. So, the final data, after
reduction, consist either of a spatially continuous distribution of
spectra (3D cube), or a set of individual spectra arranged across the
FoV in certain fixed positions. It is beyond the scope of this review
to enter in the details of the IFS technique, to explain the
differences between the several Integral Field Units (IFUs), and to
explain the data acquisition, reduction and analysis of this
particular type of data. For these details I refer to
\citet{bershady09}, and more extensively to the book published by
R. Bacon (Optical 3D-Spectroscopy for Astronomy). I include in this
review a brief description of the technique since most of the reviewed
results are based on IFS.

IFS is not a totally new technique, with early experiments already
presented in the 80's \citep[e.g.][]{vander87}. For years it was used
to explore in detail the spatially resolved spectroscopic properties
of a single object (or a handful of them). In the particular case of
the study of galaxy properties, the stellar and gas kinematics,
together with the exploration of ionization conditions in different
regions, were more frequent uses of this technique
\citep[e.g.][]{bego97}. The computational complexity of the data
reduction and analysis hampered the development of IFS as common-user
technique for decades, excluding its use for large samples of
galaxies. Collective efforts \citep[like the Euro3D RTN,][]{euro3d},
inspiring small-group initiatives \citep[like the SAURON
project,][]{bacon01}, and the increase of the computational
capabilities along the years, allowed the implementation of IFS as a
common-user technique and extended its use as a survey mode tool.

The {\tt SAURON} project is usually considered the first attempt to
perform an IFS galaxy survey \citep{de-zeeuw02}. It explored the
central regions of a representative sample of 72 early-type galaxies
(E/S0/Sa) in the nearby universe ($<$42 Mpc) \citep{de-zeeuw02}. The
path opened by this project was continued by {\tt Atlas3D}
\citep{cappellari10}, the real first IFS-GS. It uses the same
instrument as {\tt SAURON} to study the central regions (R$<1.5$Re) of
a volume-limited sample of 260 galaxies of the same morphological type
and at the same cosmological distances. Both projects explored mostly
the properties of the stellar populations in early-type and bulges of
early-spirals \citep[e.g.][]{emsellem04,ganda:2006p3135},
studying their dynamics, stellar composition, and the ionized gas
properties \citep[e.g.][]{kuntschner10,sarzi10}. Most of the results
from both surveys were recently reviewed by \citet{cappellari16}.

\begin{table}[]
\caption{Summary of the main properties of different IFS-GS}
\label{tab:comp_IFUs}      
\begin{center}
\begin{tabular}{lcccc}        
\hline                 
Parameter & MaNGA & SAMI & CALIFA & AMUSING++ \\
\hline
Current sample size               & 6850 & 2400 & 974 & 540 \\
Selection                         & M$_*$ flat distribution& Volume limited & Diameter & Compilation \\
Redshift range                     & 0.01-0.15 & 0.01-0.1 & 0.005-0.03 & 0.001-0.1 \\
Mean redshift                      & 0.03      & 0.04     & 0.015      & 0.017 \\
  Coverage                         & 1.5 R$_e$ (2/3), 2.5 R$_e$ (1/3)& $\sim$1 R$_e$ & $>$2.5 R$_e$& $>$2 R$_e$\\
S/N at 1$r_e$ per spaxel           & 20 & 10 & 50 & 5\\
S/N at 1$r_e$ per arcsec           & 40 & 20 & 50 & 25\\
Wavelength range (\AA)             & 3600-10300 & 3700-5700/6250-7350 & 3700-7500& 4650-9300\\
Original sampling elements         & 3$\times$127$^a$ & n$\times$61$^b$ & 3$\times$331 & 90,000 \\
Spectral resolution ($\sigma$)     & 80 km/s& 75/28 km/s& 85/150 km/s&  40 km/s\\
Spatial resolution (FWHM)          & 2.5$''$ & 2.3$''$& 2.5$''$ & $\sim$0.7$''$ \\
Physical spatial res. (kpc)  & 2.5 [0.5,6.5] & 2.2 [0.4,4.6] & 0.8 [0.3,1.5] & 0.3 [0.1,1.3] \\ 
Telescope size                     & 2.5m  & 3.6m & 3.5m & 8.2m \\
\hline                                   
  Gal. in sample                 & 4655  & 910  & 2222 & 447 \\
  Gal. with morphology           & 2796  & 910  & 1062 & 343 \\
  Gal. in {\it resolved} sample  &  728  & 534  &    0$^c$ & 256 \\
\hline
\end{tabular}
\end{center}
\begin{tabnote}
  $^a$ This corresponds to the largest MaNGA bundle. Each MaNGA plate provides with 17 bundles of different amount of fibers: 2x19; 4x37; 4x61; 2x91; 5x127.
  $^b$ The dithering scheme in SAMI is repeated until a certain S/N is reached. Therefore the number of independent sampling elements is variable target to target.
  $^c$ Excluded due to the FoV of the SAMI IFU.
\end{tabnote}
\end{table}

The exploration of late type galaxies by IFS-GS started later. One
major problem that delayed these studies was the
presence of ionized gas along the disk of these objects. This involves
an additional technical complication since it is required to separate
the stellar and gas components in the analysis
\citep[e.g.][]{patri11}. Pioneering projects in the exploration of
spiral-galaxies by IFS surveys were (i) the Disk Mass Survey
\citep[DMS,][]{bershady10}; (ii) the {\tt PINGS} survey
\citep{rosales-ortega10}; (iii) the {\tt VENGA} project
\citep{blanc10}; and (iv) the CALIFA pilot survey
\citep{marmol-queralto11}. Most of them were focused on the study of
the ionized gas, and, in some cases, of the stellar populations, more
than on the stellar/gas dynamics. I do not discuss in here
explorations at high-redshift that are beyond the scope of this
review.

None of those pioneering surveys explored a wide range of galaxy
properties, including all morphological types and/or covering a wide
mass/luminosity/color regime, and in an statistically large sample of
galaxies (e.g., {\tt Atlas3D} observed only E/S0 and red
early-spirals).  The first IFS-GS fully representative of the
population of galaxies in the near universe was the CALIFA survey,
started in 2010 \citep[][]{sanchez12}. Moreover, it was the first
IFS-GS designed as a legacy survey, i.e., with the main and primary
goal to deliver the data freely to the community, easily accessible,
and fully documented. Soon after, there were started the two major
on-going IFS-GS, following the same phylosophy: MaNGA
\citep[][]{manga} and SAMI \citep[][]{sami}. It is beyond the scope of
this review to make a detailed comparison between them \citep[for a
referencee read ][]{2015IAUS..309...85S,sanchez17a}.  I briefly
discuss of their main properties, listed in Table \ref{tab:comp_IFUs}.
From them, it is appreciated that CALIFA offers the best
compromise between spatial resolution, covered fraction of the galaxy
extension, and number of sampled elements. However, this comes at a
cost, sampling much lower number of galaxies than MaNGA (1/10th) and
SAMI (1/3th). Thus, for detailed analysis of the spatial resolved
properties CALIFA may present an advantage. However, for galaxy
statistics, MaNGA is far more superior. It also offers a wider
wavelength range and a slightly better spectral resolution. Finally,
it is worth noticing that although SAMI offers the smaller FoV (and
covered area of each galaxy), it has by far the highest spectral
resolution (in the red wavelength range). Thus, for kinematics (in
particular gas one) it provides with more precise estimations. In
general the three IFS-GS are pretty complementary, compensating
somehow the limitations and strengths of each other.

\begin{figure}[h]
\includegraphics[trim=5 25 3 25,clip,width=5in]{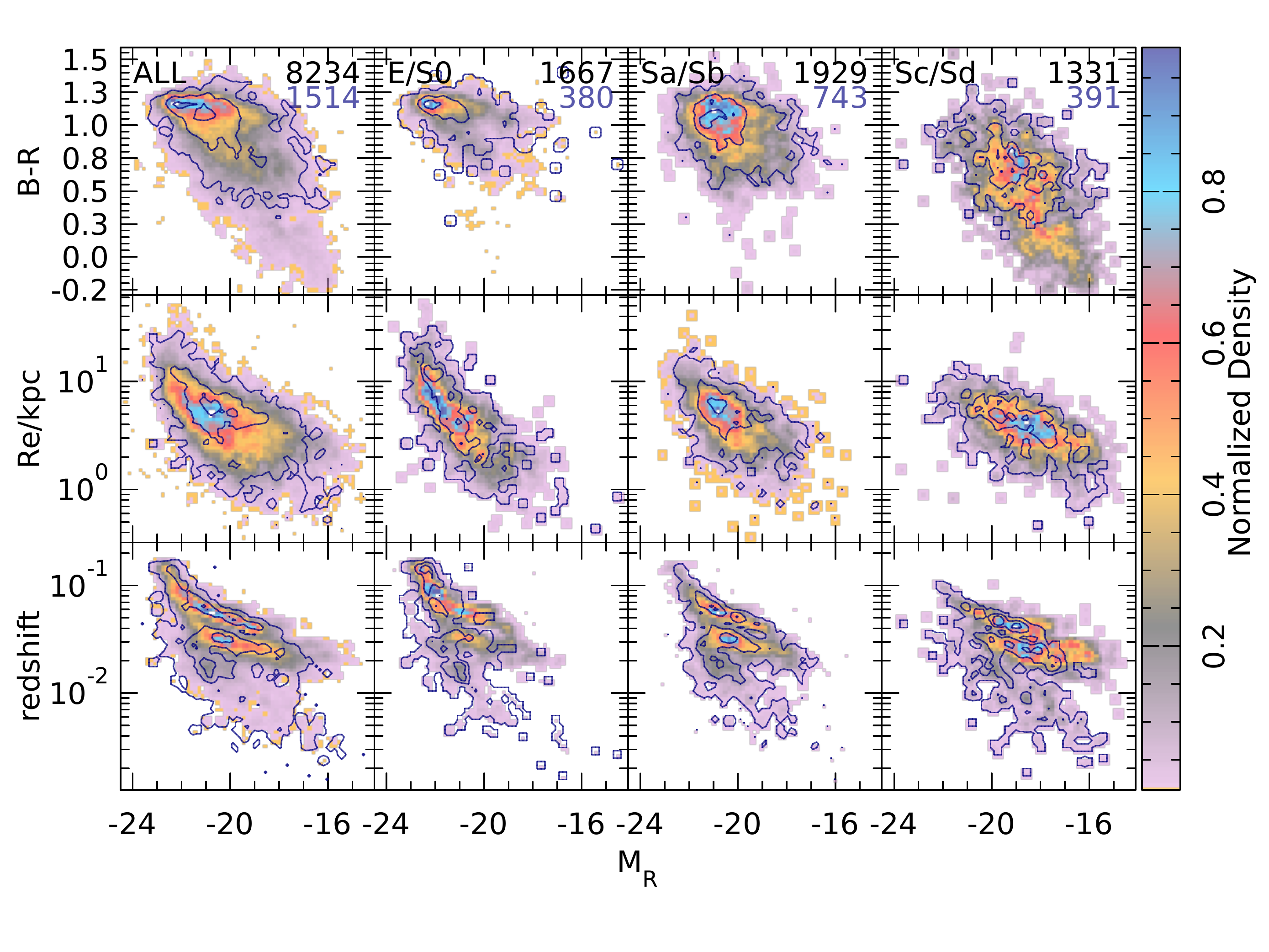}
\caption{Distribution colors ($B-R$, {\it top panels}), effective radii ({\it central panels}), and redshifts ({\it bottom panels}) along the $R$-band absolute magnitudes for the collection of galaxies with IFS observations used in this review. {\it Left hand panels} show the distribution for the full sample, and each consecutive panel from left to right show similar distributions segregated by morphology (with the morphology indicated in the top panels). Color-maps show the density distribution normalized to the peak density for the full compilation, while contours show the same distribution for the sub-sample of {\it well resolved} galaxies (see the text). Each contour includes a 90\%, 50\% and 10\% of the points. Finally, the number of galaxies included panel is indicated in the top panels, in black for the full sample and in color for the {\it well resolved} one.}
\label{fig:SAMPLE}
\end{figure}

In addition to this three major IFS-GS I included in
Tab. \ref{tab:comp_IFUs} a large compilation of IFS observations of
galaxies in the nearby universe performed with MUSE \citep{muse}. This
instrument is by far the IFS that offers the best compromise between
spatial sampling, resolution, FoV, spectral resolution and collecting
power.  It presents the only drawback of a limited wavelength range in
the blue optical range (cut at 4650\AA), what strongly affects any
analysis of the stellar populations at low redshift. The current
compilation, named {\tt AMUSING++} (Lopez-Coba et al. in prep.),
involves a diameter selection (1 R$_e <$FoV$<$ 3R$_e$) of galaxies
included in the ESO archive corresponding to different projects
\citep[e.g., MAD, GASP, TIMER, PHANGS, and
AMUSING,][]{erroz19,GASP,TIMER,kreckel17,galbany16b}. Two thirds of
the objects are extracted from the {\tt AMUSING} survey (PI:
J. Anderson), that was already used in different studies
\citep[e.g.][]{censushii,laura16,carlos17,laura18,galbany18}. It is by
far the largest compilation of IFU data with the best possible spatial
resolution. It was included here for completeness and to highlight
what would be achieved in the near future when such new instruments would
be used in a survey mode.

I compiled all the public available data from all those IFS-GS (and
compilations) to reproduce the main results reviewed along this
article (as indicated in Sec. \ref{sec:data}). This compilation
comprises a total number of 8234 galaxies (and datacubes). This way I
expect to minimize the possible biases introduced by the
particularities of sample selection and instrumental details of each
IFS-GS. The number of galaxies corresponding to each survey included
in this compilation are listed in Tab. \ref{tab:comp_IFUs}. In
addition, I obtained the morphological classification for 2/3rd of
them, using public information by different sources \citep[Hyperleda
for MUSE,][Cortesse private communications for
SAMI]{walcher14,sanchez18}. For the particular purpose of this review
I selected the best resolved and sampled galaxies, as indicated in
Sec. \ref{sec:data} (i.e., the {\it well resolved} sub-sample).

It is clear that the current compilation does not comprise a well
defined sample of galaxies (in principle). However, it is broadly
representative of the general population of galaxies in the considered
redshift range, since it was created from samples that are indeed
representatives themselves. Figure \ref{fig:SAMPLE} shows the
distribution of colors ($B-R$), effective radii and redshifts along
the $R$-band absolute magnitudes for the full compilation of galaxies
(and segregated by morphology).
Features as
the red-sequence (for E/S0) and the blue-cloud (for Sc/Sd) are nicely
recovered. Indeed, the morphological distribution is similar to the
one recovered by volume limited surveys, with a $\sim$33\%\ of
early-type galaxies (E/S0), a $\sim$40\% of early-spirals (Sa/Sb) and
a $\sim$27\% of late-spirals (Sc/Sd) \citep[e.g.][]{bamford+2009}. The
fraction of SFGs ($\sim$50\%), RGs ($\sim$46\%), and AGN hosts
($\sim$4\%) is also consistent with the most recent estimations at the
considered redshift range \citep[e.g.][, and references therein]{mariana16}. The
distribution along the R$_e$-M$_{abs,R}$ plane follows the expected trends
for both the full population and the different morphological types
\citep[e.g.][]{cappellari16}. Finally, the redshift-M$_{abs,R}$
distribution (and the derived redshift-R$_e$ one) resembles that of a
diameter-selected sample. This may or may not be the primarily
selection criteria of the different IFS-GS included in this
compilation, but all of them seek that their targets fit somehow
within the FoV of their IFUs. In this later diagram the two
distinctive distributions correspond to the CALIFA/AMUSSING++ and the
MaNGA/SAMI subsamples, that cover two clearly different redshift
ranges in general (see
Tab. \ref{tab:comp_IFUs}).

Fig. \ref{fig:SAMPLE} shows also the distributions for the {\it well
  resolved} sub-sample (defined in Sec. \ref{sec:data}). In general it
presents a similar distribution than the full compilation for the
parameters analyzed here. The strongest difference is the cut at
low-luminosity and blue-colors, that corresponds in general to the
tail towards low-mass galaxies presents in the SAMI sample, not well
covered by other IFS-GS \citep[e.g.][]{sanchez17a}. However, excluding
dwarf galaxies (not included in the current review), the two samples
cover a similar range of parameters. Moreover, both of them seem to be
representative of the global population of galaxies in the nearby
universe (at a first order). As expected, the number of galaxies in
the considered compilation is clearly dominated by the publicly
available data from the MaNGA IFS-GS \citep[DR15,][]{sdss_dr15},
comprising $\sim$57\%\ of them (Tab. \ref{tab:comp_IFUs}). However,
only half of them have publicly available morphology, and just
$\sim$15\%\ fulfil our severe criteria for the {\it well resolved}
sub-sample. These cuts exclude all the SAMI data too, since I impose a
minimum galaxy size larger than the FoV of their IFU. These are
somehow arbitrary criteria, aimed to recover just the best sampled and
resolved IFS data, and to homogenize (somehow) the different
datasets. They do not imply any judgement on the quality of the
different IFS-GS datasets considering their different goals and
scopes. Further details of the compilation and its main global
properties would be provided elsewhere (S\'anchez et al. in prep.).





\section{Ionized gas: A practical classification scheme}
\label{sec:class}

Based on the different results discussed along this review I present
the following practical classification scheme to distinguish between
different kinds of ionizing sources, in particular for IFS data
at $\sim$1 kpc resolution:

\begin{marginnote}[]
\entry{\HII\ region}{a gas cloud composed mostly by hydrogen ionized by young and hot stars.}
\end{marginnote}
\begin{itemize}

\item {\bf A star-forming or \HII\ region:} It would be a clumpy area
  (clustered) in a galaxy in which the ionized gas emission line
  ratios are below the \citet{kewley01} demarcation line in at least
  one of the classical diagnostic diagrams involving
  [\ION{O}{iii}/H$\beta$] vs. [\ION{N}{ii}]/H$\alpha$,
  [\ION{S}{ii}]/H$\alpha$ or [\ION{O}{i}/H$\alpha$], with
  EW(H$\alpha$) above 6\AA, and with a fraction of young stars
  ($age<100Myr$) in flux in the visual-band of at least a 4-10\%.

\begin{marginnote}[]
\entry{AGN}{active galactic nucleus, a compact region in the center of a galaxy, powered by the gas accreation to a super-massive black hole, with a blue and hard ionizing spectrum.}
\end{marginnote}
\item {\bf An AGN ionized region:} It would be that central ionized
  region, clearly more intense than the diffuse ionized gas, in which the emission
  line ratios are above the \citet{kewley01} demarcation line in the
  three diagrams indicated before, and with EW(H$\alpha$) above
  3\AA. Below that limit it is not possible to determine whether the
  ionization is due to an AGN or to other processes. The line ratios
  decreases with respect to the center of the galaxy, and the flux decays
  at r$^{-2}$ or faster (nor following the light surface-brightness of the continuum).

\item {\bf Diffuse gas ionized by old stars (HOLMES, post-AGBs):} It
  is that smooth ionized structure that follows the light distribution
  of the old stellar population in galaxies, presenting an
  EW(H$\alpha$) clearly below 3\AA. The fraction of young stars for
  the underlying stellar population is never larger than a 4\%. It
  does not present any clumpy or filamentary distribution. In the BPT
  diagram could be located covering the LINER-like region (or AGN locus) towards the
  location of metal-rich star-forming regions. It is present in
  galaxies with old stellar populations (massive, earlier types) or
  regions in galaxies with the same characteristics (bulges), and it
  shares the same kinematic structure as the old stellar
  population.

\begin{marginnote}[]
  \entry{DIG}{diffuse ionized gas, a low intensity and smooth gas spread through most of the optical
  extension of galaxies.}
\end{marginnote}

\item {\bf Diffuse gas due to photon-leaking by \hii\ regions:} They are
  smooth ionized structures present in galaxies with young stellar
  populations (in general, low mass and late-type galaxies) or regions
  in galaxies with the same characteristics (disks), presenting an
  EW(H$\alpha$) below 3\AA, with fraction of young stars never larger
  than a 4\%. It shares the same location of classical \hii\ regions in
  the diagnostic diagrams, and should be included in the photon budget
  to derive the SFR in galaxies. Its kinematics is not
  fundamentally different of that of the disk. In high spatial
  resolution data (10-100 pc) it may present some shells or
  bubble-like structures, not visible at kpc scales.

\item {\bf A high-velocity shock ionized region:} It would be a
  filamentary or biconical ionized gas structure, with an intensity
  stronger than that of the diffuse ionized gas, in which the emission
  line ratios are above the indicated demarcation line in the three
  diagrams and with an EW(H$\alpha$) above 3\AA, that shows
  asymmetrical lines extended through the filamentary structure. It
  presents a clear increase of the considered line ratios (in
  particular [\ION{O}{i}/H$\alpha$]) with the velocity dispersion and
  the distance from the source of the outflow (the central region in
  the case of star-forming or AGN host galaxies). AGN and star-forming
  driven outflows could be separated, in principle, using the
  demarcation line proposed by \citet{bland95}.

\item {\bf A low-velocity shock ionized region:} It shares many of the
  characteristics of the diffuse gas ionized by old stars, and indeed
  it is considered by different authors as diffuse
  \citep[e.g.][]{dopita96,monreal10}. However, it presents a clear
  filamentary structure, and a velocity distribution not following the
  general rotational pattern of the galaxy. Recent proposed red-geysers most
  probably are ionized by this kind of process \citep{kehrig12,cheung18}.
\begin{marginnote}[]
\entry{Red Geyser}{elliptical galaxy with filamentary ionized gas most probably due to shocks, but not clearly related with high-speed outflows.}
\end{marginnote}

\end{itemize}

Other sources of ionization, like super-nova remnants or planetary
nebulae, are in general not resolved at kiloparsec scales, and they
are sources of contamination that may complicate the classification
even more. Moreover, in many cases it is not easy to clearly define
which kind of region is observed, when different types are on the same
aperture/line of sight. For a recent review on the properties of the
ISM and how to recover them from emission lines I refer to
\citet{kewley19}.

\section*{DISCLOSURE STATEMENT}
The author is not aware of any affiliations, memberships, funding, or financial holdings that
might be perceived as affecting the objectivity of this review.

\section*{ACKNOWLEDGMENTS}

I thank my wife, K. Hansen, for allowing me to work in this review when I should not.

I am grateful for their comments and corrections on the content of
this review to E. P\'erez, L. S\'anchez-Menguiano, 
P. S\'anchez-Blazquez, C.J. Walcher, R.C. Kennicutt,
J.K. Barrera-Ballesteros, C. Morisset, B. Hussemann, A. Bolatto, R. Cid-Fernandes, E. Lacerda, A. Mejia and C. L\'opez-Cob\'a.

I thank M.M. Roth and L. Wisotski for recovering me for astronomy.  It
is an honor to have such friends. I thank the full CALIFA
collaboration for their incredible work along the years, and in
particular to D. Mast, B. Husemann and R. Garcia-Benito. The CALIFA
survey was my personal excuse to justify being around of an
exceptional group of astronomers from whom I learned and developed most of
what it is included in this review. I also thank too the members of
the MaNGA, SAMI and AMUSING collaborations for allowing me to continue
developing new ideas with new amazing datasets and incredible
people. I would like to thanks the EDGE collaboration and in
particular A. Bolatto, for opening a new branch of opportunities with
the inclusion of spatial resolve CO data. Finally, I thank all the
Calar Alto Observatory staff that allowed me to develop me
professionally and personally. When I am there, I know I am at home.

I used along this review data from the CALIFA, SAMI and MaNGA IFS-GS.
I also use data from the MUSE instrument installed at the ESO VLT facility.

I am grateful for the support of a CONACYT grant CB-285080 and
FC-2016-01-1916, and funding from the PAPIIT-DGAPA-IA101217 (UNAM)
project.

\bibliographystyle{ar-style2}
\bibliography{ARAA}


\end{document}